\documentclass[12pt]{article}
\usepackage{setspace}
\RequirePackage[vmarginratio=2:3, outer=24mm, inner=24mm, lines=26]{geometry}
\usepackage{amssymb, amsmath, amsthm}
\usepackage{graphicx}
\usepackage{lscape}
\usepackage{multirow}

\newcommand{\argmin}{\operatorname*{arg \ min}}
\usepackage{natbib}
\usepackage{}
\usepackage{amsmath}
\usepackage[table]{xcolor}
\usepackage{algorithm2e}
\RestyleAlgo{boxruled}
\usepackage{float}
\usepackage{xr-hyper}
\usepackage{hyperref}

\newcolumntype{L}{>$l<$}
\theoremstyle{plain}

\newtheorem{theorem}{Theorem}
\newtheorem{prop}{Proposition}
\newtheorem{remark}{Remark}
\newtheorem{condition}{Condition}

\newtheorem{lemma}{Lemma}
\newtheorem{corol}{Corollary}

\newtheorem*{McD*}{McDiarmid's Inequality}

\title{A likelihood-based approach for multivariate categorical response regression in high dimensions}
\author{Aaron J. Molstad\footnote{Correspondence: amolstad@ufl.edu}\hspace{3pt} and Adam J. Rothman$^\dagger$\\
Department of Statistics and Genetics Institute, University of Florida$^*$\\
School of Statistics, University of Minnesota$^\dagger$\vspace{-10pt}
\bigskip}
\date{}
\begin{document}
\maketitle
\vspace{-20pt}
\begin{abstract}
We propose a penalized likelihood method to fit the bivariate categorical response regression model. 
Our method allows practitioners to estimate which predictors are irrelevant, which predictors only affect the marginal distributions of the bivariate response, and which predictors affect both the marginal distributions and log odds ratios. 
To compute our estimator, we propose an efficient algorithm which we extend to settings 
where some subjects have only one response variable measured, i.e., a semi-supervised setting. 
We derive an asymptotic error bound which illustrates the performance of our estimator in high-dimensional settings.
Generalizations to the multivariate categorical response regression model are proposed. 
Finally, simulation studies and an application in pan-cancer risk prediction demonstrate the 
usefulness of our method in terms of interpretability and prediction accuracy.\medskip

\end{abstract}
\doublespacing
\vspace{-15pt}
\section{Introduction}

In many regression applications, 
the response is multivariate.  If all of the
components of the response are numerical, then the 
standard multivariate response linear regression
model can be used. If some response components are categorical, then
it is unclear what should be done. In this article, we develop a method for multivariate response
regression when all of the components of the response are categorical.
For example, given the gene expression profile of a patient with cancer originating in the kidney, a practitioner may want to predict both the cancer type (chromophobe, renal clear cell carcinoma, or renal papillary cell carcinoma) and five-year mortality risk (high or low). To simplify matters, we will focus on the bivariate categorical response regression model, but as discussed in a later section, our developments can be generalized to settings with arbitrarily many categorical response variables. 
\vspace{-10pt}

\subsection{Bivariate categorical response regression model}
Let $(Y_1,Y_2|x)$ be the random bivariate categorical response
with numerically-coded support $\{1,\ldots, J\}\times \{1,\ldots, K\}$
when the explanatory variables have values in the vector 
$x \in \mathbb{R}^{p}$ with its first entry set to one.
Existing work on this problem proposed and analyzed 
links between $x$ and the multivariate distribution of the response \citep{mccullaghnelder89, glonek1995multivariate}.  
For reasons to be discussed, we consider the simple link defined by
\begin{equation}\label{eq:full_model}
P(Y_1 = j, Y_2 = k|x) = \frac{\exp(x'\boldsymbol{\beta}^*_{:,j,k})}{\sum_{s=1}^J \sum_{t=1}^{K}  \exp(x'\boldsymbol{\beta}^*_{:,s,t})}, \quad (j,k)\in\{1,\ldots, J\}\times\{1,\ldots, K\}, 
\end{equation}
where $\boldsymbol{\beta}^{*} \in \mathbb{R}^{p \times J \times K}$ 
is the three-way tensor (three-dimensional array) of unknown regression coefficients
and $\boldsymbol{\beta}^*_{:, j,k} \in \mathbb{R}^{p}$ 
is the regression coefficient vector corresponding to the 
response category pair $(Y_1 = j, Y_2 = k)$. 
This model can be expressed as a univariate multinomial logistic regression model
for the categorical response $(\tilde Y|x)$ 
where $\tilde{Y}$ has numerically coded support $\left\{1, \dots, JK\right\}$ so that
\begin{equation}
P(\tilde{Y} = f(j,k) \mid x) = P(Y_1 = j, Y_2 = k\mid x), 
\quad (j,k)\in\{1,\ldots, J\}\times \{1,\ldots, K\},
\end{equation}
where $f(j,k) = (k - 1)J + j$. To simplify notation, for the remainder of this article we will use $[n]$ to denote the set $\{1, \dots, n\}$ for all $n \in \mathbb{N}$, and will use $0_n$ $(0_{n \times n}$) to denote an $n$-dimensional vector ($n \times n$ matrix) of zeroes.
 
Many methods exist for penalized (univariate response) multinomial logistic regression. For example, 
\citet{zhu2004classification} proposed a ridge-penalized multinomial logistic regression model, and later, \citet{vincent2014sparse} proposed to use a sparse group lasso penalty on rows of the unknown regression coefficient matrix. The latter approach allows for variable selection 
since $\boldsymbol{\beta}^*_{m, :, :} = 0_{J\times K}$ implies that the $m$th predictor does not affect the response category probabilities. \citet{simon2013sparse} studied the sparse group lasso from a computational perspective: the multinomial logistic regression model fits neatly into their framework. 

Other recent methods for fitting the multinomial logistic regression model rely on dimension reduction rather than variable selection.  \citet{powers2018nuclear} proposed a nuclear norm penalized multinomial logistic regression model, which could be characterized as a generalization of the stereotype model of \citet{anderson1984regression}.
\citet{Price2019Auto} penalized the Euclidean norm of pairwise differences of regression coefficient vectors for each category, which encourages fitted models for which estimated probabilities are identical for some categories.  

While these methods can perform well in terms of prediction and interpretability for multinomial logistic regression, if applied to the multivariate categorical regression model, none would account for the fact that $\tilde{Y}$ is constructed using two distinct response variables. One could fit two separate multinomial logistic regression models,
but this would fail to exploit the association between the two responses.
Thus, there is a need to develop a new penalized likelihood framework for fitting 
\eqref{eq:full_model} that exploits the multivariate response.  Our
proposed method does this, yields interpretable fitted models, and can be applied when $p$ is large.

\vspace{-10pt}

\subsection{Parsimonious parametric restrictions} 
\label{parsimonious-structure}

We propose two parametric restrictions to reduce the number of parameters 
in \eqref{eq:full_model} and incorporate the special structure of the bivariate response. The first assumes that only a subset of the predictors are relevant
in the model. Specifically, if $\boldsymbol{\beta}^*_{m, :, :} = b 1_{J \times K}$ for any constant $b \in \mathbb{R}$ and $J \times K$ matrix of ones $1_{J \times K}$, then a change in the $m$th predictor's value does not affect the response's joint probability mass function, i.e., the $m$th predictor is irrelevant. By setting $b = 0$, it is immediate to see that imposing sparsity of the form $\hat{\boldsymbol{\beta}}_{m,:,:} = 0_{J \times K}$, where $\hat{\boldsymbol{\beta}}$ is an estimator of $\boldsymbol{\beta}^*$, is a natural way to achieve variable selection of this kind. 
This restriction may be helpful when there are many predictors.

The second restriction we consider is that a subset of the predictors can only affect the 
two marginal distributions of the response: $(Y_1 \mid x)$ and $(Y_2 \mid x)$.  Specifically, 
the joint distribution of the response $(Y_1, Y_2 \mid x)$ is determined by its 
$(J-1)(K-1)$ local odds ratios:
\begin{equation} \label{eqoddsratios}
\frac{P(Y_1=j,Y_2=k \mid x) P(Y_1=j+1, Y_2 = k+1 \mid x)}{P(Y_1=j,Y_2=k+1 \mid x) P(Y_1=j+1, Y_2=k \mid x)},
\quad (j,k) \in [J-1] \times [K-1]
\end{equation}
and its two marginal distributions $(Y_1 \mid x)$ and $(Y_2 \mid x)$ \citep{agresti02}.  
We suppose that changes to a subset of the entries in $x$ do not affect the 
odds ratios in \eqref{eqoddsratios}, so they can 
only affect the marginal distributions of the response (or be irrelevant).

Suppose, for the moment, that $J=2$ and $K=2$. The log odds ratio is then
$$ 
\log  \left\{ \frac{P(Y_1=1,Y_2=1 \mid x) P(Y_1=2, Y_2 = 2 \mid x)}{P(Y_1=1,Y_2=2 \mid x) P(Y_1=2, Y_2=1 \mid x)} \right\} = x'(\boldsymbol{\beta}^*_{:, 1,1} + \boldsymbol{\beta}^*_{:, 2,2} - \boldsymbol{\beta}^*_{ :, 1,2} - \boldsymbol{\beta}^*_{ :, 2, 1}).
$$
If the $m$th element of the vector $\boldsymbol{\beta}^*_{ :, 1,1} + \boldsymbol{\beta}^*_{ :, 2,2} - \boldsymbol{\beta}^*_{:, 1,2} - \boldsymbol{\beta}^*_{ :, 2, 1}$ were zero, then changes to
the $m$th element of $x$ would not affect the odds ratio, so the $m$th predictor can only affect
the marginal distributions of the response.  
Let $\mathcal{D} = (1, -1, -1, 1)'$ and let $\beta^* \in \mathbb{R}^{p \times JK}$ be the matricized version 
of $\boldsymbol{\beta}^*$ with $\boldsymbol{\beta}^*_{m,j,k} = \beta^*_{m,f(j,k)},$ for all $(m,j,k) \in [p] \times [J] \times [K].$ Similarly, let $\beta^*_{m,:} \in \mathbb{R}^{JK}$ be the $m$th row of $\beta^*$ for all $m \in [p].$
One can see that if $\mathcal{D}'\beta^*_{m,:} = 0$, the $m$th predictor can only affect the marginal distributions of the 
response since the log odds ratio
\vspace{-10pt}
$$ x'(\boldsymbol{\beta}^*_{:, 1,1} + \boldsymbol{\beta}^*_{:, 2,2} - \boldsymbol{\beta}^*_{:, 1,2} - \boldsymbol{\beta}^*_{:, 2, 1}) = x'\beta^* \mathcal{D},
\vspace{-10pt}
$$
is not be affected by changes in the $m$th component of $x$ for all $x\in\mathbb{R}^{p}$.

When $J>2$ or $K>2$, we can express the logarithm of all 
of the local odds ratios in \eqref{eqoddsratios} 
in terms of 
a constraint matrix $\mathcal{D} \in \mathbb{R}^{JK \times (J-1)(K-1)}$. 
For example, when $J=3$ and $K=2$,
\vspace{-10pt}
$$   
\mathcal{D}' = \left( \begin{array}{r r r r r r}
1  &- 1 &0 &-1 &1& 0\\
1  &0 &-1 &-1 & 0& 1
\end{array}\right), $$ 
$$\beta^* \mathcal{D} = \left(
 \boldsymbol{\beta}^*_{:,1,1} - \boldsymbol{\beta}^*_{:,2,1} - \boldsymbol{\beta}^*_{:,1,2}  + \boldsymbol{\beta}^*_{:,2,2},  \hspace{2pt} \boldsymbol{\beta}^*_{:,1,1} - \boldsymbol{\beta}^*_{:,3,1} - \boldsymbol{\beta}^*_{:,1,2}  + \boldsymbol{\beta}^*_{:,3,2}\right),
$$
\noindent and the vector of $(J-1)(K-1)$ local 
log odds ratios is $(\beta^* \mathcal{D})'x$.
If the vector $\mathcal{D}'\beta^*_{m,:} = 0_{(J-1)(K-1)}$, then the $m$th predictor 
can only affect the marginal distributions of the response.

We propose to fit the model \eqref{eq:full_model} 
by penalized likelihood.
We add a group lasso penalty  
that is non-differentiable when the optimization
variable, $\beta \in \mathbb{R}^{p \times JK}$, is such that 
$\mathcal{D}'\beta_{m,:} = 0_{(J-1)(K-1)}$ for $m \in\{2,\ldots, p\}$.  This encourages estimates for which 
$\mathcal{D}'\hat\beta_{m,:} = 0_{(J-1)(K-1)}$ for some $m\in\{2,\ldots, p\}$, so that some predictors are estimated to only affect the marginal distributions of the response.  
We also add a second group lasso penalty that is non-differentiable when
 $\beta_{m,:} = 0_{JK}$ for $m\in\{2,\ldots, p\}$.
This has the effect of removing predictors from the model entirely.

\vspace{-10pt}

\subsection{Alternative parameterizations}
Alternative parameterizations of \eqref{eq:full_model} could be used to relate predictors to response variables.
For example, when $J=K=2$, \citet{mccullaghnelder89} 
proposed the following:
$$x'\eta_{a} = \log\left\{ \frac{P(Y_1=1 \mid x)}{P(Y_1=2 \mid x)}\right\}, \quad x'\eta_{b}  = \log\left\{ \frac{P(Y_2=1 \mid x)}{P(Y_2=2 \mid x)}\right\}, $$
\begin{equation}
x'\eta_{c}  = \log  \left\{ \frac{P(Y_1=1,Y_2=1 \mid x) P(Y_1=2, Y_2 = 2 \mid x)}{P(Y_1=1,Y_2=2 \mid x) P(Y_1=2, Y_2=1 \mid x)}  \right\},\label{ML-alt}
\end{equation}
where $\eta_{a}\in\mathbb{R}^p$, $\eta_{b}\in\mathbb{R}^p$, 
and $\eta_{c}\in\mathbb{R}^p$ are unknown coefficient vectors.
This parameterization and generalizations were also discussed by \citet{glonek1995multivariate}.
We found penalized likelihood optimization based on \eqref{ML-alt} to be much more difficult than our method based on \eqref{eq:full_model}. See \citet{qaqish2006multivariate} for more on the computational challenges involved with parameterizations like \eqref{ML-alt}. 
In addition, the parameterization \eqref{eq:full_model} allows us to establish asymptotic properties our estimator using results from \citet{bach2010self}. Nonetheless, we consider penalized likelihood methods for estimating $\eta_a, \eta_b$, and $\eta_c$ a promising direction for future research. 

\vspace{-10pt}

\subsection{Multi-label classification methods}
The problem of fitting multivariate categorical response regression models is closely related to the problem of ``multi-label'' classification. In the computer science literature, multi-label classification refers to the task of predicting many categorical (most often, binary) response variables from a common set of predictors. One of the most popular class of methods for fitting the multivariate binary response regression model is the so-called ``binary relevance'' approach, which effectively fits a separate model for each of the categorical response variables. Loss functions used for fitting these models, however, often take the classification accuracy on all response variables into account jointly, so these methods do account for the multivariate response. For a comprehensive review of binary relevance, see \citet{zhang2018binary}.

There are many extensions of the binary relevance approach which account for dependence between the response variables \citep{montanes2014dependent}. One such approach is based on ``classifier chains'' \citep{read2009classifier}, which can be described as fitting successive (univariate) categorical response regression models where in each successive model, the categorical responses from the previous models are included as predictors. For example, in the bivariate categorical response case, one would fit two models in sequence:
${\rm (i)} (Y_1\mid x)$ and  ${\rm (ii)} (Y_2 \mid x, Y_1),$
so that one could then predict $Y_1$ from some new value of $x$, say $\tilde{x}$, using (i) and then predict $Y_2$ from $\tilde{x}$ and the predicted value of $Y_1$ using (ii). ``Nested stacking'' is a similar approach, except when fitting (ii), replaces $Y_1$ with its predicted value from (i) \citep{senge2013rectifying}.

Many methods other than those based on binary relevance exist, e.g., see the review paper by \citet{tsoumakas2007multi}. However, in general, these methods are often not model-based, nor is the focus of these methods both prediction accuracy and interpretability of fitted models, as is the focus of our proposed methodology. 
\vspace{-10pt}

\section{Penalized likelihood for bivariate categorical response regression}
\label{sec:PenMaxLik}
We assume that we have observed
the result of $n$ independent multinomial experiments.  
Let $x_i = (1,x_{i2},\ldots, x_{ip})'\in\mathbb{R}^{p}$
be the values of the explanatory variables for the $i$th subject
and let 
$$
y_i=
\left( \begin{array}{ccc}
y_{i,1,1} & \cdots & y_{i,1,K} \\
\vdots & \ddots & \vdots\\
y_{i,J,1} & \cdots & y_{i,J,K}\\
\end{array} \right) \in\mathbb{R}^{J\times K}
$$
be the $i$th subject's observed response
category counts for each $i \in [n]$.
The subjects model assumes that ${\rm vec}(y_i)$ is a realization of
\vspace{-10pt}
\begin{equation}\label{eq:dataGen}
{\rm vec}(\mathcal{Y}_i) \sim {\rm Multinomial}\{n_i, \pi^*_{1,1}(x_i),\pi^*_{2,1}(x_i),\ldots, \pi^*_{J,K}(x_i)\},
\vspace{-5pt}
\end{equation}
where $n_i$ is the number of trials for the $i$th subject and 
$$
\pi^*_{j,k}(x_i) =  \frac{\exp(x_i'\boldsymbol{\beta}^*_{:, j, k})}{\sum_{s=1}^J \sum_{t=1}^{K}  \exp(x_i'\boldsymbol{\beta}^*_{:, s, t})}, \quad (i,j,k)\in [n] \times [J]\times [K].
$$
Ignoring constants, the (scaled by $1/n$) negative log-likelihood function evaluated at $\boldsymbol{\beta} \in \mathbb{R}^{p \times J \times K}$ is
\begin{align*}
\mathcal{G}(\boldsymbol{\beta}) & = - \frac{1}{n} \sum_{i=1}^n \left[ \sum_{j=1}^J \sum_{k=1}^K y_{i,j,k} (x_{i}'\boldsymbol{\beta}_{:, j,k}) 
-   n_i \log\left\{ \sum_{s=1}^J \sum_{t=1}^{K}  \exp\left(x_i'\boldsymbol{\beta}_{:, s, t}\right) \right\} \right].
\end{align*}
Without loss of generality, we set $n_i = 1$ for all $i\in[n]$.

To discover the parsimonious structure described in Section \ref{parsimonious-structure}, 
we propose the penalized maximum likelihood estimator 
\begin{equation}
\argmin_{\beta \in \mathbb{R}^{p\times JK}} \left\{ \mathcal{G}(\boldsymbol{\beta}) + \lambda \sum_{m=2}^{p} \|D'\beta_{m,:}\|_{2} + \gamma \sum_{m=2}^{p} \|\beta_{m,:}\|_{2} \right\}, \label{eq:estimator}
\end{equation}
where $(\lambda, \gamma) \in (0, \infty) \times (0, \infty)$ are user-specified tuning parameters; 
and $\|\cdot\|_2$ is the Euclidean norm of a vector. As $\lambda \to \infty$, the estimator in \eqref{eq:estimator} becomes equivalent to fitting separate multinomial logistic regression models to each of the categorical response variables. Conversely, as $\lambda \to 0$, \eqref{eq:estimator} tends towards the group lasso penalized multinomial logistic regression estimator for response $\tilde{Y}$. Throughout, $\hat\beta$ will be used to denote \eqref{eq:estimator}. 

The matrix $D$ used in \eqref{eq:estimator} is distinct from the matrix $\mathcal{D}$ introduced in Section \ref{parsimonious-structure}. Specifically, the matrix $D \in \mathbb{R}^{JK \times \binom{J}{2}\binom{K}{2}}$ is constructed by appending additional, linearly dependent columns to $\mathcal{D}$ so that all $\binom{J}{2}\binom{K}{2}$ 
log odds ratios are penalized. If we had instead used the matrix $\mathcal{D}$ in \eqref{eq:estimator}, our estimator would depend on which $(J-1)(K-1)$ log odds ratios the columns of $\mathcal{D}$ correspond to. Thus, using the matrix $D$ avoids this issue and penalizes all possible log odds ratios equivalently. In the case that $J=K =2$, it is trivial to see $\mathcal{D} = D$. In the case that $J = 3$ and $K=2$, for example, the additional column of $D$ would be the second column minus the first column of $\mathcal{D}$. Using this matrix $D$, $D'\beta_{m, :} = 0_{\binom{J}{2}\binom{K}{2}}$ implies $ \mathcal{D}'\beta_{m, :} = 0_{\binom{J}{2}\binom{K}{2}}$, 
and thus, $D'\beta_{m, :} = 0_{\binom{J}{2}\binom{K}{2}}$ implies that the $m$th predictor can only affect the 
marginal distributions of the response variables. We further discuss using $D$ instead of $\mathcal{D}$ in Supplementary Material Section \ref*{subsec:D_discussion}.

In addition to encouraging variable selection, 
the second penalty on the $\beta_{m,:}$ leads to a (practically) unique solution. 
If $\gamma = 0$, the solution to \eqref{eq:estimator} is not unique because for any 
vector $a \in \mathbb{R}^{p}$ and any minimizer of \eqref{eq:estimator}, say  
$\hat\beta$, $\hat\beta - a 1_{JK}'$ has the same objective function value 
as $\hat\beta$. When $\gamma > 0$, a minimizer $\hat\beta$ is (practically) unique: 
$\hat\beta_{1,:}$, the intercept, is non-unique, but the $(p -1) \times JK$ submatrix 
excluding the intercept is unique. This follows from the fact that 
for a minimizer $\hat\beta$, for all $m \in \{2,\ldots, p\}$, 
$\|\hat\beta_{m, :}\|_2 = \min_{a_m \in \mathbb{R}}\|\hat\beta_{m,:} - a_m 1_{JK}\|_2$, 
otherwise $\hat\beta$ cannot be the solution to \eqref{eq:estimator}.
Making the intercept unique is trivial: one could simply impose the 
additional constraint that $\sum_{j=1}^{J}\sum_{k=1}^K \hat\beta_{1,f(j,k)} = 0$, in which case 
\eqref{eq:estimator} would be entirely unique. A similar argument about uniqueness in penalized multinomial logistic regression models was used in \citet{powers2018nuclear}. 

There are also situations in which we recommend penalizing the intercept. Specifically, when $J$ and $K$ are large relative to $n$, we suggest replacing
$\sum_{m=2}^p \|D'\beta_{m,:}\|_2$ with $\sum_{m=1}^p \|D'\beta_{m,:}\|_2.$ This way, for sufficiently large values of $\lambda$, one can obtain a minimizer of the objective function even when $\sum_{i=1}^n y_{i,j,k} = 0$ for some pairs $(j,k)$, as long as $\sum_{i=1}^n (1_K'y_{i,j,:}) > 0$ for all $j \in [J]$ and $\sum_{i=1}^n (1_J'y_{i,:,k}) > 0$ for all $k \in [K].$ We discuss this further in a later section where we describe applying \eqref{eq:estimator} in settings with more than two categorical response variables. 


\section{Statistical properties}\label{sec:Theory}
We study the statistical properties of \eqref{eq:estimator} with $n$, $p$, $J$, and $K$ varying. We focus on settings where the predictors are non-random. To simplify notation, we study the properties of a version of our estimator where the intercept is included in both penalties. Define the matrix $X = (x_1, \dots, x_n)' \in \mathbb{R}^{n \times p}$. For a set $\mathcal{C} \subset [p]$, let $\beta_{\mathcal{C}, :} \in \mathbb{R}^{|\mathcal{C}| \times JK}$ denote the submatrix of $\beta$ including only rows whose indices belong to $\mathcal{C}$ where $|\mathcal{C}|$ denotes the cardinality of $\mathcal{C}$. Finally, let $\|A\|_F^2 = {\rm tr}(A'A)$ denote the squared Frobenius norm of a matrix $A$.

To establish an error bound, we must first define our estimand, i.e., the value of $\beta^*$ from \eqref{eq:full_model} for which our estimator is consistent. As described in the previous section, for any $\beta^* \in \mathbb{R}^{p \times JK}$ which leads to \eqref{eq:full_model}, $\beta^* - a 1_{JK}'$ also leads to $\eqref{eq:full_model}$ for any $a \in \mathbb{R}^{p}$. Let the set $\mathcal{F}_\pi$ denote the set of all $\beta^*$ which lead to \eqref{eq:full_model}, i.e.,
$$  \mathcal{F}_\pi = \left\{ \beta^* \in \mathbb{R}^{p \times JK}: 
P(Y_1 = j, Y_2 = k\mid x) = \frac{{\rm exp}(x'\beta^*_{:, f(j, k)})}{\sum_{s,t} {\rm exp}( x'\beta^*_{:, f(s, t)})},~~~(x,j,k) \in \mathbb{R}^p \times [J]\times [K] \right\}.$$
Then, with $\mathcal{F}_\pi$, we define our estimation target as 
$\beta^\dagger = \argmin_{\beta \in \mathcal{F}_\pi} \|\beta\|_{1,2}$ where $\|A\|_{1,2} = \sum_{m}\|A_{m, :}\|_2$. Notice, we could equivalently write $
\beta^\dagger = \argmin_{\beta \in \mathcal{F}_\pi} \left\{ \mathcal{G}(\boldsymbol{\beta}) + \lambda\|\beta D\|_{1,2} + \gamma\|\beta\|_{1,2}\right\}.$
Since the optimization for $\beta^\dagger$ is over feasible set $\mathcal{F}_\pi$, and because $\mathcal{G}(\boldsymbol{\beta})$ and $\lambda\|\beta D\|_{1,2}$ are equivalent for all elements of $\mathcal{F}_\pi$, $\beta^\dagger$ is simply the element of $\mathcal{F}_\pi$ which minimizes $\|\beta\|_{1,2}$, i.e.,  $\beta^\dagger$ does not depend on the data or tuning parameters. By the same argument used to describe the uniqueness of \eqref{eq:estimator} in the previous section, $\beta^\dagger$ is unique. Given any $\beta \in \mathcal{F}_\pi$, $\beta^\dagger$ is simply $\beta - (JK)^{-1}(\beta 1_{JK})1_{JK}'$, i.e., $\beta^\dagger$ is the row-wise average zero version of the given $\beta$. We prove this in Supplementary Material Section \ref*{subsec:beta_dagger}.

We will require the following assumptions.
\begin{itemize}
\item[] \textbf{A1.} The responses $\mathcal{Y}_i \in \mathbb{R}^{J \times K}$ are independent and generated from \eqref{eq:dataGen} for all $i \in [n].$
\item[] \textbf{A2.} The predictors are normalized so that $\|X_{:,j}\|_2^2 \leq n$ for each $j \in [p]$.
\end{itemize}
We will also require the definition of a number of important sets. 
Let $S_L$ denote the subset of $[p]$ such that $\beta_{l, :}^\dagger \neq 0_{JK}$ and $D'\beta^\dagger_{l, :}\neq 0_{\binom{J}{2}\binom{K}{2}}$ for each $l \in S_L$; let $S_M$ denote the subset of $[p]$ such that $\beta_{m, :}^\dagger \neq 0_{JK}$ and $D'\beta^\dagger_{m, :}= 0_{\binom{J}{2}\binom{K}{2}}$ for each $m \in S_M$; and let $S_I = [p] \setminus S_L \cup S_M$. The sets $S_L$, $S_M$, and $S_I$ denote the set of predictors which affect the log odds ratios, affect only the marginal probabilities, and are irrelevant, respectively.  Let $\mathcal{S}$ denote this partition of $[p]$ into $S_L, S_M,$ and $S_I$. Next, with $\phi = (\phi_1, \phi_2) \in (1, \infty) \times (0, \infty)$, we define the set 
\begin{align*}
\mathbb{C}(\mathcal{S}, \phi) = &\left\{ \Delta \in \mathbb{R}^{p \times JK}: \Delta \neq 0_{p \times JK},  
(\phi_1 + 1)\|\Delta_{S_L \cup S_M, :}\|_{1,2} + \phi_1 \phi_2\|\Delta_{S_L, :}D\|_{1,2} \geq \right. \\
& \hspace{170pt} \left. (\phi_1 - 1) \|\Delta_{S_I, :}\|_{1,2} + \phi_1\phi_2\|\Delta_{S_I \cup S_M, :}D\|_{1,2} \right\}
\end{align*}

\noindent In the Supplementary Material, we show that when the tuning parameters are chosen as prescribed in Theorem \ref{thm:consistency}, $\hat\beta - \beta^\dagger$ belongs to the set $\mathbb{C}(\mathcal{S}, \phi)$ with high probability. This set $\mathbb{C}(\mathcal{S}, \phi)$ is needed establish our third assumption, A3. Let $\tilde{\mathcal{G}}: \mathbb{R}^{p \times JK} \to \mathbb{R}$ denote the version of $\mathcal{G}$ that takes matrix variate inputs, and let $\nabla^2 \tilde{\mathcal{G}}: \mathbb{R}^{p \times JK} \to \mathbb{R}^{pJK \times pJK}$ denote the Hessian of $\tilde{\mathcal{G}}$ with respect to the vectorization of its argument.
\begin{itemize}
\item[] \textbf{A3.} (Restricted eigenvalue) For all $\phi_1 > 1$ and $\phi_2 > 0$, there exists a constant $k$ such that $\kappa(\mathcal{S}, \phi) \geq k > 0$ where
\begin{equation}
\kappa(\mathcal{S}, \phi) = \inf_{\Delta \in \mathbb{C}(\mathcal{S}, \phi)} \frac{{\rm vec}(\Delta)' \nabla^2 \tilde{\mathcal{G}}(\beta^\dagger){\rm vec}(\Delta)}{\|\Delta\|_F^2}.\notag
\end{equation}
\end{itemize}
Assumption A3 is effectively a restricted eigenvalue condition, which often appear in the penalized maximum likelihood estimation literature \citep{raskutti2010restricted}. We can express $\nabla^2 \tilde{\mathcal{G}}(\beta^\dagger)$ as $n^{-1}\sum_{i=1}^n \{P_{\beta^\dagger}^*(x_i) \otimes x_i x_i'\}$ where the form of $P_{\beta^\dagger}^*(x_i),$ a positive semidefinite matrix, is given in the Supplementary Material and $\otimes$ denotes the Kronecker product. If the $P_{\beta^\dagger}^*(x_i)$'s were replaced with $I_{JK}$, then $\kappa(\mathcal{S}, \phi)$ would be equivalent to the restricted eigenvalue for least squares estimators, i.e., $\inf_{\Delta \in \mathbb{C}(\mathcal{S}, \phi)} \|X\Delta\|_F^2/(n \|\Delta\|_F^2)$. 

We must also define the following \textit{subspace compatibility constant} \citep{negahban2012} which we write as
$$ \Psi_{J,K}(S) = \hspace{-10pt}\sup_{M \in \mathbb{R}^{p \times JK}, M \neq 0_{p \times JK}}\frac{\|M_{S, :}D\|_{1,2}}{\|M\|_F}.$$
The quantity $\Psi_{J,K}(S)$ measures the magnitude of the log odds penalty over the set of $p \times JK$ matrices $M$ with Frobenius norm equal to one, where $M_{S^c, :}D = 0_{|S^c| \times \binom{J}{2}\binom{K}{2}}$ with $S^c = [p] \setminus S$ for a set $S \subseteq [p]$. Importantly, only the cardinality of $S$ affects $\Psi_{J,K}$. In the following remark, we provide an upper bound on $\Psi_{J,K}(S)$ for the bivariate response setting.
\begin{remark}
For all $J$ and $K$, and every set $S \subseteq [p]$, $\Psi_{J,K}(S) \leq \sqrt{|S|JK}.$ 
\end{remark}
With assumptions A1--A3, we are ready to state our main result, which will depend on Condition \ref{cond1}, detailed below. 
\begin{theorem}\label{thm:consistency}
Suppose assumptions A1--A3 hold and let $c > 2$, $\alpha \in (0,1)$, $\phi_1 > 1$, and $\phi_2 > 0$ be fixed constants. Define $\bar{\phi}_c = c(\phi_1 + 1)$ and $\underline{\phi}_c = c \hspace{1pt}\phi_1 \phi_2$. If $\gamma = \phi_1 [\{JK/(4n)\}^{1/2} + \{\log(p/\alpha)/n\}^{1/2}]$, $\lambda = \phi_2 \gamma$, and Condition \ref{cond1} holds, then 
$$ \|\hat\beta - \beta^\dagger\|_F  \leq  \frac{\bar{\phi}_c \sqrt{|S_L| + |S_M|} + \underline{\phi}_c \Psi_{J,K}(S_L)}{ \kappa(\mathcal{S}, \phi)}\left( \sqrt{\frac{JK}{4n }} +  \sqrt{\frac{\log (p/\alpha)}{n}}\right)$$
with probability at least $1 - \alpha$.
\end{theorem}
The proof of Theorem \ref{thm:consistency}, which can be found in the Supplementary Material, relies on a property of the multinomial negative log-likelihood closely related to self-concordance \citep{bach2010self}. In our proof, we have a precise condition on the magnitude of $n$ needed for the result of Theorem \ref{thm:consistency} to hold.

\begin{condition}\label{cond1} Given fixed constants $c > 2$, $\alpha \in (0,1)$, $\phi_1 > 1,$ and $\phi_2 > 0$, with  $\Phi_n = (\{\bar{\phi}_c(|S_L| + |S_M|)^{1/2} + \underline{\phi}_c \Psi_{J,K}(S_L)\}[\{JK/(4n)\}^{1/2} + \{\log(p/\alpha)/n\}^{1/2}])/\kappa(\mathcal{S}, \phi)$ and \\$d_n = \sqrt{6}\max_{i \in [n]}\|X_{i,:}\|_2$, $n$ is sufficiently large (with respect to $c$, $\alpha$, $\phi$, $d_n$, $p$, $JK$, and the cardinality of the sets $S_L$ and $S_M$) such that $e^{-\omega_n} + \omega_n - \omega_n^2/c - 1 > 0$, where $\omega_n = d_n \Phi_n$.
\end{condition}

The bound in Theorem \ref{thm:consistency} illustrates the effects of both the group lasso penalty and the penalty corresponding to the log odds ratios. In particular, $(|S_L| + |S_M|)^{1/2}$ corresponds to having to estimate $\left|S_L \cup S_M\right|$ total nonzero rows of $\beta^{\dagger}$, whereas the additional term $\Psi_{J,K}(S_L)$ comes from estimating the $|S_L|$ rows of $\beta^\dagger$ which do not satisfy $D'\beta^{\dagger}_{m, :} = 0_{\binom{J}{2}\binom{K}{2}}$. 

The constants $\phi = (\phi_1, \phi_2)$ balance the magnitude of $\lambda$ and $\gamma$ relative to $[\{JK/(4n)\}^{1/2} + (\log p/n)^{1/2}]$. In doing so, they affect the error bound by scaling $(|S_L| + |S_M|)^{1/2}$ and $\Psi_{J,K}(S_L)$, but also by controlling the restricted eigenvalue $\kappa(\mathcal{S}, \phi)$. Specifically, $\phi$ controls the set $\mathbb{C}(\mathcal{S}, \phi)$: for example, if $|S_L|$ were small, a larger $\phi_2$ would mean a larger $\kappa(\mathcal{S}, \phi)$.  Hence, the optimal choice of $(\phi_1, \phi_2)$ would be that which increases $\kappa(\mathcal{S}, \phi)$ relative to the magnitude of $\Psi_{J,K}(S_L).$ 


Using that $\hat\beta - \beta^\dagger \in \mathbb{C}(\mathcal{S}, \phi)$ with high probability under the choices of $\lambda$ and $\gamma$ as prescribed in Theorem \ref{thm:consistency}, we are also able to establish an error bound in the $\|\cdot\|_{1,2}$-norm. 
\begin{corol}\label{prop:L_12_norm}
Let $\Phi_n$ be as defined in Condition \ref{cond1}. If the conditions of Theorem \ref{thm:consistency} hold, then it follows that
$\|\hat\beta - \beta^\dagger\|_{1,2} \leq \Phi_n\{2\phi_1(|S_L| + |S_M|)^{1/2}  + \phi_1\phi_2 \Psi_{J,K}(S_L)\}/(\phi_1 - 1)$ with probability at least $1 - \alpha.$
\end{corol}
Finally, in the Supplementary Material Section \ref*{subsec:add_replicates}, we discuss how our theoretical results would change if additional replicates were available for at least one subject (i.e., $n_i > 1$ for at least one $i \in [n]$). In brief, error bounds can be improved by introducing additional replicates with the number of distinct subjects fixed at $n.$


\section{Computation}\label{sec:Computing}

In this section, we propose a proximal gradient descent algorithm (\citet{parikh2014proximal}, Chapter 4) to compute \eqref{eq:estimator}. Throughout, we treat $\gamma$ and $\lambda$ as fixed. We let $\mathcal{F}_{\lambda, \gamma}(\beta)$ denote the objective function from \eqref{eq:estimator} evaluated at $\beta$ with tuning parameter pair $(\lambda, \gamma)$ and recall $\tilde{\mathcal{G}}(\beta) = \mathcal{F}_{0, 0}(\beta):\mathbb{R}^{p \times JK} \to \mathbb{R}$ denotes the negative log-likelihood. In the following subsection, we describe our proposed proximal gradient descent algorithm at a high-level, and in the subsequent section, we describe how to solve the main subproblem in our iterative procedure.


\subsection{Accelerated proximal gradient descent algorithm}
Proximal gradient descent is a first-order iterative algorithm which generalizes gradient descent.
As in gradient descent, to obtain the $(t+1)$th iterate of our algorithm, we must compute the gradient of $\tilde{\mathcal{G}}$ evaluated at the $(t)$th iterate $\beta^{(t)}$. Letting  $W^{(t)} \in \mathbb{R}^{n \times JK}$ where 
$$W^{(t)}_{i,f(j,k)} = \frac{{\rm exp}(x_i'\beta_{:,f(j,k)}^{(t)})}{\sum_{l=1}^{J}\sum_{m=1}^K {\rm exp}(x_i' \beta_{:, f(l, m)}^{(t)})}-  y_{i,j,k}, ~~~ (i,j,k) \in [n] \times [J] \times [K],$$
the gradient can be expressed as 
$ \nabla \tilde{\mathcal{G}}(\beta^{(t)}) = n^{-1} X'W^{(t)}.$
One way to motivate our algorithm is through an application of the majorize-minimize principle. Specifically, since the negative log-likelihood is convex and has Lipschitz continuous gradient \citep{powers2018nuclear}, we know
\begin{equation}\label{eqMajorizer}
\tilde{\mathcal{G}}(\beta) \leq \tilde{\mathcal{G}}(\beta^{(t)}) +  {\rm tr}\left\{\nabla \tilde{\mathcal{G}}(\beta^{(t)})'(\beta - \beta^{(t)})\right\} + \frac{1}{2s^{(t)}}\|\beta - \beta^{(t)}\|_F^2
\end{equation}
for all $\beta$ and $\beta^{(t)}$ with some sufficiently small step size $s^{(t)}$. Letting $\mathcal{M}_{s^{(t)}}(\beta; \beta^{(t)})$ denote the right hand side of the inequality in \eqref{eqMajorizer}, it follows that
$$\mathcal{F}_{\lambda, \gamma}(\beta) \leq \mathcal{M}_{s^{(t)}}(\beta; \beta^{(t)}) + \lambda \sum_{m=2}^{p} \|D'\beta_{m,:}\|_{2} + \gamma \sum_{m=2}^{p} \|\beta_{m,:}\|_{2},$$
for all $\beta$ with equality when $\beta = \beta^{(t)}.$ That is, the right hand side of the above is a majorizing function of $\mathcal{F}_{\lambda, \gamma}$ at $\beta^{(t)}$. Hence, if we obtain the $(t+1)$th iterate of $\beta$ with
\begin{equation} \label{eq:majorizing_update}
\beta^{(t+1)} = \argmin_{\beta \in \mathbb{R}^{p \times JK}} \left\{ \mathcal{M}_{s^{(t)}}(\beta; \beta^{(t)}) + \lambda \sum_{m=2}^{p} \|D'\beta_{m,:}\|_{2} + \gamma \sum_{m=2}^{p} \|\beta_{m, :}\|_{2}\right\},
\end{equation}
the majorize-minimize principle \citep{lange2016mm} ensures that 
$\mathcal{F}_{\lambda, \gamma}(\beta^{(t+1)}) \leq \mathcal{F}_{\lambda, \gamma}(\beta^{(t)}).$ Thus, to solve \eqref{eq:estimator}, we propose to iteratively solve \eqref{eq:majorizing_update}. It is well known that the sequence of objective function values at the iterates generated by an accelerated version of this procedure (see Algorithm \ref{alg1}) converges to the optimal value at a rate of $O(1/t^2)$ when $s^{(t)}$ is chosen via backtracking line search (see 10. of Algorithm \ref{alg1}). For example, see \citet{beck2009fast} or Section 4.2 of \citet{parikh2014proximal} and references therein.

After some algebra, we can write \eqref{eq:majorizing_update} as  
\begin{equation}
\beta^{(t+1)} = \argmin_{\beta \in \mathbb{R}^{p \times JK}} \left\{ \frac{1}{2s^{(t)}}\|\beta - \beta^{(t)} + s^{(t)} \nabla \tilde{\mathcal{G}}(\beta^{(t)})\|_F^2 + \lambda \hspace{-2pt}\sum_{m=2}^{p} \|D'\beta_{m,:}  \|_{2} + \gamma \hspace{-2pt} \sum_{m=2}^{p} \|\beta_{m,:}\|_{2} \right\}. \label{eq:prox_update}
\end{equation}
Fortunately, \eqref{eq:prox_update} can be solved efficiently row-by-row of $\beta$. In particular, this problem can be split into $p$ separate optimization problems since for $m=2, \dots, p$,
\begin{equation}
\beta^{(t+1)}_{m,:} = \argmin_{\eta \in \mathbb{R}^{JK}} \left\{ \frac{1}{2}\|\eta - \beta_{m, :}^{(t)} + s^{(t)}[\nabla \tilde{\mathcal{G}}(\beta^{(t)})]_{m, :}\|_2^2 +   s^{(t)} \lambda \|D'\eta\|_{2} + s^{(t)} \gamma \|\eta\|_{2}\right\} \label{eq:Prox},
\end{equation}
where $[\nabla \tilde{\mathcal{G}}(\beta^{(t)})]_{m,:} \in \mathbb{R}^{JK}$ denotes the $m$th row of $\nabla \tilde{\mathcal{G}}(\beta^{(t)}).$
For the intercept (i.e., $\beta_{m,:}$ with $m=1$), the solution has a simple closed form:
$\beta_{1,:}^{(t+1)} = \beta_{1,:}^{(t)} - n^{-1} s^{(t)} \{1_n'W^{(t)}\}'.$
Then, it is straightforward to see that for $m \in \{2, \dots, p\}$, each of the subproblems in \eqref{eq:Prox} can be expressed  
\begin{equation}\label{eq:ProxOperator}
\hat{\eta}_{\bar\lambda, \bar\gamma} = \argmin_{\eta \in \mathbb{R}^{JK}} \left\{ \frac{1}{2}\|\eta - \nu\|_2^2 + \bar{\lambda} \|D'\eta\|_2 + \bar\gamma \|\eta\|_2 \right\},
\end{equation}
where $\nu$ corresponds to a row of $\beta^{(t)} - s^{(t)}\nabla \tilde{\mathcal{G}}(\beta^{(t)})$, $\bar{\lambda} = s^{(t)}\lambda$, and $\bar{\gamma} = s^{(t)}\gamma$. In the next subsection, we show that \eqref{eq:ProxOperator} can be solved very efficiently.


\subsection{Efficient computation of subproblem \eqref{eq:ProxOperator}}
Our first theorem reveals that $\hat{\eta}_{\bar\lambda,\bar\gamma}$ from \eqref{eq:ProxOperator} can be obtained in closed form. Throughout, let $A^{-}$ denote Moore-Penrose pseudoinverse of a matrix $A$. 
\begin{theorem}\label{proxSolutions} For arbitrary $J$ and $K$, \eqref{eq:ProxOperator} can be solved in a closed form. Specifically,
\begin{itemize}
\item[] (i) If $\|\nu\|_2 < \bar\gamma$, then $\hat{\eta}_{\bar\lambda, \bar\gamma} = 0_{JK}$. 
\item[](ii) If $\|\nu\|_2 \geq  \bar\gamma$ and $\|(D'D)^{-}D'\nu\|_2 \leq \bar\lambda$, then 
$\hat{\eta}_{\bar\lambda, \bar\gamma} =  \max\left(1 - \frac{\bar\gamma}{\|\mathcal{P}^{\perp}_{D,0} \nu\|_2}, 0\right)\mathcal{P}^{\perp}_{D,0} \nu,$
where $\mathcal{P}^{\perp}_{D,0} = I - D(D'D)^{-}D'$.
\item[] (iii) If  $\|\nu\|_2 \geq  \bar\gamma$ and $\|(D'D)^{-}D'\nu\|_2 > \bar\lambda$, then 
$\hat{\eta}_{\bar\lambda, \bar\gamma} = \max\left(1 - \frac{\bar\gamma}{\|\mathcal{P}^{\perp}_{D, \tau} \nu\|_2}, 0\right)\mathcal{P}^{\perp}_{D, \tau} \nu ,$
where $\mathcal{P}^{\perp}_{D, \tau} = I - D(D'D + \tau I)^{-1}D'$ for $\tau > 0$ such that $\|(D'D + \tau I)^{-1}D'\nu\|_2 = \bar{\lambda}.$
\end{itemize}
\end{theorem}

A proof of Theorem \ref{proxSolutions} can be found in the Supplementary Material. The results suggest that we can first screen all rows of $\beta^{(t)} - s^{(t)}\nabla \tilde{\mathcal{G}}(\beta^{(t)})$, as we know that those Euclidean norm less than $\bar\gamma$ will have minimizer $\hat\eta_{\bar\lambda, \bar\gamma} = 0_{JK}$. Of the rows that survive this screening, we need either apply the result from (ii) or (iii). Based on the statement of Theorem \ref{proxSolutions}, (ii) is immediate and does not require any optimization.  Regarding (iii), we do not have an analytic expression for $\tau$ which would satisfy $\|(D'D + \tau I)^{-1}D'\nu\|_2 = \bar{\lambda}$ for arbitrary $D$ and $\nu$. However, it turns out that the structure of our $D$ yields a closed form expression for $\tau$, which we detail in the following proposition. 
\begin{prop}\label{lemmaiii}
Let $r = (J-1)(K-1)$,  and let $u_l \in \mathbb{R}^{JK}$ be the left singular vector of $D$ corresponding to $\sigma_l \geq 0$, the $l$th largest singular value of $D$, for each $l \in [r]$. Then, 
$ \sum_{l=1}^{r} \{ w_l^2 \sigma_l^2/(\sigma_l^2 + \tau)^2\} = \bar\lambda^2$
implies $\|(D'D + \tau I)^{-1}D'\nu\|_2 = \bar{\lambda}$
where $w_l = u_l'\nu$ for $l \in [r]$. Consequently, since $\sigma_l^2 = JK$ for each $l \in [r]$, it follows that the condition in (iii) of Theorem \ref{proxSolutions} is satisfied when $\tau = \{(JK \sum_{l=1}^r w_l^2)^{1/2}/\bar\lambda\} - JK.$
\end{prop}
Together, Theorem \ref{proxSolutions} and Proposition \ref{lemmaiii} verify that we can solve \eqref{eq:ProxOperator} in a closed form. Since the singular value decomposition of $D$, $\mathcal{P}^{\perp}_{D,0}$, and $D'D$ can be precomputed and stored, these updates are extremely efficient to compute. To provide further intuition about the result of Theorem \ref{proxSolutions}, we present the closed form solution for this setting which covers (i), (ii),  and (iii) in the case where $J=K=2.$
\begin{theorem}\label{theorem:ExactSol} Suppose $J=K=2$ so that $\nu = (\nu_1, \nu_2, \nu_3, \nu_4)' \in \mathbb{R}^{4}$. Let $\ddot{\nu} = \nu_1 - \nu_2 - \nu_3 + \nu_4$. Then, $\hat{\eta}_{\bar\lambda,\bar\gamma} = \max\left(1 - \frac{\bar\gamma}{\|\hat{\eta}_{\bar\lambda,0}\|_2}, 0 \right)\hat{\eta}_{\bar\lambda,0},$
where 
$$\hat{\eta}_{\bar\lambda,0} = \left\{\begin{array}{l l}
(\nu_1 - \ddot{\nu}/4, \nu_2 + \ddot{\nu}/4, \nu_3 + \ddot{\nu}/4, \nu_4 - \ddot{\nu}/4)' & :|\frac{\ddot{\nu}}{4\bar{\lambda}}| \leq 1\\
(\nu_1 - \bar{\lambda}, \nu_2 + \bar{\lambda}, \nu_3 + \bar{\lambda}, \nu_4 - \bar{\lambda})' & :\ddot{\nu} > 4\bar{\lambda}\\
 (\nu_1 + \bar{\lambda}, \nu_2 - \bar{\lambda}, \nu_3 - \bar{\lambda}, \nu_4 + \bar{\lambda})' & :\ddot{\nu} < -4\bar{\lambda}
 \end{array} \right..$$

\end{theorem}
\textcolor{black}{Before concluding this section, we note that were one to penalize the intercept as discussed in Section \ref{sec:PenMaxLik}, $\beta^{(t+1)}_{1,:}$ would be obtained by applying the result of Theorem 2 (iii) and Proposition \ref{lemmaiii} with $\bar\gamma = 0.$}

\subsection{Summary and extensions}
We propose to iteratively update $\beta$ using \eqref{eq:majorizing_update} where we solve the $p$ subproblems using the result of Theorem \ref{proxSolutions}. This approach is especially efficient for large $p$ and moderately sized $J$ and $K$ since each of the subproblems involves a $JK$-dimensional optimization variable.  In practice, when the tuning parameter $\gamma$ is relatively large, (i) of Theorem \ref{proxSolutions} serves as a simple but exact screening heuristic: we often need only solve \eqref{eq:Prox} using (ii) or (iii) from Theorem \ref{proxSolutions} for a small number of the $p$ predictors.

To further reduce the required computing time, we employ an accelerated variation of the proximal gradient descent algorithm described above. Briefly, this approach extrapolates a search point for the next iterate based on the previous two iterates, e.g., see \citet{beck2009fast}. We summarize our complete algorithm in Algorithm \ref{alg1}. An implementation of this algorithm, along with a number of auxiliary functions, is available for download in the R package \texttt{BvCategorical} at \url{https://github.com/ajmolstad/BvCategorical}. 

In Section \ref*{sec:SemiSupervised} of the Supplementary Material, we extend our method and algorithm to settings where only one of the two response variables is observed, i.e., a semi-supervised setting. To estimate $\boldsymbol\beta^*$ in this scenario, we propose to minimize a penlized version of the observed data negative log-likelihood. Fortunately, we need not rely on an expectation-maximization algorithm: 
we can solve the corresponding optimization problem using a modified version of Algorithm \ref{alg1} based on the procedure proposed by \citet{li_nonconvex}. 

Finally, we discuss how we determine candidate tuning parameters for \eqref{eq:estimator} in Section \ref*{subsec:candidate_tuning} of the Supplementary Material.

 \begin{algorithm}[t]
 \caption{Accelerated proximal gradient descent for \eqref{eq:estimator}}\label{alg1}
 1. Initialize $\beta^{(0)} = \beta^{(1)}\in \mathbb{R}^{p \times JK}$, $\alpha^{(0)} = \alpha^{(1)} = 1$, $s^{(1)} > 0$, $\rho \in (0,1),$ and $t = 1$.
  \begin{tabbing}
  2. $\Gamma^{(t)} \leftarrow \beta^{(t)} + \left(\frac{\alpha^{(t-1)} - 1}{\alpha^{(t)}}\right)(\beta^{(t)} - \beta^{(t-1)})$.\\
  3. $U^{(t)} \leftarrow \Gamma^{(t)} - s^{(t)}\nabla \tilde{\mathcal{G}}(\Gamma^{(t)})$\\
  4. $\tilde{\beta}_{1,:} \leftarrow U_{1, :}^{(t)}$\\
  5. $\mathcal{A} \leftarrow \left\{m: m\in \{2, \dots, p\}, \|U_{m,:}^{(t)}\|_2 \geq s^{(t)}\gamma\right\}$\\
  6. For each $k_1 \in \{2, \dots, p\} \setminus \mathcal{A} $ \\
     \quad\quad\quad \texttt{(i):}$\tilde{\beta}_{k_1, :} \leftarrow 0_{JK}$\\
  7. $\mathcal{A}_1 \leftarrow \left\{m: m\in \mathcal{A} , \|(D'D)^{-}D'U_{m, :}^{(t)}\|_2 \leq s^{(t)}\lambda\right\}$\\
  8. For each $k_2 \in \mathcal{A}_1$\\
   \quad\quad\quad \texttt{(i):} $\tilde{\beta}_{k_2,:} \leftarrow \max\left(1 - s^{(t)}\gamma/\|\mathcal{P}^\perp_{D,0}U_{k_2,:}^{(t)}\|_2, 0\right)\mathcal{P}^\perp_{D,0} U_{k_2, :}^{(t)} $\\
  9. For each $k_3 \in \mathcal{A} \setminus \mathcal{A}_1$\\
  \quad\quad\quad \texttt{(i):} Compute $\tau$ according to Proposition \ref{lemmaiii}\\
  \quad\quad\quad \texttt{(ii):}$\tilde{\beta}_{k_3,:} \leftarrow \max\left(1 - s^{(t)}\gamma/\|\mathcal{P}^\perp_{D,\tau}U_{k_3,:}^{(t)}\|_2, 0\right)\mathcal{P}^\perp_{D,\tau}U_{k_3,:}^{(t)} $\\
  10. If $\tilde{\mathcal{G}}(\tilde{\beta}) \leq \tilde{\mathcal{G}}(\Gamma^{(t)}) + {\rm tr}\left\{\nabla \tilde{\mathcal{G}}(\Gamma^{(t)})'(\tilde{\beta} - \Gamma^{(t)})\right\} + \frac{1}{2s^{(t)}}\|\tilde{\beta} - \Gamma^{(t)}\|_F^2$\\
  \quad\quad\quad \texttt{(i):} $\beta^{(t+1)} \leftarrow \tilde{\beta}$\\
  \quad \hspace{4pt}Else\\
 \quad\quad\quad \texttt{(i):} $s^{(t)} \leftarrow \rho s^{(t)}$ and return to 3.\\
  11. $\alpha^{(t+1)} \leftarrow (1 + \sqrt{1 + 4[\alpha^{(t)}]^2})/2$ \\
  12. $s^{(t+1)} \leftarrow s^{(t)}$\\
  13. If not converged, set $t \leftarrow t + 1$ and return to 2. Otherwise, return $\beta^{(t+1)}.$
\end{tabbing}
  \end{algorithm}


\section{Generalization to more than two categorical responses}\label{sec:3MvMult}
Next, we describe the generalization of our method to arbitrarily many categorical response variables. In this setting, our method could be used to identify predictors that are irrelevant, that affect only the marginal distributions, and affect all higher-order log odds ratios. 
\subsection{Construction of multivariate constraint matrix}\label{sec:D_Mult}
To begin, consider the case where there are three categorical response variables with $J$, $K$, and $L$ response categories, respectively. Then, for the sake of example, suppose $p=1$ and the intercept is omitted. Under this scenario, for the predictor to affect only the marginal distributions, it must be that for all $(j,k,l) \in [J] \times [K] \times [L]$ and for all $x$, 
\begin{equation}\label{eq:threeVarIndep}
P(Y_1 = j, Y_2 = k, Y_3 = l \mid x) = P(Y_1 = j\mid x) P(Y_2 = k \mid x) P(Y_3 = l\mid x).
\end{equation}
This structure to can be achieved by our framework. Specifically, we can impose constraints enforcing two levels of conditional independence:
\begin{enumerate} 
	\item[]\text{a)} $P(Y_1 = j , Y_2 = k \mid x, Y_3 = l) =  P(Y_1 = j  \mid x, Y_3 = l)P(Y_2 = k  \mid x, Y_3 = l),$ 
	\item[]\text{b)} $P(Y_1 = j  \mid x, Y_3 = l) = P(Y_1 = j \mid x)~~ \text{ and }~~ P(Y_2 = k \mid x, Y_3 = l) = P(Y_2 = k \mid x).$
\end{enumerate}
It is easy to show that a) and b) together imply \eqref{eq:threeVarIndep}. 
To enforce a) and b) via linear constraints on the regression coefficients is less straightforward: we establish such combinations in the following lemma. 

\begin{lemma}\label{theorem:threeMN}
Let 
$\pi^*_{j,k,l}(x) = {\rm exp}(x'\boldsymbol{\beta}^*_{:, j,k,l}) /\sum_{s=1}^J \sum_{t=1}^K\sum_{u=1}^L {\rm exp}(x'\boldsymbol{\beta}^*_{:, s,t,u}).$ If
\begin{equation}\label{eq:a_sufficient_cond}
\log\left\{\frac{\pi^*_{j,k,l} (x)\pi^*_{j+1,k+1,l}(x)}{\pi^*_{j+1,k,l}(x) \pi^*_{j,k+1,l}(x)} \right\} = 0, \quad (j,k, l) \in [J-1] \times [K-1] \times [L]
\end{equation}
for all $x \in \mathbb{R}^p$, then a) holds. If \eqref{eq:a_sufficient_cond}, and in addition, for all $x \in \mathbb{R}^p$
$$ \log\left\{\frac{\pi^*_{j,1,l}(x) \pi^*_{j+1,1,l+1}(x)}{\pi^*_{j+1,1,l}(x) \pi^*_{j,1,l+1}(x)} \right\} = 0, \quad (j,l) \in [J-1] \times [L-1]$$
and 
$$ \log\left\{\frac{\pi^*_{1,k,l}(x) \pi^*_{1,k+1,l+1}(x)}{\pi^*_{1,k+1,l}(x) \pi^*_{1,k+1,l+1}(x)} \right\} = 0, \quad (k,l) \in [K-1] \times [L-1],$$
then b) also holds, and thus, \eqref{eq:threeVarIndep} holds.
\end{lemma} 

Together, this means we require  $(J-1)(K-1)L + (J-1)(L-1) + (K-1)(L-1) = 
JKL - J - K - L + 2$ linear constraints on the rows of the matricized regression coefficient tensor. This coheres with the number of combinations penalized in the bivariate categorical response setting since setting $L=1$ yields $(J-1)(K-1)$ combinations.

The matrix $\mathcal{D}$ which imposes these log odds constraints can be easily constructed by the same logic used in Section 2. For example, with $K = J = L = 2$, we can express $\beta^*$, the matricized version of $\boldsymbol{\beta}^* \in \mathbb{R}^{p \times 2 \times 2 \times 2}$, 
$\beta^* = (\boldsymbol{\beta}^*_{:,1,1,1}, \boldsymbol{\beta}^*_{:,2,1,1}, \boldsymbol{\beta}^*_{:,1,2,1},\boldsymbol{\beta}^*_{:,2,2,1}, \boldsymbol{\beta}^*_{:,1,1,2},\boldsymbol{\beta}^*_{:,2,1,2}, \boldsymbol{\beta}^*_{:,1,2,2}, \boldsymbol{\beta}^*_{:,2,2,2}) \in \mathbb{R}^{p \times8}.$
Hence, \eqref{eq:a_sufficient_cond} can be expressed as 
$$ x'(\boldsymbol{\beta}^*_{:,1,1,1} + \boldsymbol{\beta}^*_{:,2,2,1} - \boldsymbol{\beta}^*_{:,2,1,1} - \boldsymbol{\beta}^*_{:,1,2,1}) =  x'(\boldsymbol{\beta}^*_{:,1,1,2} + \boldsymbol{\beta}^*_{:,2,2,2} - \boldsymbol{\beta}^*_{:,2,1,2} - \boldsymbol{\beta}^*_{:,1,2,2}) = 0$$
and the latter two constraints from Lemma \ref{theorem:threeMN} as 
$$x'(\boldsymbol{\beta}^*_{:,1,1,1} + \boldsymbol{\beta}^*_{:,2,1,2} - \boldsymbol{\beta}^*_{:,2,1,1} - \boldsymbol{\beta}^*_{:,1,1,2}) = x'(\boldsymbol{\beta}^*_{:,1,1,1} + \boldsymbol{\beta}^*_{:,1,2,2} - \boldsymbol{\beta}^*_{:,1,2,1} - \boldsymbol{\beta}^*_{:,1,1,2}) = 0.$$
It is intuitive that four constraints are needed to impose independence: we begin with eight regression coefficient vectors, only seven of which are free since $\beta$ and $\beta - a 1_{JK}'$ yields the same probabilities for any $a \in \mathbb{R}^p$. Thus, seven free coefficients minus four linear constraints leaves three free coefficient vectors, one for each of the independent (Bernoulli) response variables. 

As discussed in Section \ref{sec:PenMaxLik}, to achieve invariance of our estimator against a particular construction of $\mathcal{D}$, we would instead use $D$, whose columns correspond to the log odds ratios 
$$\log\left\{\frac{\pi^*_{j,k,l}(x) \pi^*_{\check{j},\check{k},l}(x)}{\pi^*_{\check{j},k,l}(x) \pi^*_{j,\check{k},l}(x)} \right\}, ~~ j \neq \check{j}, k \neq \check{k}, l \in [L],\quad \log\left\{\frac{\pi^*_{j,k,l}(x) \pi^*_{j,\check{k},\check{l}}(x)}{\pi^*_{j,k,\check{l}}(x) \pi^*_{j,\check{k},l}(x)} \right\},~~  k \neq \check{k}, l \neq \check{l}, j \in [J],$$
$$\log\left\{\frac{\pi^*_{j,k,l}(x) \pi^*_{\check{j},k,\check{l}}(x)}{\pi^*_{\check{j},k,l}(x) \pi^*_{j,k,\check{l}}(x)} \right\},~~ j \neq \check{j}, l \neq \check{l}, k \in [K],$$
i.e., $D \in \mathbb{R}^{JKL \times \xi_{J,K,L}}$ where $\xi_{J,K,L} = \binom{J}{2}\binom{K}{2}L + \binom{K}{2}\binom{L}{2}J + \binom{J}{2}\binom{L}{2}K.$
It can be seen that that $D'\beta_{m, :}$ equal to the zeros vector implies $\mathcal{D}'\beta_{m, :}$ equals the zeros vector, but our penalty based on $D$ rather than $\mathcal{D}$ does not depend on the choice of the log odds ratios corresponding to its columns. From this setup, one can see that generalizing the matrix $D$ to settings with more than three response variables follows a similar logic. Given $G$ response variables, with the $l$th response having $K_l$ categories, the corresponding $D \in \mathbb{R}^{(\prod_{l=1}^G K_l) \times \xi_{\{K_j\}_{j=1}^G}}$ imposes penalties on
$ \xi_{\{K_j\}_{j=1}^G} = \sum_{j < l} \binom{K_j}{2}\binom{K_l}{2}\left(\prod_{s \neq l, j} K_s\right)$
log odds ratios.  When the number of response variables is large, the matrix $D$ will be large, but extremely sparse. In these settings, we recommend penalizing the intercept as discussed in Section \ref{sec:PenMaxLik} since it will be likely that some response category combinations are not observed in the training data. 


\subsection{Extension of theory and algorithms}
Both the theoretical results and computational approach described in Sections \ref{sec:Theory} and \ref{sec:Computing}, respectively, can be generalized to the multivariate categorical response setting. In this section, suppose that for each $i \in [n]$ we have observed realizations of $\mathcal{Y}_i \in \mathbb{R}^{K_1 \times K_2 \times \cdots \times K_G}$ independently from the version of \eqref{eq:dataGen} with $G$ categorical response variables and regression coefficient tensor $\boldsymbol{\beta}^* \in \mathbb{R}^{p \times K_1 \times K_2 \times \cdots \times K_G}.$

First, we generalize Theorem \ref{thm:consistency}: we provide a proof sketch in the Supplementary Material. 
\begin{corol}\label{corol:multivar}
Let $\check{K} = \prod_{j=1}^G K_j$ and let $\beta^\dagger \in \mathbb{R}^{p \times \check{K}}$ be the row-wise average zero version of the matricized tensor $\boldsymbol{\beta}^* \in \mathbb{R}^{p \times K_1 \times \cdots \times K_G}$. Under the conditions of Theorem \ref{thm:consistency}, if $\gamma = \phi_1 [\{\check{K}/(4n)\}^{1/2} + \{\log(p/\alpha)/n\}^{1/2}]$, $\lambda = \phi_2 \gamma$, and Condition \ref{cond1} holds (i.e., $n$ is sufficiently large), then with probability at least $1 - \alpha$, 
$$ \|\hat\beta - \beta^\dagger\|_F  \leq  \frac{\bar{\phi}_c \sqrt{|S_L| + |S_M|} + \underline{\phi}_c \Psi_{\{K_j\}_{j=1}^G}(S_L)}{\kappa(\mathcal{S}, \phi)}\left( \sqrt{\frac{\check{K}}{4n }} +  \sqrt{\frac{\log (p/\alpha)}{n}}\right)$$
where $S_L, S_I, S_M, \mathbb{C}(\mathcal{S}, \phi)$, and $\Psi_{\{K_j\}_{j=1}^G}$ (and by extension, $\kappa(\mathcal{S},\phi)$ and Condition \ref{cond1}) are all defined according to $\beta^\dagger$ and the appropriate $D$ matrix described in Section \ref{sec:D_Mult}.
\end{corol}


In the multivariate response setting, the corresponding optimization problem is effectively no different from the bivariate response version of \eqref{eq:estimator}: it requires only using a different constraint matrix $D$ and modifying the dimensions of $\beta$. Thus, Algorithm \ref{alg1} -- whose convergence properties and updating equations do not depend on the particular $D$ -- can still be used; and Theorem \ref{proxSolutions} can be applied to solve the subproblem (11). Unlike in the bivariate setting, however, Proposition \ref{lemmaiii} (which deals with Theorem \ref{proxSolutions}(iii)) does not apply to arbitrarily many categorical responses.
Fortunately, by applying the logic used in the proof of Proposition \ref{lemmaiii}(iii), one can solve for $\tau$ using a numeric univariate root-solver; this can be done very efficiently by exploiting the low-rankness of $D$. We provide a concrete example of this in Section \ref*{subsec:TrivariateGeneralization} of the Supplementary Material. In Section \ref*{sec:TrivariateApplication} of the Supplementary Material, we study the performance of our estimator in a setting with three binary response variables. 

\section{Simulation studies}\label{sec:Simulation_Studies}
To study the performance of our method in the bivariate response setting, we consider four models: at one extreme, all predictors can only affect the marginal probabilities for each response (or be irrelevant); at the other extreme, the predictors are either irrelevant or affect both log odds ratios and marginal distributions. We show that under four models along this continuum, our method dominates the competing methods.

\subsection{Data generating models and competing methods}\label{eq:comp_methods}

For 100 independent replications, we generate data from the multivariate multinomial logistic regression model with $J=3$ and $K=2$ categories. Independently for $n = 300$ training observations, we first generate $x \in \mathbb{R}^p$, a realization of $X \sim {\rm N}_p(0, \Sigma_{*X})$ where $\left[\Sigma_{*X}\right]_{s,t} = 0.5^{|s-t|}$ for $(s,t) \in [p] \times [p]$. Then, given some $\boldsymbol{\beta}^* \in \mathbb{R}^{p \times J \times K}$, we set 
$\pi_{j,k}^*(x) = {\rm exp}(x'\boldsymbol{\beta}^* _{:, j, k})/\sum_{s=1}^J \sum_{t=1}^K {\rm exp}(x'\boldsymbol{\beta}^* _{:, s, t})$
and generate the responses from \eqref{eq:dataGen} using the $\pi_{j,k}^*$ with each $n_i = 1$. This procedure is repeated to generate $500$ validation observations, and $10^4$ testing observations. In our simulation settings, we use $p \in \left\{100, 300, 500, 1000, 2000\right\}$.   

We consider four distinct structures for $\boldsymbol{\beta}^*$; recall that $\beta^* \in \mathbb{R}^{p \times JK}$ denotes the matricized version of $\boldsymbol{\beta}^*$. Note that we introduce our data generating models in the order 1, 4, 2, and 3 because Model 1 and 4 represent the two extremes, whereas Model 2 and 3 are intermediate.
\begin{itemize}
	\item\textbf{Model 1}: We randomly select 10 rows of $\beta^*$ to be nonzero. Each of the elements of these tens rows is set equal to independent realizations of a ${\rm Uniform}(-3,3)$ random variable. 
	\item \textbf{Model 4}: We randomly select 10 rows of $\beta^*$ to be nonzero. For each row independently, we generate four independent realizations of a ${\rm Uniform}(-3,3)$ random variable. Given these realizations, say $(u_1, u_2, u_3, u_4)$, we set the row of $\beta^*$ equal to 
$(-u_4 + u_3 + u_1, u_1, u_2, u_3, u_4, -u_1 + u_4 + u_2).$
Under this construction, we can see 
$D'(-u_4 + u_3 + u_1, u_1, u_2, u_3, u_4, -u_1 + u_4 + u_2) = 0_3.$
\end{itemize}

Under Model 1, each of the ten predictors corresponding to the nonzero rows of $\beta^*$ affect both marginal probabilities and log odds ratios almost surely. Under Model 4, each of the predictors corresponding to nonzero rows of $\beta^*$ affects only the marginal probabilities. Next, we consider two intermediate models which have a combination of predictors affecting only the marginal probabilities, and affecting both marginal probabilities and log odds ratios.
\begin{itemize} 
	\item \textbf{Model 2}: We randomly select six rows of $\beta^*$ to be nonzero and consist of elements which are each independent realizations of a ${\rm Uniform}(-3,3)$ random variable. Then, we select an additional four rows of $\beta^*$ to be generated in the same manner as in Model 4. 
	\item \textbf{Model 3}: We randomly select three rows of $\beta^*$ to be nonzero and consist of elements which are each independent realizations of a ${\rm Uniform}(-3,3)$ random variable. Then, we select an additional seven rows of $\beta^*$ to be generated in the same manner as in Model 4. 
\end{itemize}

Under Models 1--3, the marginal distributions alone are not sufficient to specify the distribution of $(Y_1, Y_2 \mid x)$. However, under Models 2 and 3, a decreasing number of predictors affect the log odds ratios: only six predictors under Model 2 and three predictors under Model 1. Model 4, conversely, is equivalent to generating the responses under separate multinomial logistic regression models, i.e., only $(Y_1 \mid x)$ and $(Y_2 \mid x)$ are needed to specify $(Y_1, Y_2 \mid x)$. 



We consider a number of alternative estimators in our simulation studies. For each, the tuning parameters are chosen by minimizing the joint classification error on the validation set, except for separate multinomial logistic regression models, where each model's tuning parameter is chosen to minimize classification error on the two responses marginally.
\begin{itemize}
\item \textbf{Separate multinomial} \texttt{(Sep)}: We fit two separate penalized multinomial logistic regression estimators, i.e., with tuning  parameters $(\gamma_{(1)}, \gamma_{(2)}) \in (0, \infty) \times (0, \infty)$ we fit
 \begin{equation}
\argmin_{\eta \in \mathbb{R}^{p \times J}} \left\{ - \frac{1}{n}  \sum_{i=1}^n \log \left( 
\sum_{j=1}^J \frac{\exp\left( x_i'\eta_{:, j}\right)y_{(1)i,j}}{\sum_{l=1}^J \exp\left(x_i'\eta_{:, l}\right)} 
\right) + \gamma_{(1)} \sum_{m=2}^{p} \|\eta_{m, :}\|_2\right\} \notag
\end{equation} 
for the first response and similarly for the second ($K$-category) response. 
\item \textbf{Group-penalized multivariate multinomial} \texttt{(G-Mult)}: A special case of our proposed estimator in \eqref{eq:estimator} with $\lambda = 0$ fixed and $\gamma \in (0, \infty).$
\item \textbf{Lasso-penalized multivariate multinomial} \texttt{(L-Mult)}:
The $L_1$-penalized version of the multinomial logistic regression estimator,  \texttt{G-Mult}.


\item \textbf{Overlapping group-penalized multivariate multinomial} \texttt{(OG-Mult)}: An overlapping group lasso penalized multivariate multinomial logistic regression estimator
\begin{align}
&\argmin_{\boldsymbol{\beta} \in \mathbb{R}^{p \times J \times K}} \left\{ \mathcal{G}(\boldsymbol{\beta}) + \gamma \sum_{m=2}^{p} \left(\sum_{k=1}^K\|\boldsymbol{\beta}_{m, :, k}\|_{2} + \sum_{j=1}^J\|\boldsymbol{\beta}_{m, j, :}\|_{2}\right)\right\}. \label{eq:Overlap}
\end{align}
with $\gamma \in (0,\infty)$. 
\item \textbf{Latent group-penalized multivariate multinomial} \texttt{(LG-Mult)}:
A latent group lasso penalized multivariate multinomial logistic regression estimator
\begin{equation}
\argmin_{\boldsymbol{\beta} \in \mathbb{R}^{p \times J \times K}} \left\{ \mathcal{G}(\boldsymbol{\beta}) + \gamma \sum_{m=2}^{p} \Omega_{\cup}^\mathcal{H}(\boldsymbol{\beta}_{m, :, :})\right\}, \quad \Omega_{\cup}^\mathcal{H}(\boldsymbol{\beta}_{m, :, :}) = \hspace{-3pt} \min_{v \in \mathcal{V}_{\mathcal{H}}, \sum_{h \in \mathcal{H}} v^h = \boldsymbol{\beta}_{m, :,:}} \sum_{h \in \mathcal{H}}\|v^h\|_F\label{eq:Latent}
\end{equation}
with $\gamma \in (0, \infty)$ where $\mathcal{H}$ denotes the groups penalized in \eqref{eq:Overlap} (i.e., the set of indices highlighted from each of the matrices in Figure \ref*{fig:parameters_penalized} of the Supplementary Material), $\mathcal{V}_\mathcal{H}$ denotes the set of matrices with the sparsity pattern corresponding to the groups in Figure \ref*{fig:parameters_penalized} of the Supplementary Material. 

\item \textbf{Log odds-penalized multivariate multinomial} \texttt{(LO-Mult)}: Our proposed estimator from \eqref{eq:estimator} with $(\lambda, \gamma) \in (0, \infty) \times (0, \infty)$. 

\end{itemize}

To compute both the overlapping group-penalized and latent group-penalized multivariate multinomial estimators, we use accelerated proximal gradient descent algorithms similar to those in Section \ref{sec:Computing}. We provide additional details in Section \ref*{sec:Extra_Comp} of the Supplementary Material. 

Finally, as a benchmark, we also compare to \texttt{Oracle}, the $\boldsymbol{\beta}^*$ which generated the data. 

\begin{figure}[t]
\centering
\includegraphics[width=16cm]{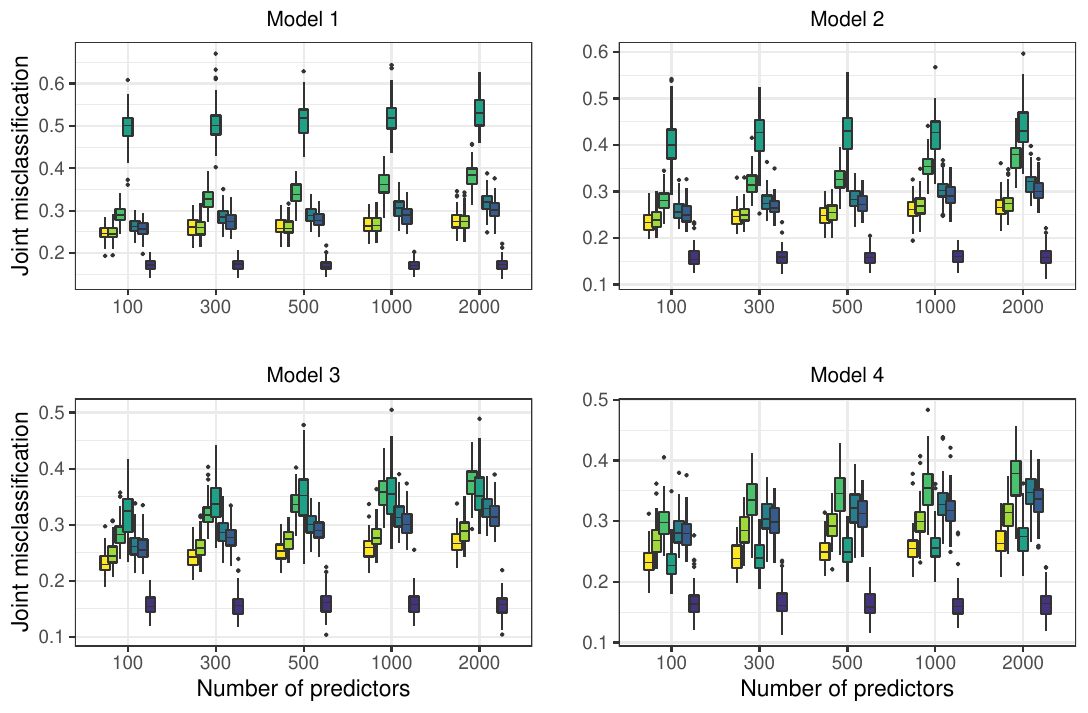}\\
\includegraphics[width=14cm]{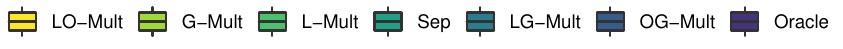}
\caption{Joint misclassification rates under Models 1--4 with $p \in \left\{100, 300, 500, 1000, 2000\right\}$. }\label{fig:joint}
\end{figure}
 
\begin{figure}[ht]
\centering
\includegraphics[width=16cm]{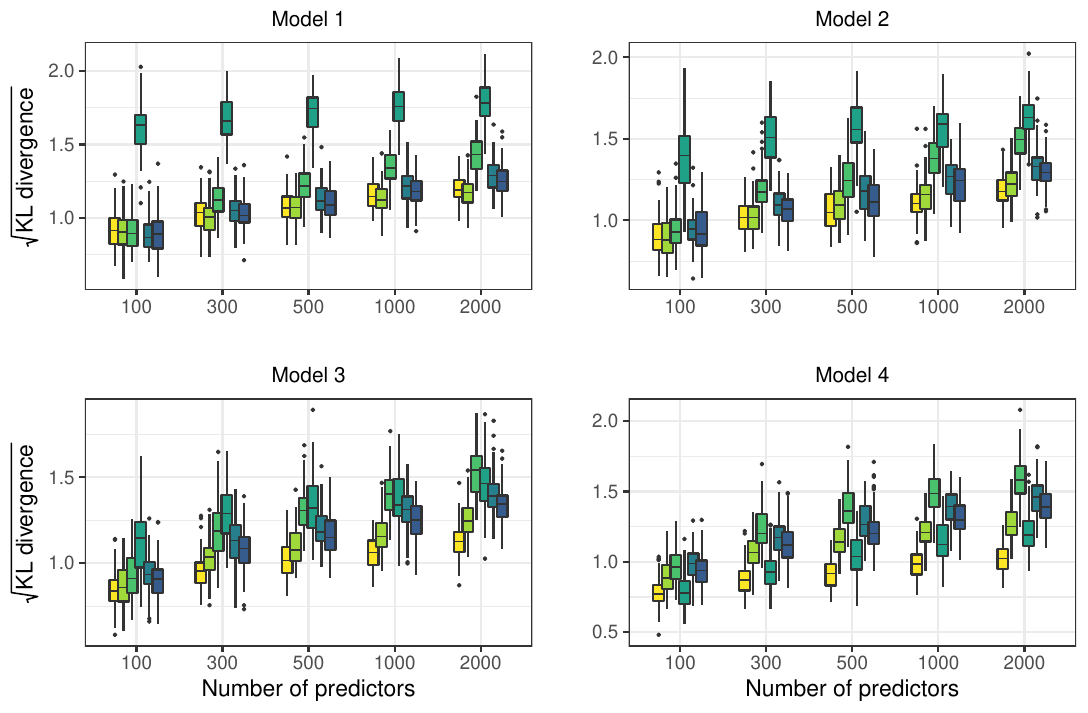}\\
\includegraphics[width=12cm]{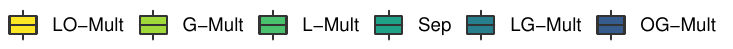}
\caption{Square-root average Kullback-Leibler divergence under Models 1--4 with $p \in \left\{100, 300, 500, 1000, 2000\right\}$. }\label{fig:KL}
\end{figure}

\subsection{Results}
Results are displayed in Figures \ref{fig:joint} and \ref{fig:KL}. Focusing first on the joint misclassification results displayed in Figure \ref{fig:joint}, we see that in every setting we considered, \texttt{LO-Mult}, our proposed estimator, performs approximately as well or better than all other considered estimators except \texttt{Oracle}, which is included to indicate the best possible misclassification rate (and thus, is implicitly omitted when we refer to ``competitors''). However, the performance of all other estimators differs dramatically across settings. Under Model 1, where predictors either only affect the log odds ratios or are irrelevant, \texttt{LO-Mult} performs similarly to \texttt{G-Mult}. This agrees with what one would expect since \texttt{G-Mult} does not assume independence; \texttt{G-Mult} assumes that predictors either affect the log odds ratios or are irrelevant. By the same reasoning, \texttt{LG-Mult} and \texttt{OG-Mult} also perform reasonably well in these settings. Conversely, \texttt{Sep} performs much worse than all competitors. This too agrees with intuition since \texttt{Sep} assumes independence of responses, which does not hold in this setting.

Turning our attention to Model 2, we again see that \texttt{LO-Mult} performs similarly to \texttt{G-Mult}, but as $p$ increases, \texttt{LO-Mult} begins to slightly outperform competitors. Under this model, four predictors only affect the marginal probabilities, whereas six affect the log odds ratios. Thus, since our approach \texttt{LO-Mult} allows for this type of variable selection, it it reasonable to expect our approach to perform best. 

Under Model 3, we see that the results are similar as under Model 2, with \texttt{LO-Mult} more clearly outperforming competitors. This is because seven of the 10 important predictors affect only the marginal probabilities: a feature which cannot be modeled by \texttt{G-Mult}, \texttt{LG-Mult}, or \texttt{OG-Mult}.  Finally, under Model 4, we see that \texttt{Sep} and \texttt{LO-Mult} perform nearly identically. Under this data generating model the two responses are independent, which agrees with the assumption made by \texttt{Sep}. We also see that \texttt{G-Mult} performs worse than \texttt{Sep} and \texttt{LO-Mult}. 

In Figure \ref{fig:KL}, we display the (square-root) average Kullback-Leibler (KL) divergence on the testing set across the four models. See Section \ref*{subsec:Alt_Tuning} of the Supplementary Material for our definition of KL divergence. To summarize briefly, the results are similar to the misclassification results displayed in Figure \ref{fig:joint}, with \texttt{LO-Mult} performing nearly as well as the best performing competitor in all four models we considered. As $p$ increases under Models 2--4, the performance of \texttt{LO-Mult} relative to competitors improves moreso than under the same settings using classification error as a performance metric. In Figure \ref*{fig:hellinger} of the Supplementary Material, we also display the average Hellinger distance for each of the methods: relative performances are similar those based on KL divergence.

Lastly, in Figure \ref*{fig:marg} of the Supplementary Material, we display the marginal misclassification rates for the response variable having $J=3$ response categories. Under Model 1 and Model 2, all methods which do not assume independence outperform \texttt{Sep} in terms of classification. Interestingly, $\texttt{LO-Mult}$ is slightly outperformed by both \texttt{L-Mult} and \texttt{G-Mult}. Under Model 3 and 4, \texttt{LO-Mult} begins to outperform the competitors, with \texttt{Sep} performing better than \texttt{G-Mult} and \texttt{L-Mult} under Model 4. 

\subsection{Additional simulation studies}\label{subsec:additional_sims}
In the Supplementary Material, we include additional simulation study results. In Section \ref*{subsec:Alt_Tuning}, we present results under Models 1--4 where instead of using classification accuracy, we selected tuning parameters by maximizing a validation likelihood. This led to better average KL divergence, but worse classification accuracy. In Section \ref*{subsec:J4K3}, we compare the methods under a similar set of data generating models as in Section \ref{sec:Simulation_Studies}, but with $J = 4$ and $K = 3$. The relative performances of the considered methods were very similar to those in the simulation settings with $J=3$ and $K=2$. Finally, in Section \ref*{sec:TrivariateApplication}, we also considered the trivariate response setting with each of the three categorical response variables being binary. In this setting, our method considerably outperformed competitors.  Notably, when the responses were truly independent, our method even outperformed \texttt{Sep}. This could be attributed to the fact that our method performs variable selection jointly across all response variables, whereas \texttt{Sep} does not. A version of \texttt{Sep} which performs variable selection across responses jointly is simply a special case of our method with $\lambda = \infty$ when the intercept is included in the first penalty.

\section{TCGA pan-kidney cancer cohort risk classification}\label{sec:DataExample}
We applied our method to the problem of risk classification in the pan-kidney cancer cohort data collected by The Cancer Genome Atlas (TCGA) project which are accessible through \href{https://www.cancer.gov.tcga}{https://cancer.gov.tcga}.  Our goal was to model 5-year survival probabilities and cancer types using gene expression profiles of $n = 420$ patients with one of three types of cancer: kidney renal clear cell carcinoma (KIRC), kidney renal papillary cell carcinoma (KIRP), and kidney chromophobe (KICH). Specifically, we hoped to identify a subset of genes which can be used to distinguish cancer types (KIRC, KIPR, or KICH) and are predictive of 5-year survival (i.e., failure before 5 years or not) simultaneously. Kaplan-Meier survival curves are displayed in Figure \ref*{fig:KapMeier} of the Supplementary Material, and counts for each cancer type are given in Table \ref*{table:Counts} of the Supplementary Material.  From Figure \ref*{fig:KapMeier}, we can see that KIRC and KIRP have similar survival curves, whereas KICH, which has the smallest sample size, appears to have lower 5-year mortality risk overall. 

\subsection{Data processing}
Starting with RNA-sequencing counts, we normalized gene expression in the following manner. First, we removed all genes whose 75th percentile count was less than 20. Then, for the $i$th subject ($i \in [420])$, we define the normalized expression for the $j$th gene as $\log\left\{ (c_{i,j} + 1)/q_{i,0.75}\right\}$ where $c_{i,j}$ is the sequencing count for the $i$th subject's $j$th gene, and $q_{i,0.75}$ is the 75th percentile of counts for the $i$th subject. We also included age and tumor stage as predictors. For simplicity, we dichotomized tumor stage into two groups representing stages i/ii and iii/iv. 

To reduce dimensionality, we performed a two-phased supervised screening before model fitting. We obtained $F$-test statistics for each gene based on the 6 category combinations, e.g., see Section 4.2 of \citet{mai2019multiclass}. In the first phase, we retained only the 2000 genes with the largest $F$-test statistics. In the second phase, we performed pruning on the retained genes so that no two genes have absolute correlation greater than 0.75. That is, starting with the gene with highest $F$-test statistic, we removed all genes with absolute correlation greater than 0.75 with this gene. Then, moving onto the gene with next largest $F$-test statistic among the remaining genes, we repeated this procedure until no two genes have absolute correlation greater than 0.75.

\subsection{Comparison to alternative methods}
To first compare the predictive accuracy of our method to four reasonable competitors, we performed leave-one-out cross-validation. That is, for each $i \in [420]$, we perform screening and fit the model using all but the $i$th subject's data; then recorded whether we correctly classify the $i$th subject based on the fitted model. 
We compared our method to  \texttt{G-Mult} and \texttt{Sep} as defined in Section \ref{eq:comp_methods}. We also compared to the $L_1$-penalized versions of each, we which call $\texttt{L-Mult}$ and $\texttt{L-Sep}$. For each method, we select tuning parameters to minimize 5-fold cross-validated classification error on the training set. 
Full results are presented in Table \ref{table:TCGAResults}.

We see that among all five methods we considered, \texttt{LO-Mult} has the lowest joint classification error at 28.81\%. The next closest, \texttt{Sep}, is more than 2\% higher. In terms of marginal classification, both \texttt{LO-Mult} and \texttt{G-Mult} have an error rate of 4.05\% for classifying cancer types, although all methods perform relatively well. In terms of classifying 5-year survival status, we see that \texttt{LO-Mult} performs best, with an error rate of 25.95\%, with the next best perform methods (\texttt{L-Mult}, \texttt{Sep}, and \texttt{L-Sep}) all misclassifying 27.38\% of subjects.  Interestingly, the models assuming independence have the lowest deviance, but among those methods which allow for response dependence, \texttt{LO-Mult} performs best. Finally, in the bottom-most row, we show that \texttt{LO-Mult}, in addition to having the lowest misclassification rates, tends to do so while selecting fewer genes as relevant than almost all other methods.  

\begin{table}
\centering
\scalebox{.95}{
\begin{tabular}{|c|  c c c c c|}
\hline
& \texttt{LO-Mult} & \texttt{G-Mult} & \texttt{L-Mult} & 
\texttt{Sep} & \texttt{L-Sep}\\
\hline
 Joint classification error & 28.81 & 32.38 & 31.19 & 30.95 & 31.67 \\ 
 Cancer type marginal error  & 4.05 & 4.05 & 5.00 & 4.52 & 5.24 \\ 
  5-year survival marginal error & 25.95 & 28.10 & 27.38 & 27.38 & 27.38 \\
  Deviance  & 1.44 & 1.48 & 1.55 & 1.38 & 1.41 \\
  Number of genes  & 64.56 & 84.85 & 76.93 & 74.60 & 39.07 \\ 
\hline
\end{tabular}
}
\caption{(Top three rows) Leave-one-out error percentages for predicting both cancer type and 5-year survival status (joint classification error), cancer type marginally, and 5-year survival status marginally. (Fourth row) Average test set deviance over the 420 subjects in the dataset. (Fifth row) The number of genes identified as relevant for either response distribution. The standard deviations of deviance (resp.\ numbers of genes) were 1.26, 1.36, 1.85, 1.30, and 1.44 (resp.\ 13.00,  5.10, 24.84, 13.00, and 12.31) for \texttt{LO-Mult}, \texttt{G-Mult}, \texttt{L-Mult},
\texttt{Sep}, and \texttt{L-Sep}, respectively. }\label{table:TCGAResults}
\end{table}

\subsection{Fitted model interpretation and insights}
To demonstrate the interpretability of our fitted models, we also performed 5-fold cross validation using the entire dataset. Our fitted model included 87 genes (of 822 considered after the two-phased screening of the entire dataset), as well as both tumor stage and age. Among these genes, 27 were estimated to affect the log odds ratios, while the remainder affect the marginal distributions only. Notably, both age and tumor stage were estimated to affect only the marginal distributions. This agrees with intuition since we may expect both of these variables to primarily be predictive of 5-year survival status marginally.

\begin{figure}
\includegraphics[width=8cm]{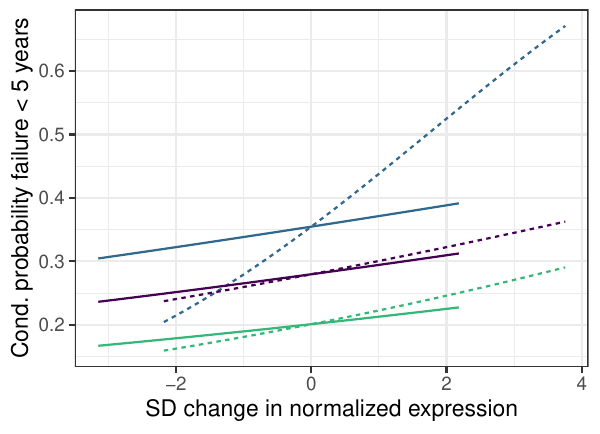}\includegraphics[width=8cm]{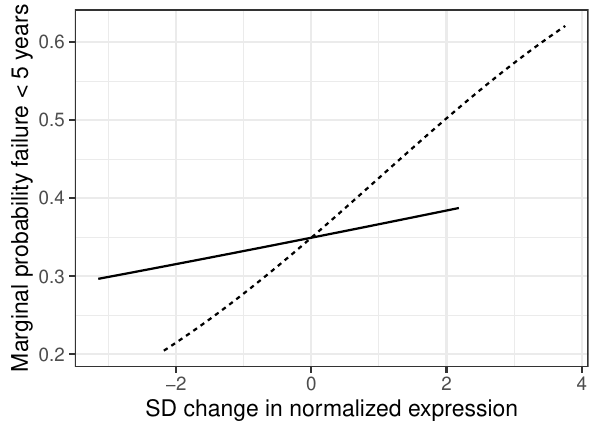}
\caption{(Left) Conditional probabilities of failure before 5 years for the three cancer types: KIRC (blue), KIRP (purple), and KICH (green); and two genes: CAV1 (solid lines) and CLN8 (dashed lines). Note that lines represent the estimated conditional probabilities as the indicated gene's expression varies with all others genes' expression is fixed at their mean, with tumor stage fixed at i/ii, and with age fixed at its mean. (Right) Marginal probabilities of failure for changes in CAV1 (solid line) or CLN8 (dashed line) with all others genes' expression fixed at their mean, with tumor stage fixed at i/ii, and with age fixed at its mean. In both panels, ranges for each gene's expression represent their observed deviations in the complete dataset.}\label{fig:fitted_probs}
\end{figure}
To visualize how changes in gene expression affect 5-year survival probabilities, we display two plots in Figure \ref{fig:fitted_probs}. These plots demonstrate how, with all other genes held fixed at their mean, a standard deviation change in expression of the given gene changes (a) conditional (on cancer type) probability of failure within five years, and (b) marginal probability of failure within five years. Of the two genes we display, CAV1 was estimated to only affect the marginal probabilities, whereas CLN8 was estimated to affect the log odds ratios. We see that in the conditional probability plot, the effect of CAV1 is effectively the same across cancer types. However, it is worth noting that these lines are not equidistant across the horizontal-axis since the intercept term does not satisfy $D'\hat{\beta}_{1,:} = 0_{3}$: if it did, then these three conditional probabilities would be equivalent for any expression value of CAV1. The effect of CLN8 across cancer types is easier to interpret: higher expression leads to a much higher probability of failure in less than five years in KIRC than in the other two cancer types with all other genes' expression fixed at the mean. In the right hand plot of Figure \ref{fig:fitted_probs}, we display the marginal probabilities of failure in less than five years under the same settings.  Overall, it would seem that overexpression of CLN8 appears to have a more substantial effect on the probability of 5-year survival than does CAV1. Interestingly, overexpression of CAV1 was found to be associated with a poor prognosis in KIRC in previous studies \citep{cav1}. Further research is necessary to determine whether these particular genes may serve as useful markers for prognoses in pan-kidney cancer.

\subsection*{Acknowledgments}
The authors thank Rohit Patra and Karl Oskar Ekvall for helpful discussions; and thank two anonymous referees and the associate editor for their helpful comments. A. J. Molstad's research was supported in part by National Science Foundation grant DMS-2113589.  A. J. Rothman's research was supported in part by the National Science Foundation grant DMS-1452068.

\bibliography{arXiv}

\newpage

\begin{center}
\vspace{60pt}
\singlespacing

\LARGE{Supplementary Material to ``A likelihood-based approach for multivariate categorical response regression in high dimensions''}
\vspace{10pt}
\end{center}
\appendix
\onehalfspacing
\section{Additional bivariate categorical response simulation studies and details}

\subsection{Alternative tuning parameter selection criterion}\label{subsec:Alt_Tuning}
In this section, we present simulation study results under exactly the data generating models described in Section \ref*{sec:Simulation_Studies}, but using a different tuning parameter selection criterion for each method. In these studies, we select tuning parameters by maximizing the log-likelihood evaluated on the validation set: for example, see equation (5) of \citet{Price2019Auto}. As in the main manuscript, we measure joint misclassification accuracy and average Kullback-Leibler divergence, the latter of which we define as 
$$ n_{\rm test}^{-1}\sum_{i=1}^{n_{\rm test}} \sum_{j=1}^J \sum_{k=1}^K \log \left(\frac{\hat{P}(Y_{i1} = j, Y_{i2} = k \mid x_i)}{P(Y_{i1} = j, Y_{i2} = k \mid x_i)}\right)\hat{P}(Y_{i1} = j, Y_{i2} = k \mid x_i)$$
where $\hat{P}(Y_{i1} = j, Y_{i2} = k \mid x_i)$ is an estimate of $P(Y_{i1} = j, Y_{i2} = k \mid x_i)$ based on some particular fitted model.  We also record and report average test set Hellinger distance, which is defined as 
$$ \frac{1}{n_{\rm test}}\sum_{i=1}^{n_{\rm test}} \left(\frac{1}{2}\sum_{j=1}^J \sum_{k=1}^K \left[ \{\hat{P}(Y_{i1} = j, Y_{i2} = k \mid x_i)\}^{1/2} - \{P(Y_{i1} = j, Y_{i2} = k \mid x_i)\}^{1/2}\right]^2\right)^{1/2}.$$

\begin{figure}[t]
\centering
\includegraphics[width=16cm]{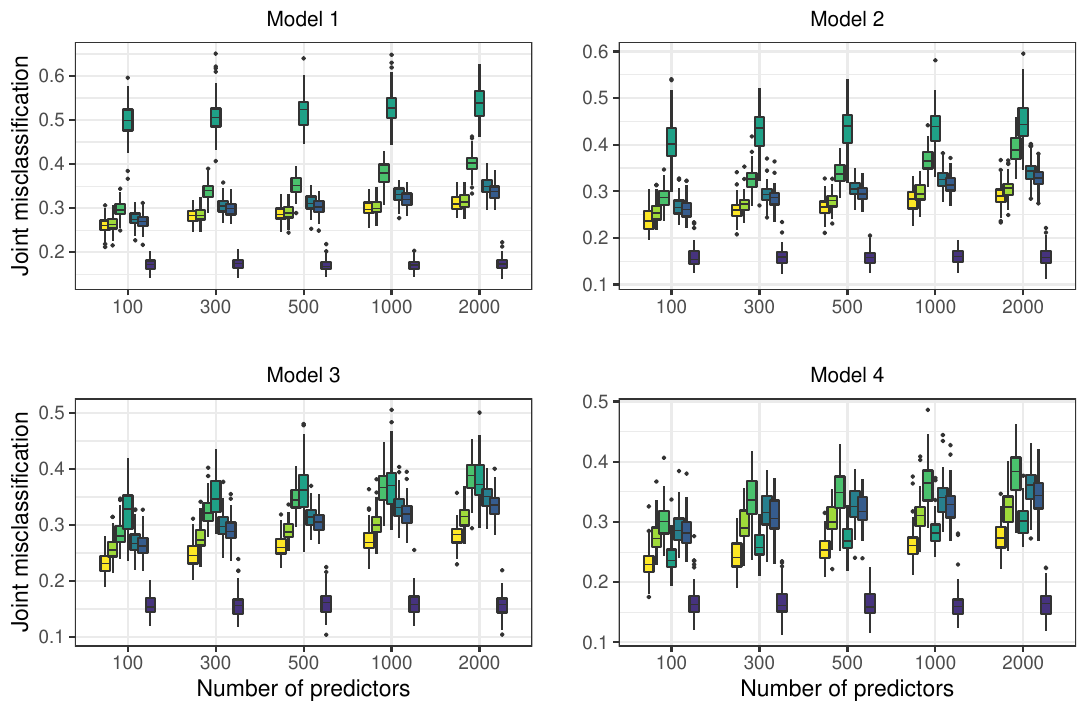}\\
\includegraphics[width=15.5cm]{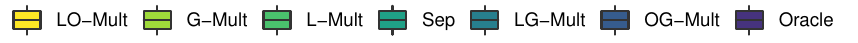}
\vspace{-15pt}
\caption{Joint misclassification rates under Models 1--4 with $p \in \left\{100, 300, 500, 1000, 2000\right\}$ and tuning parameters chosen to maximize the validation likelihood. }\label{fig:joint_Val}
\end{figure}

\begin{figure}[t]
\centering
\includegraphics[width=16cm]{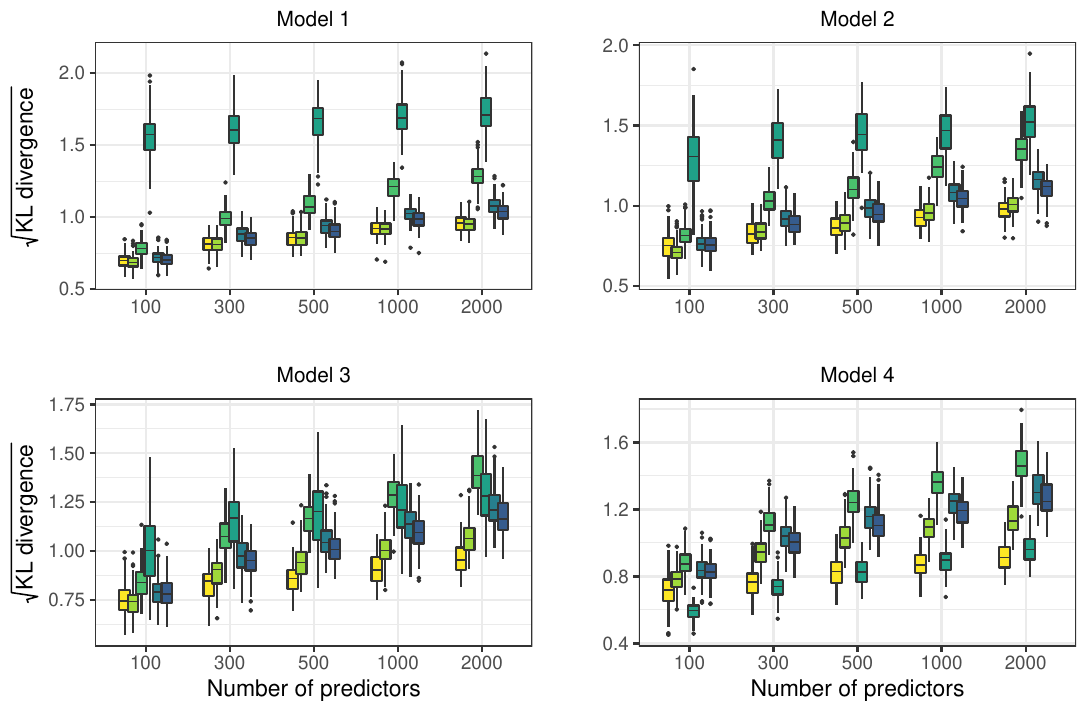}\\
\includegraphics[width=14.5cm]{Plots/Legend_KL.pdf}
\vspace{-15pt}
\caption{Square-root average Kullback-Leibler diverence under Models 1--4 with $p \in \left\{100, 300, 500, 1000, 2000\right\}$ and tuning parameters chosen to maximize the validation likelihood. }\label{fig:KL_Val}
\end{figure}

\begin{figure}[t]
\centering
\includegraphics[width=16cm]{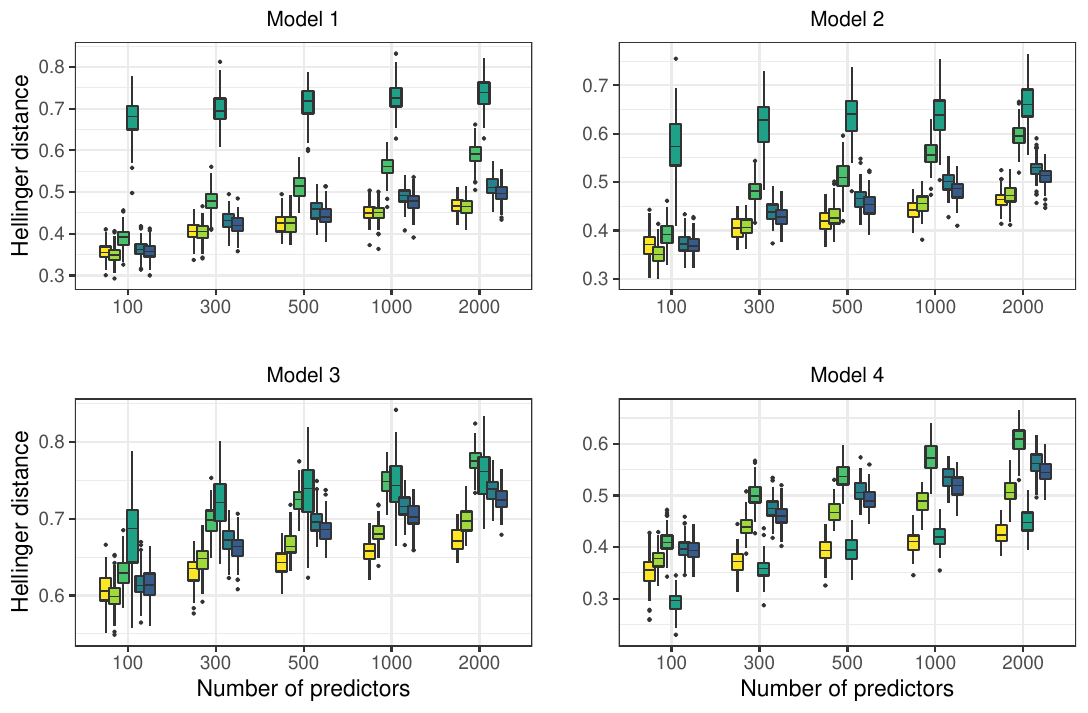}\\
\includegraphics[width=14cm]{Plots/Legend_KL.pdf}
\vspace{-15pt}
\caption{Average Hellinger distance under Models 1--4 with $p \in \left\{100, 300, 500, 1000, 2000\right\}$ and tuning parameters chosen to maximize the validation likelihood. }\label{fig:HELL_Val}
\end{figure}

In Figures \ref{fig:joint_Val}, \ref{fig:KL_Val}, and \ref{fig:HELL_Val} we display the joint misclassification rates, average KL divergence, and average Hellinger distance under exactly the data generating models in Section \ref*{sec:Simulation_Studies}, but with tuning parameters chosen to maximize the validation likelihood. As can be seen comparing these results to those from Section \ref*{sec:Simulation_Studies}, the metric used to select tuning parameters does have an effect on the results. While, the relative performances of each methods is essentially unchanged; and the classification accuracy decreases whereas the KL divergence and Hellinger distances are larger than when selecting tuning parameters by minimizing the validation misclassification rate.

\subsection{Additional performance metrics and details}
\begin{figure}[t]
\centering{}
\includegraphics[width=16cm]{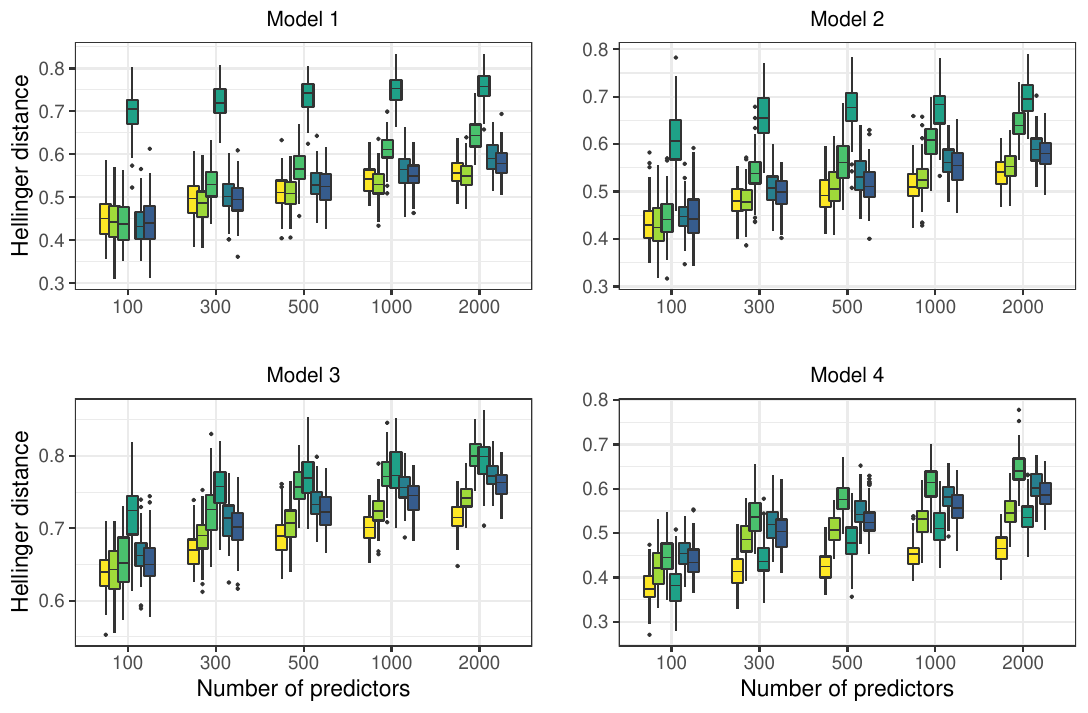}\\
\includegraphics[width=14.5cm]{Plots/Legend_KL.pdf}
\vspace{-15pt}
\caption{Average Hellinger distance under Models 1--4 with $p \in \left\{100, 300, 500, 1000, 2000\right\}$ and tuning parameters chosen as in Section \ref{sec:Simulation_Studies}. }\label{fig:hellinger}
\end{figure}

\begin{figure}[t]
\centering{}
\includegraphics[width=16cm]{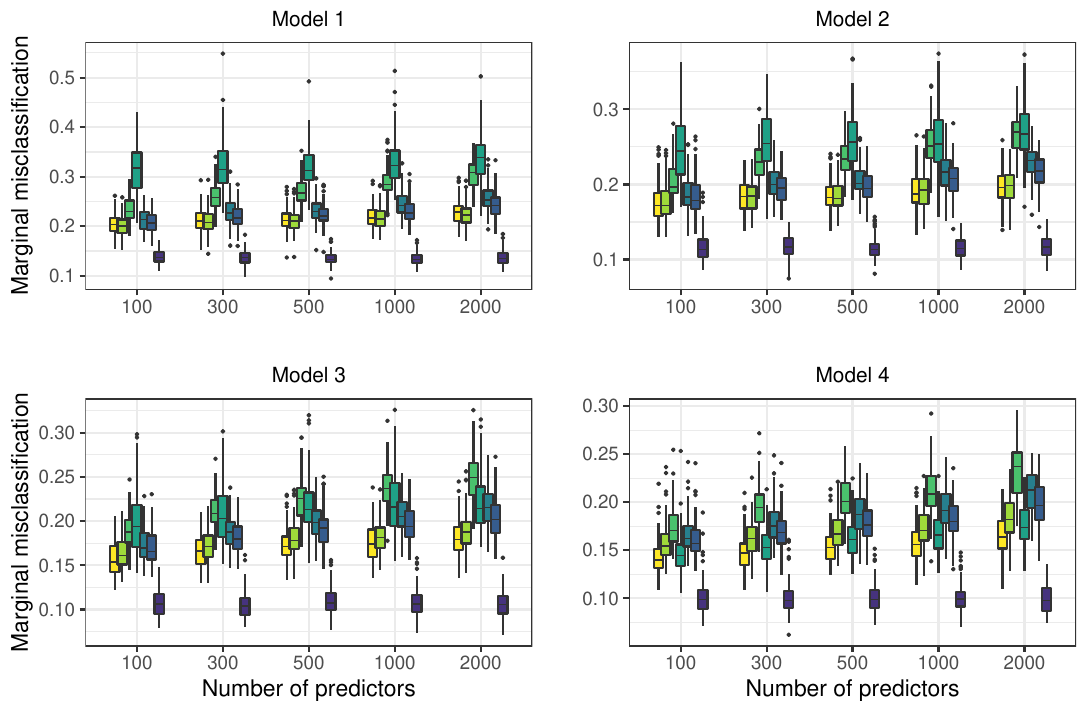}\\
\includegraphics[width=15.5cm]{Plots/Legend_Misclass2.pdf}
\vspace{-15pt}
\caption{Marginal misclassification rates (for the $J$-category response variable) under Models 1--4 with $p \in \left\{100, 300, 500, 1000, 2000\right\}$. }\label{fig:marg}
\end{figure}

In Figure \ref{fig:hellinger} and \ref{fig:marg}, we display the average test set Hellinger distances and marginal misclassification rates, respectively, under the same data generating models and tuning parameter selection criterion as in Section \ref*{sec:Simulation_Studies}. In Figure \ref{fig:parameters_penalized}, we provide a visualization of the groups being penalized by both the overlapping group lasso (\texttt{OG-Mult}) and latent group lasso (\texttt{LG-Mult}) estimators described in Section \ref*{sec:Simulation_Studies}.

\begin{figure}[t!]
\centering
\begin{equation*}
  \left(\begin{array}{>{\columncolor{gray!40}}ccc}
    \beta_{j,1,1}  & \beta_{j,1,2}  & \beta_{j,1,3} \\
 \beta_{j,2,1}  & \beta_{j,2,2}  & \beta_{j,2,3} 
  \end{array}\right)\quad \quad 
   \left(\begin{array}{c>{\columncolor{gray!40}}cc}
    \beta_{j,1,1}  & \beta_{j,1,2}  & \beta_{j,1,3} \\
 \beta_{j,2,1}  & \beta_{j,2,2}  & \beta_{j,2,3} 
  \end{array}\right) \quad \quad 
    \left(\begin{array}{cc>{\columncolor{gray!40}}c}
    \beta_{j,1,1}  & \beta_{j,1,2}  & \beta_{j,1,3} \\
 \beta_{j,2,1}  & \beta_{j,2,2}  & \beta_{j,2,3} 
  \end{array}\right)
\end{equation*}
\begin{equation*}
  \left(\begin{array}{ccc}
  \rowcolor{gray!40}
    \beta_{j,1,1}  & \beta_{j,1,2}  & \beta_{j,1,3} \\
 \beta_{j,2,1}  & \beta_{j,2,2}  & \beta_{j,2,3} 
  \end{array}\right)
  \quad\quad 
  \left(\begin{array}{ccc}
    \beta_{j,1,1}  & \beta_{j,1,2}  & \beta_{j,1,3} \\
     \rowcolor{gray!40}
 \beta_{j,2,1}  & \beta_{j,2,2}  & \beta_{j,2,3} 
  \end{array}\right)
\quad\quad 
\end{equation*}
\caption{The groups of parameters which are penalized by both the overlapping and latent group-penalized multivariate multinomial estimators in (\ref*{eq:Overlap}) and (\ref*{eq:Latent}) with $J=2$ and $K=3$ for $j= 2, \dots, p$. }\label{fig:parameters_penalized}
\end{figure}

\section{Results with $J= 4$ and $K=3$}\label{subsec:J4K3}
In this section, we present simulation studies essentially identical to those from Section \ref*{sec:Simulation_Studies}, but with $J = 4$ and $K=3$. The data generating models differ only in how $\beta^*$ is constructed under Models 2--4. In this setting, we simply find a $V$ such that $V \in {\rm Null}(D')$ and set the rows of $\beta_*$ corresponding to predictors affecting only marginal distributions to be equal to $V u \in \mathbb{R}^{12}$ where $u \in \mathbb{R}^{6}$ with each element drawn independently from ${\rm Uniform}(-3, 3).$ This way, for each $\beta_{j,:} = Vu $, we have that $\beta_{j, :} \neq 0_{12}$, but $D'\beta_{j,:} = 0_{12}.$

\begin{figure}[t]
\centering
\includegraphics[width=16cm]{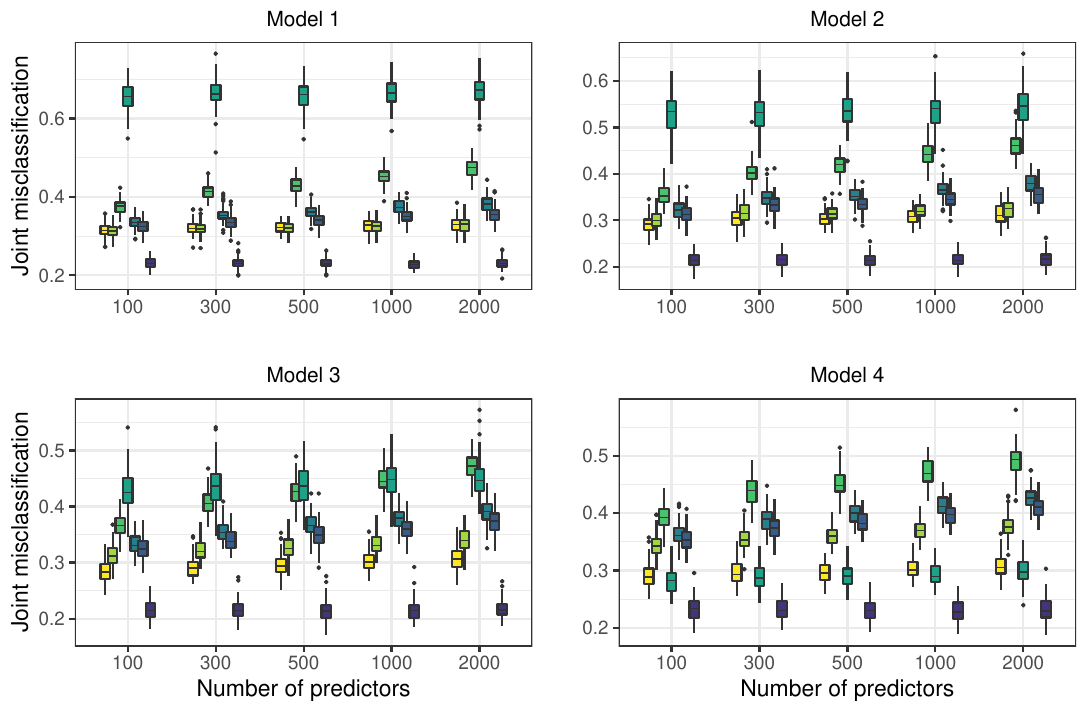}\\
\includegraphics[width=15.5cm]{Plots/Legend_Misclass2.pdf}
\vspace{-15pt}
\caption{Joint misclassification rates under Models 1--4 with $p \in \left\{100, 300, 500, 1000, 2000\right\}$ with $J = 4$ and $K=3$.}\label{fig:joint_J4_K3}
\end{figure}

\begin{figure}[t]
\centering
\includegraphics[width=16cm]{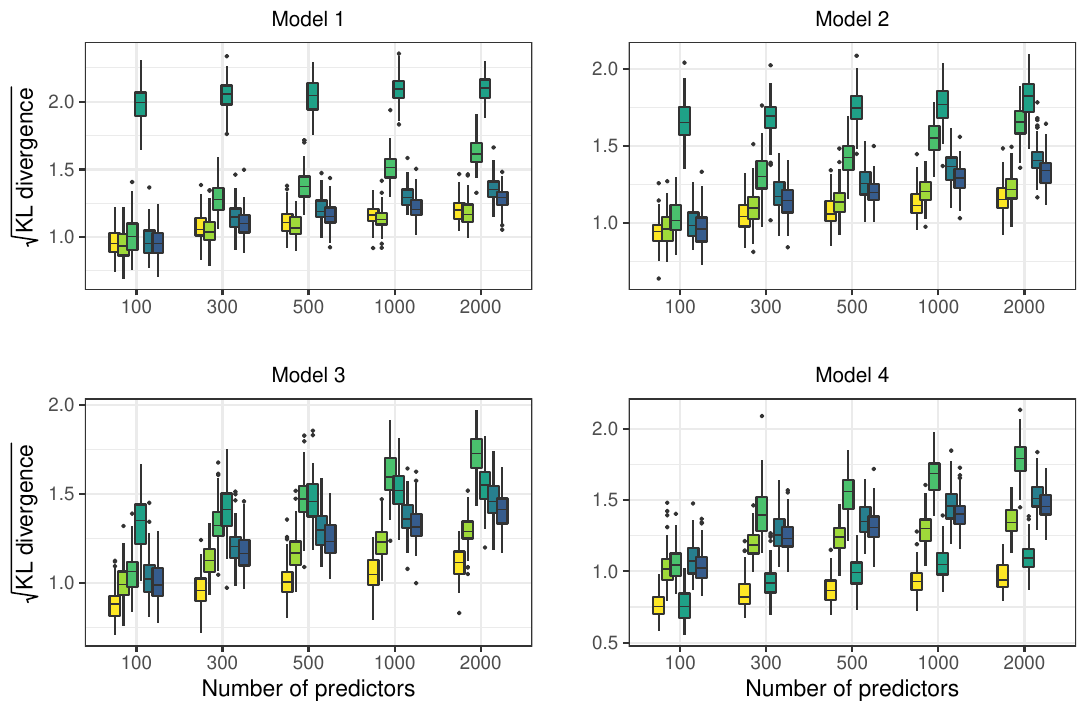}\\
\includegraphics[width=14.5cm]{Plots/Legend_KL.pdf}
\vspace{-10pt}
\caption{Square-root average Kullback-Leibler divergence under Models 1--4 with $p \in \left\{100, 300, 500, 1000, 2000\right\}$ with $J = 4$ and $K=3$. }\label{fig:KL_J4_K3}
\end{figure}

Misclassification rates and average KL divergences are displayed in Figures \ref{fig:joint_J4_K3} and \ref{fig:KL_J4_K3}. The performance of the 
methods relative to one another is quite similar to the settings where $J = 3$ and $K = 2$. In general, each method performs slightly worse, which can be easily explained by the fact that with more response categories, lower classification accuracy (even for the oracle) is expected. 

\section{Trivariate categorical response simulations}\label{sec:TrivariateApplication}
In this section, we present results from a simulation study in which we considered a trivariate response. That is, we have three response variables with $J = K = L = 2$ categories each and 
$$ P(Y_1 = j, Y_2 = k, Y_3 = l \mid x) = \frac{{\rm exp}(x'\boldsymbol{\beta}^*_{:,j,k,l})}{\sum_{s=1}^J \sum_{t=1}^K \sum_{u=1}^L {\rm exp}(x'\boldsymbol{\beta}^*_{:,s,t,u})}$$
for $(j,k,l) \in \{1,2\} \times  \{1,2\} \times \{1,2\}.$ Define the matricized version of $\boldsymbol{\beta}^*$ as $\beta^* \in \mathbb{R}^{p \times JKL}$ where $\boldsymbol{\beta}^*_{:,j,k,l} = \beta^*_{:,h(j,k,l)}$ where $h(j,k,l) = (k-1)J + j + (l-1)JK.$
We will compare four methods for estimating the mass function of $(Y_1, Y_2, Y_3 \mid x):$ \texttt{LO-Mult}, \texttt{G-Mult}, \texttt{L-Mult}, and \texttt{Sep}. 

\subsection{Implementation}\label{subsec:TrivariateGeneralization}
In order to implement \texttt{LO-Mult}, we must first construct $D$ as described in Section \ref{sec:3MvMult}. Recalling that under the mapping $h$, $$\beta = (\boldsymbol{\beta}_{:,1,1,1}, \boldsymbol{\beta}_{:,2,1,1}, \boldsymbol{\beta}_{:,1,2,1},\boldsymbol{\beta}_{:,2,2,1}, \boldsymbol{\beta}_{:,1,1,2},\boldsymbol{\beta}_{:,2,1,2}, \boldsymbol{\beta}_{:,1,2,2}, \boldsymbol{\beta}_{:,2,2,2}) \in \mathbb{R}^{p \times JKL},$$
so that we have
\begin{equation} \label{eq:D_trivariate}
D' = \left(\begin{array}{rrrrrrrr}
1 & -1 & -1 & 1 & 0 & 0 & 0 & 0\\
0 & 0 & 0 & 0 & 1 & -1 & -1 & 1 \\
1 & -1  & 0 & 0 & -1 & 1 & 0 & 0 \\  
0 & 0 & 1 & -1 & 0 & 0 & -1 & 1\\
1 & 0 & -1 & 0 & -1 & 0 & 1& 0\\
0 & 1 & 0 & -1 & 0 & -1 & 0 & 1\\
\end{array}\right).
\end{equation}
Note that this is $D$ matrix is constructed according to the discussion on Section \ref*{sec:D_Mult}.
To apply Algorithm 1 to the trivariate setting, we need only consider how to solve (\ref*{eq:ProxOperator}) with $D$ as defined above. For this purpose, we can straightforwardly apply Theorem \ref*{proxSolutions}; however, the closed form solution for (iii) in Proposition \ref*{lemmaiii} no longer holds. In this setting, to obtain a $\tau$ which satisfies Theorem \ref*{proxSolutions} (iii), we resort to a numeric root-solver to find $\tau$. Note that the $D$ in \eqref{eq:D_trivariate} has $4$ non-zero singular values: their values are $(\sigma_1, \sigma_2, \sigma_3, \sigma_4) =  (\sqrt{12}, 2, 2, 2).$  Hence, by the same logic as in the proof of Proposition \ref*{lemmaiii}, letting $w_l = u_l'\nu$ (where $u_l$ is the $l$th left singular vector of $D$), we need $\tau$ such that 
$$ \sum_{l=1}^{{\rm rank}(D)} \frac{w_l^2 \sigma_l^2}{(\sigma_l^2 + \tau)^2} = \lambda^2 \implies 12 \frac{w_1^2}{(12 + \tau)^2} + 4 \sum_{l=2}^4 \frac{w_l^2}{(4 + \tau)^2} - \lambda^2 = 0.$$
Under the conditions of Theorem 2 (iii), such a $\tau > 0$ always exists and can be found using a numeric root-solver in R, e.g., \texttt{rootSolve}. For problems with moderately sized $J$, $K$, and $L$, this is can be done with reasonable efficiency. 

\subsection{Data generating models}
To compare the various methods in the trivariate categorical response setting, we consider four data generating models similar to those from Section \ref*{sec:Simulation_Studies}. Just as in Section \ref{subsec:J4K3}, we first obtain $V \in {\rm Null}(D')$ for the $D$ defined in \eqref{eq:D_trivariate}. Then, we consider Models 5--8.
\begin{figure}[t]
\centering
\includegraphics[width=16cm]{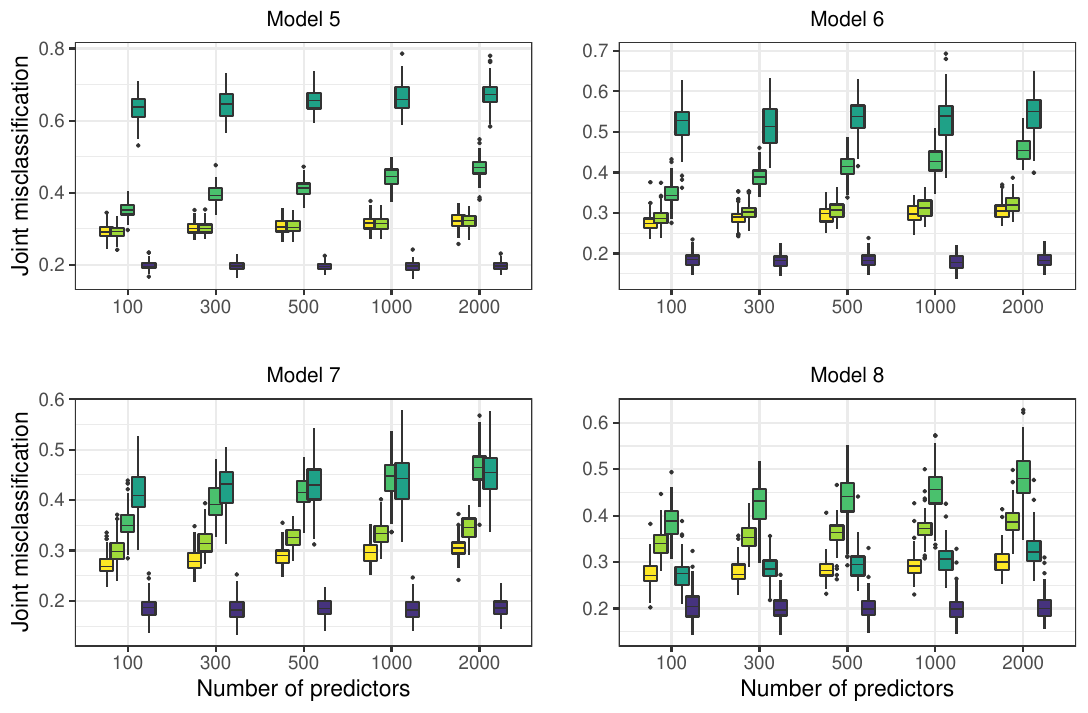}\\
\includegraphics[width=11cm]{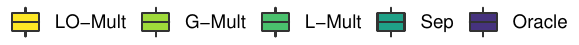}
\vspace{-15pt}
\caption{Joint misclassification rates under Models 5--8 with $p \in \left\{100, 300, 500, 1000, 2000\right\}$ with $J = K = L = 2$.}\label{fig:joint_Mult}
\end{figure}
\begin{itemize}
    \item\textbf{Model 5}: We randomly select 10 rows of $\beta^* \in \mathbb{R}^{p \times JKL}$ to be nonzero. Each of the elements of these tens rows is set equal to independent realizations of a ${\rm Uniform}(-3,3)$ random variable. 
    \item \textbf{Model 8}: We randomly select 10 rows of $\beta^*$ to be nonzero. For each row independently, we generate four independent realizations of a ${\rm Uniform}(-3,3)$ random variable. Given these realizations, say $(u_1, u_2, u_3, u_4)$, we set the row of $\beta^*$ equal to 
$Vu$
Under this construction, we can see 
$D'Vu = 0_6.$
\end{itemize}

Just as in Section \ref*{sec:Simulation_Studies}, Models 6 and 7 are, in a sense, intermediate to Models 6 and 7. 
\begin{itemize} 
    \item \textbf{Model 6}: We randomly select six rows of $\beta^*$ to be nonzero and consist elements which are each independent realizations of a ${\rm Uniform}(-3,3)$ random variable. Then, we select an additional four rows of $\beta^*$ to be generated in the same manner as Model 4. 
    \item \textbf{Model 7}: We randomly select three rows of $\beta^*$ to be nonzero and consist elements which are each independent realizations of a ${\rm Uniform}(-3,3)$ random variable. Then, we select an additional seven rows of $\beta^*$ to be generated in the same manner as Model 4. 
\end{itemize}
As mentioned, in these simulation studies, we only consider the estimators \texttt{LO-Mult}, \texttt{G-Mult}, \texttt{L-Mult}, \texttt{Sep}, and when appropriate, \texttt{Oracle}. 

\subsection{Results}
In this section, we discuss results under Models 5--8. In Figure \ref{fig:joint_Mult}, we present the joint (i.e., trivariate) misclassification rates for each of the considered methods.  Relative performances are essentially the same as in the various bivariate settings considered previously. Under Model 5, \texttt{LO-Mult} and \texttt{G-Mult} perform similarly -- which is to be expected for the same reasons as described in Section \ref{sec:Simulation_Studies}. As we move from Model 5 to Models 6--8, we see that \texttt{LO-Mult} starts to outperform \texttt{G-Mult}. Meanwhile, \texttt{Sep} begins to perform better as we move from Model 5 towards Model 8: in Model 8, \texttt{Sep} -- which correctly assumes the responses are independent -- performs nearly as well as \texttt{LO-Mult}.

\begin{figure}[t]
\centering
\includegraphics[width=16cm]{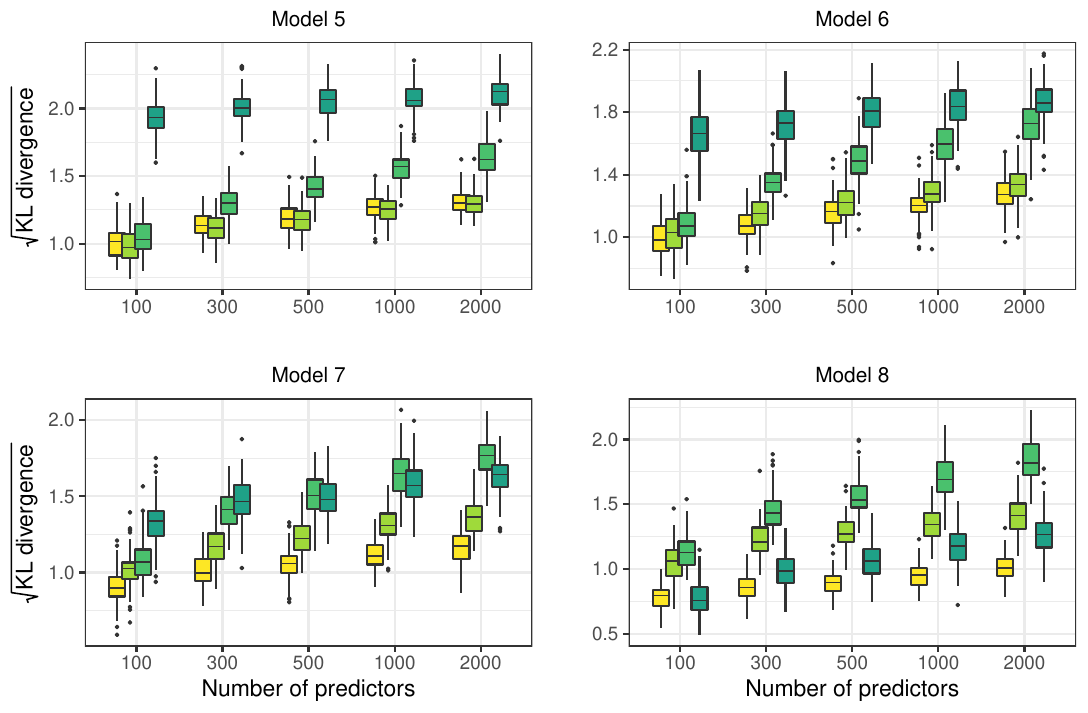}\\
\includegraphics[width=9cm]{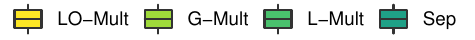}\\
\vspace{-10pt}
\caption{Square-root average Kullback-Leibler divergence under Models 5--8 with $p \in \left\{100, 300, 500, 1000, 2000\right\}$ with $J = K = L = 2$. }\label{fig:KL_Mult}
\end{figure}

\begin{figure}[t]
\centering
\includegraphics[width=16cm]{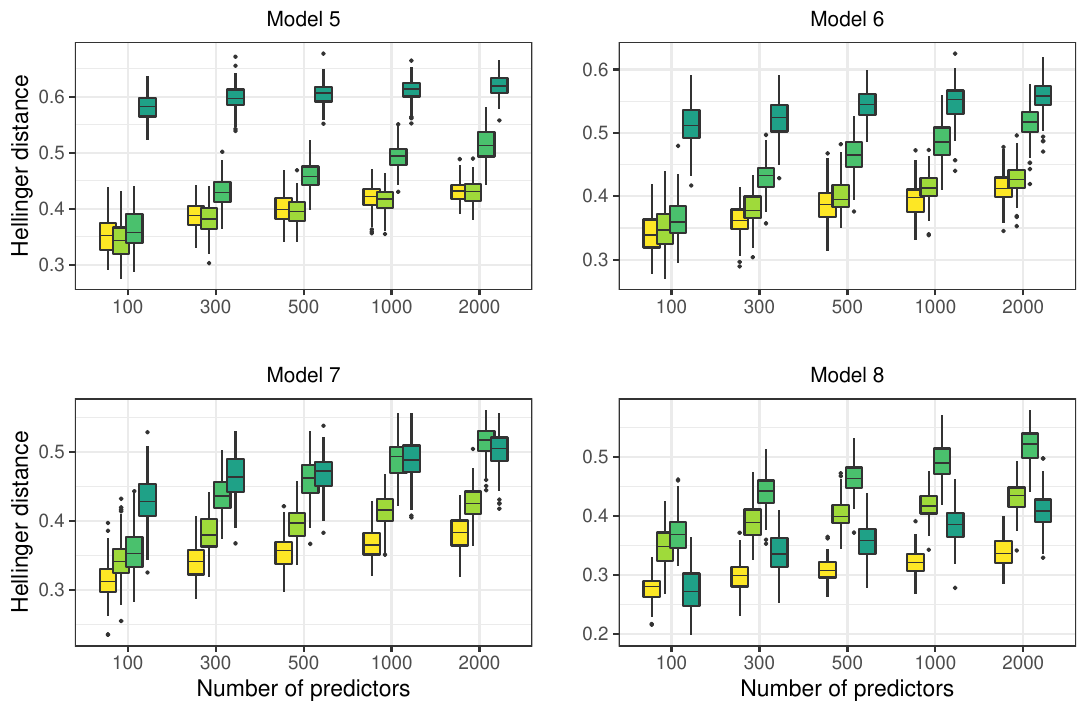}\\
\includegraphics[width=9cm]{Plots/Legend_KL_Mult.pdf}\\
\vspace{-10pt}
\caption{Average Hellinger distance under Models 5--8 with $p \in \left\{100, 300, 500, 1000, 2000\right\}$ with $J = K = L = 2$. }\label{fig:HELL_Mult}
\end{figure}

In Figure \ref{fig:KL_Mult} and \ref{fig:HELL_Mult}, we display both average Kullback-Leibler divergence and average Hellinger distances for the various methods. Just as with classification accuracy, performances largely agree with the bivariate setting. Of particular note is that as $p$ grows, \texttt{LO-Mult} tends to outperform competitors more relative to when, say, $p= 100$. 

\section{Proofs of results in Section \ref*{sec:Computing}}

In this and the following sections, for ease of display, we omit the subscript on $0$ when refering to a matrix or vector of zeros. 
The key to proving Theorem \ref{proxSolutions} is the following lemma, which reveals that we need only concern ourselves with computing $\hat{\eta}_{\bar\lambda,0}$.
\begin{lemma}\label{lemma:TwoStepSolution}
Let $\hat{\eta}_{\bar\lambda,\bar\gamma}$ be a minimizer of (\ref*{eq:ProxOperator}) and let $\hat{\eta}_{\bar\lambda,0}$ be the minimizer of (\ref*{eq:ProxOperator}) with $\bar\gamma = 0$. Then
\begin{equation}\label{eq:rowSoft}
\hat{\eta}_{\bar\lambda, \bar\gamma} = \left\{ \begin{array}{cc} 
\left(1 - \frac{\bar\gamma}{\|\hat{\eta}_{\bar\lambda,0}\|_2}\right)\hat{\eta}_{\bar\lambda,0} & :\|\hat{\eta}_{\bar\lambda,0}\|_2 > \bar\gamma\\
0 & :\|\hat{\eta}_{\bar\lambda,0}\|_2 \leq \bar\gamma
\end{array}\right..
\end{equation} 
\end{lemma}

\noindent \textbf{Proof of Lemma \ref{lemma:TwoStepSolution}.} To prove Lemma \ref{lemma:TwoStepSolution}, we show that first-order conditions for $\hat\eta_{\bar\lambda, 0}$ imply the first-order conditions for $\hat\eta_{\bar\lambda, \bar\gamma}$ as defined in \eqref{eq:rowSoft}. First, recall that the zero subgradient equation for $\hat{\eta}_{\bar\lambda, 0}$ is 
\begin{equation}\label{eq:first_order_eta0}
0 = -\nu  + \hat{\eta}_{\bar\lambda, 0} + \bar\lambda D\tilde\phi 
\end{equation}
for some $\tilde{\phi}$ 
such that $\tilde\phi = D'\hat{\eta}_{\bar\lambda, 0}/\|D'\hat{\eta}_{\bar\lambda, 0}\|_2$ if $[D'\hat{\eta}_{\bar\lambda, 0}] \neq 0$ and $\|\tilde\phi\|_2 \leq 1$ otherwise (i.e., $\tilde{\phi}$ is a subgradient of $\eta \mapsto \|D'\eta\|_2$ at $\hat{\eta}_{\bar\lambda, 0}$). Then, recall that the zero subgradient equation for $\hat\eta_{\bar\lambda, \bar\gamma}$ is 
\begin{equation} \label{eq:first_order_gamma}
0 = -\nu + \hat\eta_{\bar\lambda, \bar\gamma} + \bar\lambda D\phi + \bar\gamma v,
\end{equation}
for $(v, \phi) \in \mathbb{R}^{JK} \times \mathbb{R}^{\binom{J}{2}\binom{K}{2}} $ such that $v = \hat\eta_{\bar\lambda, \bar\gamma}/\|\hat\eta_{\bar\lambda, \bar\gamma}\|_2$ if $\hat\eta_{\bar\lambda, \bar\gamma} \neq 0$ and $\|v\|_2 \leq 1$ otherwise; and $\phi = D'\hat{\eta}_{\bar\lambda, \bar\eta}/\|D'\hat{\eta}_{\bar\lambda, \bar\eta}\|_2$ if $D'\hat{\eta}_{\bar\lambda, \bar\eta} \neq 0$ and $\|\phi\|_2 \leq 1$ otherwise.

We will consider three cases: (i) $\|\hat{\eta}_{\bar\lambda, 0}\|_2 > \bar\gamma$, (ii) $0 < \|\hat{\eta}_{\bar\lambda, 0}\|_2 \leq \bar\gamma,$ and (iii) $\hat{\eta}_{\bar\lambda, 0} = 0$.\\ 

\noindent Case (i): We know from \eqref{eq:first_order_eta0} that there exists a subgradient $\tilde\phi$ such that
\begin{equation}\label{eq:zero_subgrad1}
0 = -\nu  + \hat{\eta}_{\bar\lambda, 0} + \bar\lambda D\tilde\phi.
\end{equation}
We assume that  $\|\hat{\eta}_{\bar\lambda, 0}\|_2 > \bar\gamma$ so that $\hat{\eta}_{\bar\lambda, \bar\gamma} = \hat{\eta}_{\bar\lambda, 0}(1 - \bar\gamma/\|\hat{\eta}_{\bar\lambda, 0}\|_2).$ We will show that this $\hat{\eta}_{\bar\lambda, \bar\gamma}$ satisfies the first-order conditions \eqref{eq:first_order_gamma}. In particular, from \eqref{eq:zero_subgrad1}, we have
\begin{align}
& 0 = -\nu  + \hat{\eta}_{\bar\lambda, 0} + \bar\lambda D \tilde\phi \notag\\
\implies& 0 = -\nu  + \hat{\eta}_{\bar\lambda, 0} + \bar\lambda D \tilde\phi  + \bar\gamma\hat{\eta}_{\bar\lambda, 0}/\|\hat{\eta}_{\bar\lambda, 0}\|_2 - \bar\gamma\hat{\eta}_{\bar\lambda, 0}/\|\hat{\eta}_{\bar\lambda, 0}\|_2\notag\\
\implies& 0 = -\nu  + \hat{\eta}_{\bar\lambda, 0}(1 - \bar\gamma/\|\hat{\eta}_{\bar\lambda, 0}\|_2) + \bar\lambda D \tilde\phi  + \bar\gamma\hat{\eta}_{\bar\lambda, 0}/\|\hat{\eta}_{\bar\lambda, 0}\|_2\notag\\
\implies& 0 = -\nu  + \hat{\eta}_{\bar\lambda, 0}(1 - \bar\gamma/\|\hat{\eta}_{\bar\lambda, 0}\|_2) + \bar\lambda D \tilde\phi  + \bar\gamma\hat{\eta}_{\bar\lambda, 0}(1 - \bar\gamma/\|\hat{\eta}_{\bar\lambda, 0}\|_2)/\|\hat{\eta}_{\bar\lambda, 0}(1 - \bar\gamma/\|\hat{\eta}_{\bar\lambda, 0}\|_2)\|_2\notag\\
\implies& 0 = -\nu  + \hat{\eta}_{\bar\lambda, \bar\gamma} + \bar\lambda D \tilde\phi  + \bar\gamma\hat{\eta}_{\bar\lambda, \bar\gamma}/\|\hat{\eta}_{\bar\lambda, \bar\gamma}\|_2\label{eq:line_line_casei}
\intertext{
Since $\|\hat{\eta}_{\bar\lambda, \bar\gamma}\|_2 > 0$ by assumption on $\hat{\eta}_{\bar\lambda, 0}$, we can take $v = \hat{\eta}_{\bar\lambda, \bar\gamma}/\|\hat{\eta}_{\bar\lambda, \bar\gamma}\|_2$. It only remains to check that $\tilde{\phi} = \phi$ where $\phi = D'\hat\eta_{\bar\lambda, \bar\gamma}/\|D'\hat\eta_{\bar\lambda, \bar\gamma}\|_2$ if $D'\hat\eta_{\bar\lambda, \bar\gamma} \neq 0$ and $\|\phi\|_2 \leq 1$ otherwise. However, this is trivial since $\hat\eta_{\bar\lambda, \bar\gamma}$ is a scalar multiple of $\hat\eta_{\bar\lambda, 0}$, so $D'\hat\eta_{\bar\lambda, 0}$ is a scalar multiple of $D'\hat\eta_{\bar\lambda, \tilde\gamma}$. Thus, if $D'\hat\eta_{\bar\lambda, 0} \neq 0$, then $D'\hat\eta_{\bar\lambda, \bar\gamma} \neq 0$, whereas if $D'\hat\eta_{\bar\lambda, 0} = 0$, then $D'\hat\eta_{\bar\lambda, \bar\gamma} = 0$. In either case, we can take $\phi = \tilde{\phi}$ 
so that finally, from \eqref{eq:line_line_casei}, } 
& 0 = -\nu  + \hat{\eta}_{\bar\lambda, \bar\gamma} + \bar\lambda D \tilde\phi  + \bar\gamma\hat{\eta}_{\bar\lambda, \bar\gamma}/\|\hat{\eta}_{\bar\lambda, \bar\gamma}\|_2 \implies 0 = -\nu  + \hat{\eta}_{\bar\lambda, \bar\gamma} + \bar\lambda D\phi  + \bar\gamma v \notag
\end{align}
which verifies that $\hat\eta_{\bar\lambda, \bar\gamma}$ as defined in \eqref{eq:rowSoft} satisfies the first-order optimality conditions for (\ref*{eq:ProxOperator}) when $\|\hat\eta_{\bar\lambda, 0}\|_2 > \bar\gamma$.\\

\noindent Case (ii): Assume $0 < \|\hat{\eta}_{\bar\lambda, 0}\|_2 \leq \bar\gamma$. We will show that $\hat\eta_{\bar\lambda, \bar\gamma} = 0$ satisfies the first-order conditions for (\ref*{eq:ProxOperator}) given in \eqref{eq:first_order_gamma}. Recall that by definition, there exists a subgradient $\tilde\phi$ such that
\begin{equation}\label{zero:firstOrder1}
0 = -\nu  + \hat{\eta}_{\bar\lambda, 0} + \bar\lambda D\tilde{\phi}. 
\end{equation}
Since $\|\hat{\eta}_{\bar\lambda, 0}\|_2 \leq \bar{\gamma}$, $1 \leq \bar\gamma/\|\hat{\eta}_{\bar\lambda, 0}\|_2$, so we can write $ 1 = \bar\gamma/\|\hat{\eta}_{\bar\lambda, 0}\|_2 - z_1$ for some $z_1 \geq 0$ and thus, \eqref{zero:firstOrder1} implies
$$ 0 = -\nu + \hat{\eta}_{\bar\lambda, 0}\left(\frac{\bar\gamma}{\|\hat{\eta}_{\bar\lambda, 0}\|_2} - z_1\right) + \bar\lambda D \tilde{\phi} $$
which in turn implies
\begin{equation}\label{zero:firstOrder2}  0 = -\nu  + \hat\eta_{\bar\lambda, \bar\gamma}  + \bar\lambda D \tilde{\phi} + \bar\gamma \left (\frac{\hat{\eta}_{\bar\lambda, 0}}{\|\hat{\eta}_{\bar\lambda, 0}\|_2} - \frac{z_1\hat{\eta}_{\bar\lambda, 0}}{\bar{\gamma}}\right)
\end{equation}
since $\hat\eta_{\bar\lambda, \bar\gamma} = 0$ by assumption.
Then, because we must have $\|\phi\|_2 \leq 1$, we can simply take $\phi = \tilde{\phi}$ since $\|\tilde{\phi}\|_2 \leq 1$ regardless of whether $D'\hat\eta_{\bar\lambda, 0} = 0$ or $D'\hat\eta_{\bar\lambda, 0} \neq 0$. Thus, \eqref{zero:firstOrder2}
suggets that $\hat\eta_{\bar\lambda, \bar\gamma} = 0$ satisfies the first-order conditions for (\ref*{eq:ProxOperator}) as long as 
$$\left\Vert \frac{\hat{\eta}_{\bar\lambda, 0}}{\|\hat{\eta}_{\bar\lambda, 0}\|_2} - \frac{z_1\hat{\eta}_{\bar\lambda, 0}}{\bar{\gamma}}\right\Vert_2 \leq 1.$$
Letting $z_2 = \hat{\eta}_{\bar\lambda, 0}/\|\hat{\eta}_{\bar\lambda, 0}\|_2$ so that $\|z_2\|_2 = 1$, we have 
$$\left\Vert 
\frac{\hat{\eta}_{\bar\lambda, 0}}{\|\hat{\eta}_{\bar\lambda, 0}\|_2} - \frac{z_1\hat{\eta}_{\bar\lambda, 0}}{\bar{\gamma}}\right\Vert_2 = \| z_2(1  -  \bar\gamma^{-1} z_1 \|\hat{\eta}_{\bar{\lambda}, 0}\|_2)\|_2 = \|z_2\|_2 \left(1  -  \frac{z_1 \|\hat{\eta}_{\bar\lambda, 0}\|_2}{\bar{\gamma}}\right) = \left(1  -  \frac{z_1}{1 + z_1}\right) \leq 1.$$
Therefore, with $v = \frac{\hat{\eta}_{\bar\lambda, 0}}{\|\hat{\eta}_{\bar\lambda, 0}\|_2} - \frac{z_1\hat{\eta}_{\bar\lambda, 0}}{\bar{\gamma}}$, from \eqref{zero:firstOrder2} we can conclude,  
$$0 = -\nu  + \hat\eta_{\bar\lambda, \bar\gamma} + \bar\lambda D \tilde{\phi} + \bar\gamma \left (\frac{\hat{\eta}_{\bar\lambda, 0}}{\|\hat{\eta}_{\bar\lambda, 0}\|_2} - \frac{z_1\hat{\eta}_{\bar\lambda, 0}}{\bar{\gamma}}\right)  \implies 0 = -\nu + \hat\eta_{\bar\lambda, \bar\gamma}+ \bar\lambda D\phi + \bar\gamma v   $$
for a $(v, \phi)\in \mathbb{R}^{JK} \times \mathbb{R}^{\binom{J}{2}\binom{K}{2}} $ such that $\|v\|_2 \leq 1$ and $\|\phi\|_2 \leq 1$, which is exactly the zero subgradient equation when $\hat\eta_{\bar\lambda,\bar\gamma} = 0$. \\

\noindent Case (iii): This case is trivial: to see that zero subgradient equation for $\hat\eta_{\bar\lambda, 0} = 0$ implies the zero subgradient equation for $\hat\eta_{\bar\lambda, \bar\gamma} = 0$, simply take $\phi = \tilde\phi$ and $v = 0$. \quad\quad $\blacksquare$\\

With Lemma \ref{lemma:TwoStepSolution} in place, we are ready to prove Theorem \ref*{proxSolutions}.\\

\noindent \textbf{Proof of Theorem \ref*{proxSolutions}.} Recall that the zero subgradient equation for $\hat\eta_{\bar\lambda, \bar\gamma}$ is
\begin{equation}\label{eq:first_order_again}
 0 = -\nu + \hat\eta_{\bar\lambda, \bar\gamma} + \bar\lambda D\phi + \bar\gamma v,
 \end{equation}
where $$v \in \{v \in \mathbb{R}^{JK}: v = \hat\eta_{\bar\lambda, \bar\gamma}/\|\hat\eta_{\bar\lambda, \bar\gamma}\|_2 \text{ if }\hat\eta_{\bar\lambda, \bar\gamma} \neq 0 \text{ and }\|v\|_2 \leq 1 \text{ otherwise}\},$$ and
$$\phi \in \{\phi \in \mathbb{R}^{\binom{J}{2}\binom{K}{2}}: \phi = D'\hat{\eta}_{\bar\lambda, \bar\eta}/\|D'\hat{\eta}_{\bar\lambda, \bar\eta}\|_2\text{ if }D'\hat{\eta}_{\bar\lambda, \bar\eta} \neq 0\text{ and }\|\phi\|_2 \leq 1\text{ otherwise}\}.$$ We consider each of the three cases set out in the statement of Theorem \ref{proxSolutions}.  To deal with cases (ii) and (iii), we focus on the solution for $\hat\eta_{\bar\lambda, 0}$ and then apply Lemma \ref{lemma:TwoStepSolution}.\\

\noindent Case (i): If $\|\nu\|_2 \leq \bar\gamma$, we can set $\hat\eta_{\bar\lambda, \bar\gamma} = 0$, $\phi = 0$, and $v = \nu/\bar\gamma$, so that $\|v\|_2 \leq 1$, and thus, $\hat\eta_{\bar\lambda, \bar\gamma} = 0$ would satisfy the first-order conditions \eqref{eq:first_order_again}. \\

\noindent Case (ii): We consider the dual problem of (\ref*{eq:ProxOperator}) with $\bar\gamma = 0$ (e.g., see the derivation of a related dual problem in Section 4 of \citet{tibshirani2011solution}):
$$ \hat{u} \in \argmin_{u} \|\nu - Du\|_2^2, \quad \|u\|_2 \leq \bar\lambda,$$
where $\hat{\eta}_{\bar\lambda, 0} = \nu - D\hat{u}$. 
Hence, if $\|(D'D)^{-}D'\nu\|_2 \leq \bar\lambda$, $\hat{u} = (D'D)^{-}D'\nu$, so it would follow that $\hat{\eta}_{\bar\lambda, 0} = \nu - D(D'D)^{-}D'\nu = \mathcal{P}^\perp_{D,0} \nu$. An application of Lemma \ref{lemma:TwoStepSolution} yields the second result.\\

\noindent Case (iii): We again consider the dual problem of (\ref*{eq:ProxOperator}) with $\bar\gamma = 0$. If $\|(D'D)^{-}D'\nu\|_2 > \bar{\lambda}$, it must be that the minimizer $\hat{u}$ is only the boundary of the constraint set $\{u:\|u\|_2 \leq \bar\lambda\}$, or equivalently, $\|\hat{u}\|_2^2 = \bar\lambda^2$. Then, because there is a one-to-one correspondence between the constrained version of ridge regression and its Lagrangian form when the constraint is active, we know there exists a $\tau > 0$ such that for every $\bar\lambda$ satisfying the condition of (iii),
$$ \hat{u} = \argmin_{u: \|u\|_2^2 \leq \bar\lambda^2} \|\nu - Du\|_2^2  = \argmin_{u } \|\nu - Du\|_2^2 + \tau \|u\|_2^2,$$
and thus, since $(D'D + \tau I)^{-1}D'\nu$ minimizes the rightmost objective function above, if
$ \|(D'D + \tau I)^{-1}D'\nu\|_2^2 = \bar\lambda^2$, we know $\hat{u} = (D'D + \tau I)^{-1}D'\nu$. The result then follows from $\nu - D(D'D + \tau I)^{-1}D'\nu = \mathcal{P}^\perp_{D, \tau} \nu$ and Lemma \ref{lemma:TwoStepSolution}. $\blacksquare$\\

Next, we provide a sketch of the proof of Proposition \ref*{lemmaiii}. \\

\noindent \textbf{Proof of Proposition \ref*{lemmaiii}.}
Let $U  {\rm Diag}\left(\{\sigma_l\}_{l=1}^{k} \right) V'$ be the singular value decomposition of $D$ where $k = \min(JK, \binom{J}{2}\binom{K}{2})$, $U'U = I_k$, $V'V = I_k$, and $\sigma_l \geq 0$ for $l \in [k]$. Note that by construction, only the first $r = (J-1)(K-1)$ singular values of $D$ are nonzero (e.g., see discussion of $D$ versus $\mathcal{D}$ in Section 2). Then, letting $\Sigma = {\rm Diag}\left(\{\sigma_l\}_{l=1}^{k} \right)$, we can write
\begin{align*}
(D'D + \tau I)^{-1}D'\nu & = V (\Sigma^2 + \tau I)^{-1}\Sigma U'\nu
\end{align*}
so that $$\|(D'D + \tau I)^{-1}D'\nu\|_2 = \bar\lambda \iff \nu'U\Sigma  (\Sigma^2 + \tau I)^{-2}\Sigma U'\nu = \bar\lambda^2.$$
Letting $u_l$ denote the $l$th column of $U$, we can define $w = (w_1, \dots, w_k)' \in \mathbb{R}^k$ where $w_l = u_l'\nu \in \mathbb{R}$ so that we may write
$$ \nu'U \Sigma(\Sigma^2 + \tau I)^{-2}\Sigma U'\nu = w' A w,$$
where $A$ is diagonal with $(l,l)$th entry 
$(\sigma_l^2 + \tau)^{-2}\sigma_l^2$. Thus, it follows that $$w'A w = \sum_{l=1}^{r} \frac{w_l^2 \sigma_l^2 }{(\sigma_l^2 + \tau)^2},$$  
which yields the first result. Then because for each $l \in [r]$,  $\sigma_l = \sqrt{JK}$, it further follows that
$$ \sum_{l=1}^{r} \frac{w_l^2 \sigma_l^2 }{(\sigma_l^2 + \tau)^2} = \lambda^2 \implies JK \sum_{l=1}^{r} \frac{ w_l^2  }{(JK + \tau)^2} = \lambda^2.$$
And thus, the previous equality implies 
\begin{align*} 
\tau = \frac{\sqrt{JK\sum_{l=1}^{r} w_l^2}}{\bar\lambda} - JK.
\end{align*}
It is easy to check that under the conditions of (iii), this $\tau$ must be positive. $~~\blacksquare$\\

\noindent \textbf{Proof of Theorem \ref*{theorem:ExactSol}.} 
We again appeal to Lemma \ref{lemma:TwoStepSolution}, which will give us the result for $\hat\eta_{\bar\lambda, \bar\gamma}$ once we have obtained the expression for $\hat\eta_{\bar\lambda, 0}$. We thus focus on the solution for $\hat\eta_{\bar\lambda, 0}$. Recall that when $J = K = 2$, $D'\hat\eta_{\bar\lambda, 0} \in \mathbb{R}$ and $\tilde\phi \in \mathbb{R}$, where $\tilde{\phi} = {\rm sign}(D'\hat\eta_{\bar\lambda, 0})$ if $D'\hat\eta_{\bar\lambda, 0} \neq 0$ and $\tilde{\phi} \in [-1,1]$ otherwise.  We consider all three cases enumerated in the statement of Theorem \ref*{theorem:ExactSol}. Let $\ddot{\nu} = \nu_1 - \nu_2 - \nu_3 + \nu_4$ and recall in this setting, $D = (1, -1, -1, 1)'.$\\

\noindent Case (iii): Suppose $\ddot{\nu} < - 4\bar\lambda$. If we let $\hat\eta_{\bar\lambda, 0} = (\nu_1 + \bar\lambda, \nu_2 - \bar\lambda, \nu_3 - \bar\lambda, \nu_4 + \bar\lambda)'$, then the subgradient of the objective is
$$ -\nu + \hat\eta_{\bar\lambda, 0} + \bar\lambda D {\rm sign}(D'\hat\eta_{\bar\lambda, 0}) = -\left(\begin{array}{c}\nu_1 \\ \nu_2 \\\nu_3 \\ \nu_4\end{array}\right) + \left(\begin{array}{c}\nu_1 + \bar\lambda \\ \nu_2 + \bar\lambda \\ \nu_3- \bar\lambda \\ \nu_4+ \bar\lambda\end{array}\right) -\bar\lambda D$$
since  $${\rm sign}(D'\hat\eta_{\bar\lambda, 0})  = {\rm sign}(\nu_1 + \bar\lambda - ( \nu_2 - \bar\lambda) - (\nu_3 - \bar\lambda) + \nu_4 + \bar\lambda) = {\rm sign}(\ddot{\nu} + 4 \bar\lambda) = -1$$
by our assumption $\ddot{\nu} < - 4\bar\lambda$. Hence, because 
$$ -\left(\begin{array}{c}\nu_1 \\ \nu_2 \\\nu_3 \\ \nu_4\end{array}\right) + \left(\begin{array}{c}\nu_1 + \bar\lambda \\ \nu_2 - \bar\lambda \\ \nu_3- \bar\lambda \\ \nu_4+ \bar\lambda\end{array}\right) -\bar\lambda \left(\begin{array}{r}1 \\ -1 \\-1\\1\end{array}\right) = 0,$$ when $\ddot{\nu} < - 4\bar\lambda$, the first-order conditions
$$-\nu + \hat\eta_{\bar\lambda, 0} + \bar\lambda D {\rm sign}(D'\hat\eta_{\bar\lambda, 0}) = 0$$
are satisfied with $\hat\eta_{\bar\lambda, 0} = (\nu_1 + \bar\lambda, \nu_2 - \bar\lambda, \nu_3 - \bar\lambda, \nu_4 + \bar\lambda)'.$\\

\noindent Case (ii): When $\ddot{\nu} > 4\bar\lambda$, the result follows from a nearly identical proof as in case (iii). \\

\noindent Case (i): Suppose $|\ddot{\nu}| \leq 4\bar\lambda$. Let $\hat\eta_{\bar\lambda, 0} = (\nu_1 - \ddot{\nu}/4, \nu_2+ \ddot{\nu}/4, \nu_3 +  \ddot{\nu}/4, \nu_4 - \ddot{\nu}/4)'$. We want to show that 
\begin{equation}\label{eq:firstOrderCasei}
-\nu + \hat\eta_{\bar\lambda, 0} + \bar\lambda D u = 0
 \end{equation}
for some $u \in [-1,1]$. Notice, 
$$ -\nu + \hat\eta_{\bar\lambda, 0} + \bar\lambda D u = -\left(\begin{array}{c}\nu_1 \\ \nu_2 \\\nu_3 \\ \nu_4\end{array}\right) + \left(\begin{array}{c}\nu_1 + \ddot{\nu}/4 \\ \nu_2 - \ddot{\nu}/4 \\ \nu_3 - \ddot{\nu}/4 \\ \nu_4+ \ddot{\nu}/4\end{array}\right) + \bar\lambda \left(\begin{array}{r}1 \\ -1 \\-1\\1\end{array}\right) u =  \left(\begin{array}{r}\ddot{\nu}/4 \\ - \ddot{\nu}/4 \\ - \ddot{\nu}/4 \\ \ddot{\nu}/4\end{array}\right) + \bar\lambda \left(\begin{array}{r}1 \\ -1 \\-1\\1\end{array}\right) u.$$
Therefore, if we set $u = -\ddot{\nu}/(4\bar\lambda)$, we know $u \in [-1,1]$ by assumption and thus,
$$-\nu + \hat\eta_{\bar\lambda, 0} + \bar\lambda D u =\left(\begin{array}{r}\ddot{\nu}/4 \\ - \ddot{\nu}/4 \\ - \ddot{\nu}/4 \\ \ddot{\nu}/4\end{array}\right) - \bar\lambda \left(\begin{array}{r}1 \\ -1 \\-1\\1\end{array}\right) \ddot{\nu}/(4\bar\lambda) = 0$$
so that the first-order conditions \eqref{eq:firstOrderCasei} are satisfied. $\blacksquare$\\

\section{Proofs of results in Section \ref*{sec:3MvMult}}

\noindent \textbf{Proof of Lemma \ref*{theorem:threeMN}.}
It is straightforward to show, e.g., see \citet{agresti02}, that (\ref*{eq:threeVarIndep}) implies a). To show that the latter two log odds constraints imply b), notice with a) holding, 
$$P(Y_1 = j, Y_2 = 1 \mid x, Y_3 = l) = P(Y_1 = j \mid x, Y_3 = l) P(Y_2 = 1 \mid x, Y_3 = l), ~~ (j,l) \in [J] \times [L],$$
so that we can write, for all $(j, l) \in [J-1] \times [L-1]$, 
\begin{align*}
& \frac{P(Y_1 = j \mid x, Y_3 = l)P(Y_1 = j + 1 \mid x, Y_3 = l +1)}{P(Y_1 = j +1 \mid x, Y_3 = l)P(Y_1 = j \mid x, Y_3 = l +1)} \\
&~~ =  \frac{P(Y_1 = j \mid x, Y_3 = l)P(Y_1 = j + 1 \mid x, Y_3 = l +1)}{P(Y_1 = j +1 \mid x, Y_3 = l)P(Y_1 = j \mid x, Y_3 = l +1)}  \frac{P(Y_2 = 1\mid x, Y_3 = l)P(Y_2 = 1\mid x, Y_3 = l+1)}{P(Y_2 = 1\mid x, Y_3 = l)P(Y_2 = 1\mid x, Y_3 = l+1)}\\
& \quad\quad\quad = \frac{P(Y_1 = j, Y_2 = 1 \mid x, Y_3 = l)P(Y_1 = j+1, Y_2 = 1 \mid x, Y_3 = l+1)}{
P(Y_1 = j+1, Y_2 = 1 \mid x, Y_3 = l)P(Y_1 = j, Y_2 = 1 \mid x, Y_3 = l+1)}
\end{align*}
and thus, 
$$\log\left(\frac{\pi^*_{j,1,l}(x) \pi^*_{j+1,1,l+1}(x)}{\pi^*_{j+1,1,l}(x) \pi^*_{j,1,l+1}(x)} \right) = 0,\quad (j, l) \in [J-1] \times [L-1]$$ 
implies 
$$ \log\left(\frac{P(Y_1 = j \mid x, Y_3 = l)P(Y_1 = j + 1 \mid x, Y_3 = l +1)}{P(Y_1 = j +1 \mid x, Y_3 = l)P(Y_1 = j \mid x, Y_3 = l +1)}\right) = 0, ~~(j, l) \in [J-1] \times [L-1]$$which implies the left expression in b). The right expression in b) follows from the same set of arguments, reversing the roles of $Y_1$ and $Y_2.$ It is immediate that a) and b) together imply (\ref*{eq:threeVarIndep}). $\blacksquare$

\section{Proof of Theorem \ref*{thm:consistency}}
\subsection{Main proof}\label{subsec:MainProof}
We first provide a number of key lemmas which we use to establish the result in Theorem \ref*{thm:consistency}. We provide proofs of these lemmas in the subsequent subsection. 

In order to obtain our error bound, we use a property of the multinomial negative log-likelihood closely related to \textit{self-concordance}. We begin with a lemma from \citet{bach2010self}, which defines the notion of $\nu$-self-concordance and establishes an upper bound on the Taylor expansion of any function satisfying the conditions of $2$-self-concordance. 

\begin{lemma}\label{lemma:selfConcord}(Proposition 1, \citet{bach2010self})
Let $F:\mathbb{R}^{q} \to \mathbb{R}$ be a convex, three times differentiable function such that for all $w, v \in \mathbb{R}^q$, the function $g(t) = F(w + tv)$ satisfies for all $t \in \mathbb{R}$, $|\nabla^3 g(t)| \leq R \|v\|_2 \cdot [\nabla^2 g(t)]^{\nu/2}$ for some fixed constants $\nu > 0$ and $R \geq 0$. Then, if such a $R \geq 0$ exists for a given $\nu$, $F$ is said to be $\nu$-self-concordant. Moreover, if $F$ is $2$-self-concordant, then for all $w\in \mathbb{R}^q$ and $v\in \mathbb{R}^q$
$$ F(w + v) \geq F(w) + {\rm tr}\left\{v'\nabla F(w)\right\} + \frac{ v'\nabla ^2F(w)v }{R^2\|v\|_2^2}(e^{-R\|v\|_2} + R \|v\|_2 - 1),$$
for the corresponding $R \geq 0$.
\end{lemma}

Following the proof of Lemma 4 from \citet{tran2015composite}, we establish that $\mathcal{G}$, the (scaled) multinomial negative log-likelihood, is a $2$-self concordant function. For completeness, we include a proof in the next subsection. 

\begin{lemma}\label{lemma:selfConcordofG}
The function $\tilde{\mathcal{G}}:\mathbb{R}^{p \times JK} \to \mathbb{R}$ satisfies the definition of $2$-self-concordance with $R = \sqrt{6} \max_{i \in [n]}\|X_{i, :}\|_2$. 
\end{lemma}

Combining Lemma \ref{lemma:selfConcord} and Lemma \ref{lemma:selfConcordofG}, we have that for any $\beta^\dagger$ and $\Delta$, 
\begin{align}
\tilde{\mathcal{G}}(\beta^\dagger + \Delta) - \tilde{\mathcal{G}}(\beta^\dagger) \geq & \hspace{4pt}{\rm tr}\{ \Delta'\nabla \tilde{\mathcal{G}}(\beta^\dagger)\}\label{eq:TaylorExpansion}\\
&  + \frac{{\rm vec}(\Delta)'\nabla^2 \tilde{\mathcal{G}}(\beta^\dagger){\rm vec}(\Delta)}{d_n^2\|\Delta\|_F^2}\left(e^{-d_n\|\Delta\|_F} + d_n\|\Delta\|_F  - 1\right),\notag 
\end{align}
where $d_n = \sqrt{6} \max_{i \in [n]}\|X_{i, :}\|_2$. With \eqref{eq:TaylorExpansion} in hand, we then apply the proof technique outlined in \citet{negahban2012}. First, we need another lemma, Lemma \ref{lemma:cone}, which states that when the tuning parameters are chosen appropriately, the error $\hat\beta - \beta^\dagger$ belongs to the set $\mathbb{C}(\mathcal{S}, \phi).$ The proof of Lemma \ref{lemma:cone} is given in the next subsection. 

\begin{lemma}\label{lemma:cone}
If $\lambda = \phi_2 \gamma$ and $\gamma > \phi_1\|\nabla \tilde{\mathcal{G}}(\beta^\dagger)\|_{\infty, 2}$ where $\|A\|_{\infty,2} = \max_{j} \|A_{j, :}\|_2$, then
$ \hat\Delta = \hat\beta - \beta^\dagger$ belongs to the set $\mathbb{C}(\mathcal{S}, \phi)$.
\end{lemma}

\begin{lemma}\label{lemma:main}
Let $$\gamma = \frac{\phi_1\epsilon \hspace{1pt}\kappa(\mathcal{S}, \phi)}{ c \{(\phi_1 + 1)\sqrt{|S_L| + |S_M|} + \phi_1\phi_2\Psi_{J,K}(S_L)\}},$$ for some fixed constants $c  > 2$, $\phi_1 > 1$, and $\phi_2 > 0$. If $\gamma > \phi_1 \|\nabla\mathcal{G}(\beta^\dagger)\|_{\infty, 2}$ and $\epsilon > 0$ is sufficiently close to zero such that $e^{-d_n \epsilon} + d_n\epsilon - d_n^2 \epsilon^2/c - 1 > 0$, then $\|\hat\beta - \beta^\dagger\|_F \leq \epsilon$.
\end{lemma}

Finally, we need to assign a probability to the event $\gamma > \phi_1\|\nabla \tilde{\mathcal{G}}(\beta^\dagger)\|_{\infty,2}$ for a particular choice of $\gamma$. Along these lines, we have the following lemma. 

\begin{lemma}\label{lemma:ConcentrationBound}
Under assumption A1 and A2, 
\begin{align*}
P\left\{\|\nabla \tilde{\mathcal{G}}(\beta^\dagger)\|_{\infty, 2} \leq \sqrt{\frac{JK}{4n }} +  \sqrt{\frac{\log (p/\alpha)}{n}}\right\}
& \geq 1 - \alpha.
\end{align*}
\end{lemma}
With all the pieces in place, we are now ready to prove Theorem \ref*{thm:consistency}. \\

\noindent \textbf{Proof of Theorem \ref*{thm:consistency}.} To prove Theorem \ref*{thm:consistency}, we combine Lemma \ref{lemma:main} and Lemma \ref{lemma:ConcentrationBound}. 
Specifically, let $\gamma = \phi_1\{JK/(4n)\}^{1/2} +  \phi_1\{\log (p/\alpha)/n\}^{1/2}$, $\lambda = \phi_2\gamma$, and (following the first equality in the statement of Lemma \ref{lemma:main}) take 
\begin{align*}
\epsilon &= \frac{ \gamma \hspace{1pt} c \hspace{1pt}\{(\phi_1 + 1) \sqrt{|S_L| + |S_M|} + \phi_1\phi_2\Psi_{J,K}(S_L)\}}{\phi_1 \kappa(\mathcal{S}, \phi)} \\
&= \frac{c\hspace{1pt}\{(\phi_1 + 1) \sqrt{|S_L| + |S_M|} + \phi_1\phi_2\Psi_{J,K}(S_L)\}}{\kappa(\mathcal{S}, \phi)}\left\{ \sqrt{\frac{JK}{4n }} +  \sqrt{\frac{\log (p/\alpha)}{n}}\right\} 
\end{align*} where $c > 2$ is a fixed constant. Then, under Condition \ref*{cond1}, $e^{- d_n \epsilon} + d_n \epsilon - d_n^2 \epsilon^2/c - 1 > 0$ so that it follows from applications of Lemma 5 and 6 that 
$$ P(\|\hat\beta - \beta^*\|_F \leq \epsilon) \geq P\left\{\| \nabla \tilde{\mathcal{G}}(\beta^\dagger)\|_{\infty, 2}  \leq  \sqrt{\frac{JK}{4n}} + \sqrt{\frac{\log (p/\alpha)}{n}}\right\} \geq 1 - \alpha. \quad \blacksquare$$

\subsection{Proofs of results in Section \ref{subsec:MainProof}}

\noindent \textbf{Proof of Lemma \ref{lemma:selfConcordofG}.} Our proof uses the same steps as the proof of Lemma 4 from \citet{tran2015composite}, although our result is different (by a factor of $n$). Let $\tilde{g}(t) = \tilde{\mathcal{G}}(A + tB)$ for matrices $A \in \mathbb{R}^{p \times JK}$ and $B \in \mathbb{R}^{p \times JK}$. Then, we write $\tilde{g}$ as
$$ \tilde{g}(t) = - \frac{1}{n} \sum_{i=1}^n \sum_{j=1}^J \sum_{k=1}^K y_{i,j,k} [x_{i}'A_{:, j,k} + t(x_i' B_{:, j,k})]
+ \frac{1}{n}\sum_{i=1}^n  \log\left\{ \sum_{j=1}^J \sum_{k=1}^{K}  \exp\left[x_{i}'A_{\cdot, j,k} + t(x_i' B_{\cdot, j,k})\right] \right\}.$$
Our objective is to show that $\tilde{g}$ satisfies the conditions from Lemma \ref{lemma:selfConcord}. However, note that the second and third derivatives of $\tilde{g}$ depend only on the second term, so we show the conditions hold instead for 
$$ g(t) = \frac{1}{n}\sum_{i=1}^n \log\left\{ \sum_{j=1}^J \sum_{k=1}^{K}  \exp\left[x_{i}'A_{\cdot, j,k} + t(x_i' B_{\cdot, j,k})\right] \right\},$$
which would be sufficient for the desired result.  Letting $\mu_{i,j,k}(t) = {\rm exp}\left\{ x_{i}'A_{\cdot, j,k} + t(x_i' B_{\cdot, j,k})\right\}$ and $b_{i,j,k} = x_i' B_{\cdot, j,k}$, we have
$$ \nabla^2 g(t) = \frac{1}{n} \sum_{i=1}^n \left\{ \frac{\sum_{j=1}^J \sum_{k=1}^K b_{i,j,k}^2 \mu_{i,j,k}(t)}{\sum_{j=1}^J \sum_{k=1}^K \mu_{i,j,k}(t)} - \left[ \frac{\sum_{j=1}^J \sum_{k=1}^K b_{i,j,k} \mu_{i,j,k}(t)}{\sum_{j=1}^J \sum_{k=1}^K \mu_{i,j,k}(t)}\right]^2\right\}$$
and also
\begin{align*} \nabla^3 g(t) &= \frac{1}{n}\sum_{i=1}^n \left\{ \frac{\sum_{j=1}^J \sum_{k=1}^K b_{i,j,k}^3 \mu_{i,j,k}(t)}{\sum_{j=1}^J \sum_{k=1}^K \mu_{i,j,k}(t)} + 2\left[ \frac{\sum_{j=1}^J \sum_{k=1}^K b_{i,j,k} \mu_{i,j,k}(t)}{\sum_{j=1}^J \sum_{k=1}^K \mu_{i,j,k}(t)}\right]^3\right.\\
& \quad\quad\quad\quad- \left.\frac{3 \left[ \sum_{j=1}^J \sum_{k=1}^K b_{i,j,k}^2 \mu_{i,j,k}(t)\right]\left[\sum_{j=1}^J \sum_{k=1}^K b_{i,j,k} \mu_{i,j,k}(t)\right]}{\left[\sum_{j=1}^J \sum_{k=1}^K \mu_{i,j,k}(t)\right]^2}\right\}
\end{align*}
Next, we simplify $\nabla^2 g(t).$ Letting $\mu_i(t) = \sum_{j=1}^J \sum_{k=1}^K \mu_{i,j,k}(t)$, and letting $\sum_{j,k} (\text{resp.} \sum_{s,t})$ denote $\sum_{j=1}^J \sum_{k=1}^K$ (resp. $\sum_{s=1}^J \sum_{t=1}^K$) for ease of display,  
\begin{align} 
\nabla^2 g(t) &= \frac{1}{n}\sum_{i=1}^n \left\{ \frac{\mu_i(t) \left[ \sum_{j,k} b_{i,j,k}^2 \mu_{i,j,k}(t)\right] - \left[ \sum_{j,k} b_{i,j,k} \mu_{i,j,k}(t)\right]^2}{\mu_i(t)^2}\right\}\notag\\
&= \frac{1}{n}\sum_{i=1}^n \left\{ \frac{ \sum_{j,k} \sum_{s,t} (b_{i,j,k} - b_{i,s,t})^2 \mu_{i,j,k}(t)\mu_{i,s,t}(t)}{2\mu_{i}(t)^2}\right\} = \frac{1}{n}\sum_{i=1}^n \nabla^2 g_i(t)\label{gpp_pos}.
\end{align}
Based on \eqref{gpp_pos}, we can see that the second derivative is positive since the $\mu_{i,j,k}(t)$ are all positive.
It can also be verified that
\begin{align*}
\nabla^3 g(t) &= \frac{1}{n}\sum_{i=1}^n \left\{ \frac{ \sum_{j,k} \sum_{s,t} (b_{i,j,k} - b_{i,s,t})^2 \mu_{i,j,k}(t)\mu_{i,s,t}(t)\left[\sum_{l,m}(b_{i,j,k} + b_{i,s,t} - 2 b_{i,l,m}) \mu_{i,l,m}(t)\right]}{2\mu_{i}(t)^3}\right\},
\intertext{so that using the same approach from \citet{tran2015composite}, we see }
|\nabla^3g(t)| & \leq \frac{1}{n}\sum_{i=1}^n \left|\left\{ \frac{ \sum_{j,k} \sum_{s,t} (b_{i,j,k} - b_{i,s,t})^2 \mu_{i,j,k}(t)\mu_{i,s,t}(t)\left[\sum_{l,m}(b_{i,j,k} + b_{i,s,t} - 2 b_{i,l,m}) \mu_{i,l,m}(t)\right]}{2\mu_{i}(t)^3}\right\}\right|\\
& \leq  \frac{1}{n}\sum_{i=1}^n  \left\{ \frac{ \sum_{j,k} \sum_{s,t} (b_{i,j,k} - b_{i,s,t})^2 \mu_{i,j,k}(t)\mu_{i,s,t}(t)\left[\sum_{l,m}\mu_{i,l,m}(t)\sqrt{6(b_{i,j,k}^2 + b_{i,s,t}^2 + b_{i,l,m}^2)} \right]}{2\mu_{i}(t)^3}\right\}
\intertext{so that taking $b_i = (b_{i,1,1}, \dots, b_{i,J,K})' \in \mathbb{R}^{JK}$, the previous inequality implies}
|\nabla^3g(t)| & \leq \frac{1}{n}\sum_{i=1}^n  \left\{ \frac{ \sum_{j,k} \sum_{s,t} (b_{i,j,k} - b_{i,s,t})^2 \mu_{i,j,k}(t)\mu_{i,s,t}(t)\left[\sqrt{6}\|b_i\|_2 \sum_{l,m}\mu_{i,l,m}(t)\right]}{2\mu_{i}(t)^3}\right\}\\
& = \frac{1}{n}\sum_{i=1}^n  \sqrt{6}\|b_i\|_2 \left\{ \frac{ \sum_{j,k} \sum_{s,t} (b_{i,j,k} - b_{i,s,t})^2 \mu_{i,j,k}(t)\mu_{i,s,t}(t)}{2\mu_{i}(t)^2}\right\}\\
& = \frac{1}{n}\sum_{i=1}^n  \sqrt{6}\|b_i\|_2 \nabla^2 g_i(t) = \frac{\sqrt{6}}{n} \sum_{i=1}^n \|x_i'B\|_2 \nabla^2 g_i(t) \leq \frac{\sqrt{6}}{n} \sum_{i=1}^n \|X_{i,:}\|_2\|B\|_F \nabla^2 g_i(t)\\
& \leq \sqrt{6} \max_{i \in [n]} \|X_{i,:}\|_2 \|B\|_F \left( \frac{1}{n} \sum_{i=1}^n \nabla^2 g_i(t)\right) = \sqrt{6} \max_{i \in [n]} \|X_{i,:}\|_2 \|B\|_F \nabla^2 g(t)
\end{align*}
and thus, with $R = \sqrt{6} \max_{i \in [n]} \|X_{i,:}\|_2$,  we have the desired result
$$ |\nabla^3g(t)| \leq R \|B\|_F \nabla^2g(t).\quad\blacksquare$$
We prove Lemma \ref{lemma:cone} after Lemma \ref{lemma:main} since it relies on arguments outlined in  Lemma \ref{lemma:main}. \\

\noindent \textbf{Proof of Lemma \ref{lemma:main}.}
First, we define the set $\mathcal{B}_{\epsilon,\phi} = \left\{\Delta \in \mathbb{R}^{p \times JK}: \|\Delta\|_F = \epsilon, \Delta \in \mathbb{C}(\mathcal{S}, \phi) \right\}$ and the function $H(\Delta) = \mathcal{F}_{\lambda, \gamma}(\beta^\dagger + \Delta) - \mathcal{F}_{\lambda, \gamma}(\beta^\dagger)$. Following the same argument as in \citet{molstad2018shrinking}, since the objective function in (\ref*{eq:estimator}), $\mathcal{F}_{\lambda, \gamma}$, is convex and because $\hat\beta$ is its global minimizer, as long as $\gamma > \phi_1 \|\nabla \tilde{\mathcal{G}}(\beta^{\dagger})\|_{\infty, 2}$, we know that 
$\inf\left\{ H(\Delta): \Delta \in \mathcal{B}_{\epsilon,\phi}\right\} > 0$ implies $\|\hat\beta - \beta^{\dagger}\|_F \leq \epsilon.$ See Lemma 4 of \citet{negahban2012} for a proof of this fact. Hence, our goal is to show $H(\Delta) > 0$ for all $\Delta \in \mathcal{B}_{\epsilon,\phi}$ under the conditions of the lemma statement. First, we have 
\begin{align} 
H(\Delta) &= \underbrace{\mathcal{G}(\beta^\dagger + \Delta) - \mathcal{G}(\beta^\dagger)}_{T_1}   + \underbrace{\gamma( \|\beta^\dagger + \Delta\|_{1,2}   -\|\beta^\dagger\|_{1,2})}_{T_2} \label{eq:D_definition}\\
& \hspace{180pt} +  \underbrace{\gamma \phi_2( \|\beta^\dagger D + \Delta D\|_{1,2}   -\|\beta^\dagger D\|_{1,2})}_{T_3}\notag
\end{align}
We begin by bounding $T_1$. Applying Lemma \ref{lemma:selfConcordofG}, using \eqref{eq:TaylorExpansion} and assumption A2, it follows that
\begin{align} 
T_1 &\geq  {\rm tr}\left\{ \Delta'\nabla\mathcal{G}(\beta^\dagger)\right\} + \frac{{\rm vec}(\Delta)'\nabla^2 \tilde{\mathcal{G}}(\beta^\dagger){\rm vec}(\Delta)}{d_n^2\|\Delta\|_F^2}\left(e^{-d_n\|\Delta\|_F} + d_n\|\Delta\|_F  - 1\right)\notag \\
& \geq -\|\Delta\|_{1,2}\|\nabla\mathcal{G}(\beta^\dagger)\|_{\infty, 2} + \frac{{\rm vec}(\Delta)'\nabla^2 \tilde{\mathcal{G}}(\beta^\dagger){\rm vec}(\Delta)}{d_n^2\|\Delta\|_F^2}\left(e^{-d_n\|\Delta\|_F} + d_n\|\Delta\|_F  - 1\right)\label{eq:holders}
\intertext{where \eqref{eq:holders} follows from H{\"o}lder's inequality. Then, since $\Delta \in \mathcal{B}_{\epsilon, \phi}$ implies $\Delta \in \mathbb{C}(\mathcal{S}, \phi)$, by definition of $\kappa(\mathcal{S},\phi)$, the inequality \eqref{eq:holders} implies}
T_1 & \geq -\|\Delta\|_{1,2}\|\nabla \tilde{\mathcal{G}}(\beta^\dagger)\|_{\infty, 2} + \frac{\kappa(\mathcal{S}, \phi)}{d_n^2}\left(e^{-d_n\|\Delta\|_F} + d_n \|\Delta\|_F  - 1\right). \notag\\
& \geq - \frac{\gamma}{\phi_1} \|\Delta\|_{1,2} + \frac{\kappa(\mathcal{S}, \phi) }{d_n^2}\left(e^{- d_n\|\Delta\|_F} + d_n\|\Delta\|_F  - 1\right). \label{eq:I_bound}
\intertext{where \eqref{eq:I_bound} holds because $\gamma > \phi_1\|\nabla\mathcal{G}(\beta^\dagger)\|_{\infty, 2}$ by assumption. Next, we bound $T_2$ and $T_3$. Recall that $S_L$, $S_M$, and $S_I$ are sets of predictors where $\beta^\dagger_{S_L, :} \neq 0$, $\beta^\dagger_{S_M, :} \neq 0$, and $\beta^\dagger_{S_I, :} = 0$; $\beta^\dagger_{S_L, :}D \neq 0$, $\beta^\dagger_{S_M, :}D = 0$, and $\beta^\dagger_{S_I, :}D = 0.$ By the triangle inequality, we have} \notag
&~~~~~~~~~~T_2 = \gamma(\|\beta^\dagger + \Delta\|_{1,2} - \|\beta^\dagger\|_{1,2}) \\
&~~~~~~~~~~~~ = \gamma(\|\beta_{S_L \cup S_M,:}^\dagger + \Delta_{S_L \cup S_M,:}\|_{1,2}  + \|\Delta_{S_I, :}\|_{1,2}- \|\beta^\dagger_{S_L \cup S_M,:}\|_{1,2})\notag \\
&~~~~~~~~~~~~ \geq \gamma(\|\Delta_{S_I,:}\|_{1,2} - \|\Delta_{S_L \cup S_M,:}\|_{1,2})\notag
\end{align}
Similarly, for $T_3$, 
\begin{align*}
T_3 &= \gamma \phi_2( \|\beta^\dagger D + \Delta D\|_{1,2} - \|\beta^\dagger D\|_{1,2}) 
\geq \gamma \phi_2( \|\Delta_{S_I \cup S_M,:}D\|_{1,2} - \|\Delta_{S_L,:}D\|_{1,2}).
\end{align*}
Then, putting \eqref{eq:I_bound} together with the bounds for $T_2$ and $T_3$, 
\begin{align} 
H(\Delta) &\geq - \frac{\gamma}{\phi_1} \|\Delta\|_{1,2} + \frac{\kappa(\mathcal{S}, \phi)}{d_n^2}\left(e^{-d_n\|\Delta\|_F} + d_n\|\Delta\|_F  - 1\right)  +T_2 + T_3 \label{eq:BoundsForCone}\\
& \geq - \frac{\gamma}{\phi_1} (\|\Delta_{S_I,:}\|_{1,2} + \|\Delta_{S_L \cup S_M,:}\|_{1,2}) + \frac{\kappa(\mathcal{S}, \phi)}{d_n^2} \left(e^{-d_n\|\Delta\|_F} + d_n\|\Delta\|_F  - 1\right) \notag\\
& \quad\quad\quad\quad + \gamma(\|\Delta_{S_I,:}\|_{1,2} - \|\Delta_{S_L \cup S_M,:}\|_{1,2}) + T_3\notag \\
& \geq \frac{\kappa(\mathcal{S}, \phi)}{d_n^2}\left(e^{- d_n\|\Delta\|_F} + d_n\|\Delta\|_F  - 1\right) - \frac{\gamma(\phi_1 + 1)}{\phi_1} \left(\|\Delta_{S_L \cup S_M,:}\|_{1,2}\right) + T_3.\notag
\end{align}
By plugging in the bound for $T_3$, this implies
\begin{align}
H(\Delta) & \geq \frac{\kappa(\mathcal{S}, \phi)}{d_n^2}\left(e^{- d_n\|\Delta\|_F} +  d_n\|\Delta\|_F  - 1\right) -  \frac{\gamma(\phi_1 + 1)}{\phi_1} \left(\|\Delta_{S_L \cup S_M,:}\|_{1,2}\right)\notag \\
& \quad\quad\quad +  \gamma \phi_2 ( \|\Delta_{S_I \cup S_M,:}D\|_{1,2} - \|\Delta_{S_L,:}D\|_{1,2})\notag\\
& \geq \frac{\kappa(\mathcal{S}, \phi)}{d_n^2}\left(e^{- d_n\|\Delta\|_F} +  d_n\|\Delta\|_F  - 1\right) -\frac{\gamma(\phi_1 + 1)}{\phi_1} \left(\|\Delta_{S_L \cup S_M,:}\|_{1,2}\right) -  \gamma \phi_2 \|\Delta_{S_L,:}D\|_{1,2}. \notag\\
\intertext{Then, since $\Psi_{J,K}(S_L) = \sup_{M \neq 0, M \in \mathbb{R}^{p \times JK}} \|M_{S_L, :}D\|_{1,2}/\|M\|_F$, and using the fact that $\|\Delta_{S_L \cup S_M,:}\|_{1,2} \leq \sqrt{|S_L| + |S_M|}\|\Delta\|_F$, the previous inequality implies}
H(\Delta) & \geq \frac{\kappa(\mathcal{S}, \phi)}{d_n^2}\left(e^{-d_n\|\Delta\|_F} + d_n\|\Delta\|_F  - 1\right) - \gamma \|\Delta\|_F\left\{ \frac{(\phi_1 + 1)}{\phi_1} \sqrt{|S_L| + |S_M|}  + \phi_2 \Psi_{J,K}(S_L)\right\} \notag\\ 
\intertext{so that for $\Delta \in \mathcal{B}_{\epsilon,\phi}$, i.e., $\|\Delta\|_F = \epsilon$ and $\Delta \in \mathbb{C}(\mathcal{S}, \phi)$,}
& = \frac{\kappa(\mathcal{S}, \phi)}{d_n^2}\left(e^{-d_n \epsilon} + d_n\epsilon - 1 \right) - \gamma \epsilon \left\{ \frac{(\phi_1 + 1)}{\phi_1} \sqrt{|S_L| + |S_M|}  + \phi_2 \Psi_{J,K}(S_L)\right\}. \notag
\end{align}
Thus, for constant $c > 2$, with $$\gamma = \frac{\phi_1\epsilon \hspace{1pt}\kappa(\mathcal{S}, \phi)}{ c \{(\phi_1 + 1)\sqrt{|S_L| + |S_M|} + \phi_1\phi_2\Psi_{J,K}(S_L)\}},$$ it follows that
$$H(\Delta) \geq \frac{\kappa(\mathcal{S}, \phi)}{d_n^2}\left(e^{- d_n \epsilon} + d_n \epsilon - 1 \right) - \frac{\kappa(\mathcal{S}, \phi)d_n^2}{d_n^2 c}\epsilon^2\notag   =  \frac{\kappa(\mathcal{S}, \phi)}{d_n^2}\left(e^{- d_n\epsilon} + d_n\epsilon - \frac{d_n^2\epsilon^2}{c} - 1\right) $$
so that for $\epsilon$ sufficiently close to zero, 
$$\left(e^{-d_n\epsilon} + d_n\epsilon - \frac{d_n^2\epsilon^2}{c} - 1\right) > 0,$$
which yields the desired result.
\quad $\blacksquare$\\\bigskip

\noindent \textbf{Proof of Lemma \ref{lemma:cone}.}
Note that letting $\hat\Delta = \hat\beta - \beta^{\dagger}$, we know that 
$H(\hat\Delta)$ as defined in \eqref{eq:D_definition} is non-positive. Hence, because $e^{-x} + x - 1 > 0$ for all $x > 0$, by the arguments used to obtain \eqref{eq:BoundsForCone},
\begin{align*} 
0 \geq  H(\hat\Delta) & \geq - \frac{\gamma}{\phi_1} (\|\hat\Delta_{S_I,:}\|_{1,2} +\|\hat\Delta_{S_L \cup S_M,:}\|_{1,2}) + \gamma(\|\hat\Delta_{S_I,:}\|_{1,2} - \|\hat\Delta_{S_L \cup S_M,:}\|_{1,2}) \\
& \quad\quad\quad\quad + \gamma \phi_2 ( \|\hat\Delta_{S_I \cup S_M,:}D\|_{1,2}  - \|\hat\Delta_{S_L,:}D\|_{1,2}) 
\end{align*}
which implies
 $$0 \geq \frac{(\phi_1 - 1)}{\phi_1}\|\hat\Delta_{S_I, :}\|_2  - \frac{(\phi_1 + 1)}{\phi_1}\|\hat\Delta_{S_L \cup S_M, :}\|_{1,2} +  \phi_2(\|\hat\Delta_{S_I \cup S_M, :}D\|_{1,2} -  \|\hat\Delta_{S_L, :}D\|_{1,2})$$
so that $$\frac{(\phi_1 + 1)}{\phi_1}\|\hat\Delta_{S_L \cup S_M, :}\|_{1,2} + \phi_2 \|\hat\Delta_{S_L, :}D\|_{1,2} \geq \frac{(\phi_1 - 1)}{\phi_1}\|\hat\Delta_{S_I, :}\|_{1,2} + \phi_2 \|\hat\Delta_{S_I \cup S_M, :}D\|_{1,2}, $$
the desired result. \quad $\blacksquare$\\\bigskip

We prove Lemma \ref{lemma:ConcentrationBound} below. First, we state an important inequality which is key to our proof. \\

\begin{McD*}
Let $X_1, \dots, X_n$ be independent random variables each taking values in the set $\mathcal{X}$. Let $f: \mathcal{X} \times \cdots \times \mathcal{X} \to \mathbb{R}$. If for each $i \in [n]$, the function $f$ satisfies
$$ |f(X_1, \dots, X_{i-1}, X_i, X_{i +1},\dots, X_n) - f(X_1, \dots, X_{i-1}, \tilde{X}_i, X_{i +1}, \dots, X_n)| \leq c_i$$ 
for all $(X_1, \dots, X_n)$ and any $\tilde{X}_i \in \mathcal{X}$, then, for every $\epsilon > 0$, 
$$ P\left\{ f(X_1, \dots, X_n) \geq \mathbb{E} f(X_1, \dots, X_n)+ \epsilon\right\} \leq {\rm exp}\left(\frac{-2\epsilon^2}{\sum_{i=1}^n c_i^2}\right).$$
\end{McD*}

We are now ready to prove Lemma \ref{lemma:ConcentrationBound}.\\

\noindent \textbf{Proof of Lemma \ref{lemma:ConcentrationBound}.}
First, notice that $\nabla \tilde{\mathcal{G}}(\beta^\dagger) = n^{-1} X'W$ 
where $X = (x_1, \dots, x_n)' \in \mathbb{R}^{n \times p}$ and the $i$th row of $W$, $W_{i, :} \in \mathbb{R}^{JK}$, can be expressed
$W_{i,:} = {\rm vec}\{\pi^*(x_i)\} - {\rm vec}(\mathcal{Y}_i)$ for $i \in [n]$.
To simplify notation, we will let $v_{i} = {\rm vec}(\mathcal{Y}_i) \in \mathbb{R}^{JK}$ and $\pi_i^* = {\rm vec}\{\pi^*(x_i)\} = (\pi_{1,1}^*(x_i), \dots, \pi_{J,K}^*(x_i))' \in \mathbb{R}^{JK}$. We will use $v_{i,j}$ denote the $j$th element of $v_i$ and similarly for $\pi^*_i$ so that $W_{i,j} = v_{i,j} - \pi^*_{i,j}$ for each $j \in [JK].$ Note that under A1, each $W_{i, :}$ is independent but not identically distributed. 

Our objective is to find a $\gamma$ such that with high probability 
\begin{align*}
P\left(\frac{1}{n}\left\Vert X'W\right\Vert_{\infty, 2} \leq \gamma\right).
\end{align*} 
Starting with the union bound, we have 
\begin{equation}\label{eq:UnionBound}
 P\left(\frac{1}{n}\left\Vert X'W\right\Vert_{\infty, 2} \leq \gamma \right)=  1 - P\left( \frac{1}{n} \max_{j \in [p]} \left\Vert W'X_{:,j}\right\Vert_{2} > \gamma \right) \geq 1 - \sum_{j=1}^p P\left( \frac{1}{n} \left\Vert W'X_{:,j} \right\Vert_2 > \gamma\right).
 \end{equation}
To bound the probability in the final term, we apply McDiarmid's inequality. We first establish the component-wise deviation bound $c_i$. Notice, taking $f(W_1, \dots, W_n) = \|W'X_{:,j}\|_2/n$, we have that for any pair $(W_{i,:}, \tilde{W}_{i,:})$ letting $\tilde{W}$ denote $W$ with $i$th row replaced with $\tilde{W}_{i, :}$, 
$$ |\|W'X_{:,j}\|_2 - \|\tilde{W}'X_{:,j}\|_2| \leq \|(W - \tilde{W})'X_{:,j}\|_2 $$
by the reverse triangle inequality. Then, because $W_{k,:} = \tilde{W}_{k,:}$ for all $k \neq i$, 
$$\|(W - \tilde{W})'X_{:,j}\|_2  = \sqrt{x_{i,j}^2 \sum_{l=1}^{JK} 
\left(\pi^*_{i,l} - v_{i,l} - \pi^*_{i,l} + \tilde{v}_{i,l}\right)^2} 
= \sqrt{x_{i,j}^2 \sum_{l=1}^{JK} \left(\tilde{v}_{i,l}- v_{i,l}\right)^2} \leq \sqrt{2} |x_{i,j}|
$$
since $v_i$ and $\tilde{v}_i$ differ by one in at most two coordinates by definition (since each $\mathcal{Y}_i$ can have only one component equal to one and all others equal to zero). Hence, for each $i \in [n]$, we have
$$ |f(W_{1, :}, \dots, W_{i, :}, \dots,  W_{n, :})  - f(W_{1,:}, \dots,  \tilde{W}_{i,:}, \dots, W_{n,:})| \leq \frac{\sqrt{2}|x_{i,j}|}{n}$$
Therefore, by McDiarmid's inequality, 
$$ P\left(\frac{1}{n}\left\Vert W'X_{:,j}\right\Vert_2 \geq \frac{1}{n}\mathbb{E}\left\Vert W'X_{:,j}\right\Vert_2 + \epsilon\right) \leq {\rm exp}\left(\frac{-2n^2\epsilon^2}{2 \sum_{i=1}^n x_{i,j}^2} \right) \leq {\rm exp}\left(-n\epsilon^2\right),$$
where the second inequality follows from $\sum_{i=1}^n x_{i,j}^2 \leq n$, i.e., assumption A2. It remains only to bound the expectation. Notice, 
$$
\mathbb{E} \|W'X_{:,j}\|_2  = \mathbb{E}  \sqrt{ \sum_{l=1}^{JK} \left\{ \sum_{i=1}^n x_{i,j}\left(\pi^*_{i,l} - v_{i,l} \right) \right\}^2 }  \leq \sqrt{ \sum_{l=1}^{JK}  \mathbb{E} \left[ \left\{ \sum_{i=1}^n x_{i,j}\left(\pi^*_{i,l} - v_{i,l}\right)\right\}^2\right]} $$
by Jensen's inequality. Furthermore, letting  $\mathbb{V}$ denote the variance, each term under the rightmost square-root can be bounded since 
\begin{align*}
 \mathbb{E} \left[\left\{ \sum_{i=1}^n x_{i,j}\left(\pi^*_{i,l} - v_{i,l}\right)\right\}^2\right] & = \mathbb{V}\left\{\sum_{i=1}^n x_{i,j} \left(\pi^*_{i,l} - v_{i,l}\right)\right\} + \left[\mathbb{E}\left\{\sum_{i=1}^n x_{i,j}\left(\pi^*_{il} - v_{i,l}\right)\right\} \right]^2\\
& = \mathbb{V}\left\{\sum_{i=1}^n x_{i,j} \left(\pi^*_{i,l} - v_{i,l}\right)\right\}
 = \sum_{i=1}^n x_{i,j}^2  \mathbb{V}(v_{i,l})
 \leq \frac{1}{4} \sum_{i=1}^n x_{i,j}^2 \leq \frac{n}{4}
 \end{align*} 
 since $n^{-1}\mathbb{E}(v_{i,l}) = \pi^*_{i,l}$, $\mathbb{V}(v_{i,l}) = \pi^*_{il}(1-\pi^*_{il}) \leq 1/4$ and $\sum_{i=1}^n x_{i,j}^2 \leq n$ by assumption A2. Therefore, we have that $n^{-1}\mathbb{E}\left\Vert W'X_{:,j} \right\Vert_2 \leq \{JK/(4n)\}^{1/2}$
and thus 
$$ P\left(\frac{1}{n}\left\Vert W'X_{:,j}\right\Vert_2 \geq \sqrt{\frac{JK}{4n}} + \epsilon\right) \leq  {\rm exp}\left(-n\epsilon^2\right),$$
so that taking $\epsilon = \{\log (p/\alpha)/n\}^{1/2}$, it follows from \eqref{eq:UnionBound} that 
$$P\left(\|\nabla \tilde{\mathcal{G}}(\beta^\dagger)\|_{\infty, 2} \leq \sqrt{\frac{JK}{4n}} + \sqrt{\frac{\log (p/\alpha)}{n}}\right) \geq 1 - p \hspace{2pt}{\rm exp}\left(- \frac{n\log(p/\alpha)}{n}\right)  = 1 - \alpha. \quad\quad\quad\blacksquare$$

\subsection{Proofs of  Corollaries and Remarks}

\noindent \textbf{Proof of Remark 1}
By definition, $\Psi_{J,K}(S) = \sup_{M \in \mathbb{R}^{p \times JK}, M \neq 0} \frac{\|M_{S,:}D\|_{1,2}}{\|M\|_F}$. Recall that $M_{S,:}$ is the submatrix of $M$ containing only rows whose indices belong to the set $S$. By the Cauchy-Schwarz inequality,
\begin{align*}
\|M_{S,:}D\|_{1,2} & = \sum_{j=1}^p \mathbf{1}(j \in S)\|M_{j,:}D\|_2 \leq \sqrt{\sum_{j=1}^p \mathbf{1}(j \in S)^2}\sqrt{\sum_{j=1}^p \|M_{j,:}D\|_2^2} = \sqrt{|S|}\|MD\|_F.
\end{align*}
Thus, 
\begin{align*}
\sup_{M \in \mathbb{R}^{p \times JK}, M \neq 0} \frac{\|M_{S,:}D\|_{1,2}}{\|M\|_F} & \leq \sup_{M \in \mathbb{R}^{p \times JK}, M \neq 0}\frac{\sqrt{|S|} \|M D\|_{F}}{\|M\|_F}\\
& =  \sup_{U \in \mathbb{R}^{p \times JK}, \|U\|_F = 1}\sqrt{|S|}\|U D\|_{F} = \sup_{U \in \mathbb{R}^{p \times JK}, \|U\|_F = 1} \sqrt{|S|{\rm tr}(UDD'U')}.
\intertext{Letting $U_{j,:} \in \mathbb{R}^{JK}$ be the $j$th row of $U$; and letting $\varphi_1(DD')$ be the largest eigenvalue of $DD'$, we have
$$\Psi_{J,K}(S) \leq \sup_{\|U\|_F = 1}\sqrt{|S|\sum_{j=1}^p U_{j,:}'(DD') U_{j,:}}
\leq \sup_{\|U\|_F = 1}\sqrt{|S|\varphi_1(DD')\sum_{j=1}^p U_{j,:}'U_{j,:}} = \sqrt{|S|\varphi_1(DD')}.$$
The result follows from the fact that $\varphi_1(DD') = JK$ for all $J$ and $K$.$~~~~\blacksquare$}
\end{align*}

\noindent \textbf{Proof of Corollary \ref*{prop:L_12_norm}.} As before, let $\hat\Delta = \hat\beta - \beta^\dagger.$ We know that by definition of the disjoint sets $S_I, S_L,$ and $S_M$,  
\begin{align}\|\hat\beta - \beta^\dagger\|_{1,2} = \|\hat\Delta\|_{1,2} &= \|\hat\Delta_{S_I, :}\|_{1,2} + \|\hat\Delta_{S_L \cup S_M, :}\|_{1,2}.\label{eq:triangle_1}
\end{align}
Lemma \ref{lemma:cone} ensures that on the event $\gamma > \phi_1 \|\nabla \tilde{\mathcal{G}}(\beta^\dagger)\|_{\infty,2}$, $\hat\Delta \in \mathbb{C}(\mathcal{S},\phi)$, so $\gamma > \phi_1 \|\nabla \tilde{\mathcal{G}}(\beta^\dagger)\|_{\infty,2}$ equivalently implies (after some algebra)
\begin{align} \|\hat\Delta_{S_I,:}\|_{1,2} &\leq \frac{(\phi_1 + 1)\|\hat\Delta_{S_L \cup S_M, :}\|_{1,2} + \phi_1 \phi_2 (\|\hat\Delta_{S_L,:}D\|_{1,2}  - \|\hat\Delta_{S_I \cup S_M,:}D\|_{1,2})}{\phi_1 - 1}\notag\\
& \leq \frac{(\phi_1 + 1)\|\hat\Delta_{S_L \cup S_M, :}\|_{1,2} + \phi_1 \phi_2 \|\hat\Delta_{S_L,:}D\|_{1,2}}{\phi_1 - 1}.\label{eq:cone_2}
\end{align}
Thus, by \eqref{eq:triangle_1} and \eqref{eq:cone_2}, we have
\begin{align} 
\|\hat\beta - \beta^\dagger\|_{1,2} & = \|\hat\Delta_{S_I, :}\|_{1,2} + \|\hat\Delta_{S_L \cup S_M, :}\|_{1,2}\notag \\
 &\leq  \frac{(\phi_1 + 1)\|\hat\Delta_{S_L \cup S_M, :}\|_{1,2} + \phi_1 \phi_2 \|\hat\Delta_{S_L,:}D\|_{1,2}}{\phi_1 - 1} + \frac{\phi_1 - 1}{\phi_1 - 1} \|\hat\Delta_{S_L \cup S_M, :}\|_{1,2}\notag\\
&\leq  \frac{2\phi_1\|\hat\Delta_{S_L \cup S_M, :}\|_{1,2} + \phi_1 \phi_2 \|\hat\Delta_{S_L,:}D\|_{1,2}}{\phi_1 - 1}\notag \\
&\leq  \frac{2\phi_1\sqrt{|S_L| + |S_M|}\|\hat\Delta\|_F  + \phi_1 \phi_2 \Psi_{J,K}(S_L)\|\hat\Delta\|_F}{\phi_1 - 1} \notag\\
\intertext{so that the previous inequality finally implies }
\|\hat\beta - \beta^\dagger\|_{1,2} &\leq  \left\{ \frac{2\phi_1\sqrt{|S_L| + |S_M|}  + \phi_1 \phi_2 \Psi_{J,K}(S_L)}{\phi_1 - 1}\right\}\|\hat\Delta\|_F.\label{eq:last_inequality}
\end{align}
Since $\gamma > \phi_1 \|\nabla \tilde{\mathcal{G}}(\beta^\dagger)\|_{\infty,2}$ implies both \eqref{eq:last_inequality} and $\|\hat\Delta\|_F \leq \Phi_n$, the probability of
$$
\|\hat\beta - \beta^\dagger\|_{1,2} \leq  \left\{ \frac{2\phi_1\sqrt{|S_L| + |S_M|}  + \phi_1 \phi_2 \Psi_{J,K}(S_L)}{\phi_1-1}\right\}\Phi_n
$$ 
is greater than or equal to the probability of $\gamma > \phi_1 \|\nabla \tilde{\mathcal{G}}(\beta^\dagger)\|_{\infty,2}$, which under the specification in Theorem \ref*{thm:consistency}, occurs with probability at least $1- \alpha.~~~~~\blacksquare$\\

\noindent \textbf{Proof of Corollary \ref*{corol:multivar}.} The proof of Corollary \ref*{corol:multivar} follows an identical series of arguments as the Proof of Theorem \ref*{thm:consistency}. 
We simply redefine $\beta \in \mathbb{R}^{p \times \check{K}}$ and $\Psi_{\{K_j\}_{j=1}^G}$ according to the appropriate $D$ matrix. 
This modifies Condition \ref*{cond1}, which depends on the $\Psi_{\{K_j\}_{j=1}^G}$, $n$, $p$, $\phi_1$, $\phi_2$, $S_L$ and $S_M$; modifies $\mathbb{C}(\mathcal{S}, \phi)$; and modifies the restricted eigenvalue, which is based on the $p\check{K} \times p\check{K}$ Hessian of $\tilde{\mathcal{G}}$ with respect to the vectorization of its matrix-valued argument. Thus, all that is required is to determine the value of $\gamma$ such that $\gamma > \phi_1 \|\nabla \mathcal{G}(\beta^\dagger)\|_{\infty, 2}$ for (scaled) negative log-likelihood $\tilde{\mathcal{G}}:\mathbb{R}^{p \times \check{K}} \to \mathbb{R}$. It is easy to see that modifying Lemma \ref{lemma:ConcentrationBound} would require only replacing $\sum_{l=1}^{JK}$ with $\sum_{l=1}^{\check{K}}$. Thus, by an identical set of arguments as those in the proof of Lemma \ref{lemma:ConcentrationBound}, with $W \in \mathbb{R}^{n \times \check{K}}$, we would have that 
$ \mathbb{E}\|W'X_{:,j}\|_2/n \leq \{\check{K}/(4n)\}^{1/2}$, which implies 
$$ P\left(\|\nabla \tilde{\mathcal{G}}(\beta^\dagger)\|_{\infty,2} \leq \sqrt{\frac{\check{K}}{4n}} + \sqrt{\frac{\log(p/\alpha)}{n}} \right) \geq 1 - \alpha.$$
Hence, applying Lemma 4 and 5 would lead to the stated conclusion.  $~~\blacksquare$

\section{Additional details}
\subsection{Need for constraint matrix $D$}\label{subsec:D_discussion}
If instead of penalizing $\|D'\beta_{m,:}\|_{2}$, one penalized $\|\mathcal{D}_1'\beta_{m,:}\|_{2}$ or $\|\mathcal{D}_2'\beta_{m,:}\|_{2}$ (where $\mathcal{D}_1$ and $\mathcal{D}_2$ correspond to different minimal sets of odds-ratios), the solution path (i.e., set of candidate models) would depend on which sets of odds ratios are encoded in the constraint matrices $\mathcal{D}_1$ and $\mathcal{D}_2$. 
This may be problematic because at many points along the solution path $\mathcal{D}_1'\beta_{m,:} \neq 0$, but the penalty will encourage $\mathcal{D}_1'\beta_{m,:}$ to be small in Euclidean norm. This may or may not correspond to $\mathcal{D}_2'\beta_{m,:}$ being small. For this reason, selecting one particular minimal set to construct $\mathcal{D}_1$ may favor estimates with certain log odds ratios being small (but non-zero), but does not enforce (directly, at least) shrinkage of others. The use of $D$ avoids this problem entirely: all log odds-ratios are shrunken to an equal degree.

Regarding the theory, the results would be effectively unchanged if we used some $\mathcal{D}$ instead of $D$. The sets $S_I$, $S_L$, $S_M$ (and their cardinalities) would be no different: only $\mathbb{C}(\mathcal{S}, \phi)$ would have $D$ replaced with $\mathcal{D}$. In addition, we would redefine $\Psi_{J,K}$ with $\mathcal{D}$ replacing $D$ in the numerator.  However, the bound in Remark 1 would not be improved by replacing $D$ with $\mathcal{D}$. Examining the proof of Remark 1, it can be seen that the bound depends on the largest eigenvalue of $DD'$ (or $\mathcal{D}\mathcal{D}'$). It can be verified that in both cases, this is equal to $JK.$\footnote{The largest eigenvalues of $DD'$ and  $\mathcal{D}\mathcal{D}'$ match, but the second through $(J-1)(K-1)$th largest eigenvalues do not. For $DD'$ in the bivariate response case, these eigenvalues are equal to the largest: this is not true of $\mathcal{D}\mathcal{D}'$.}
\subsection{Explicit form of $\beta^\dagger$}\label{subsec:beta_dagger}
Consider that for any $\beta \in \mathcal{F}_\pi$, the matrix $\beta_a = \beta - a 1_{JK}'$ also belongs to $\mathcal{F}_\pi$ for any $a \in \mathbb{R}^p$. Hence, given any $\beta \in \mathcal{F}_\pi$ (i.e., any $\beta$ which leads to the ``true'' probabilities), our definition of $\beta^\dagger$ can be expressed
$$ \beta^\dagger  = \beta - \tilde{a} 1_{JK}', ~~~~~ \text{where }~~~~ \tilde{a} = \argmin_{a \in \mathbb{R}^p} \|\beta - a 1_{JK}'\|_{1,2}.$$
Fortunately, we can find an explicit form for $\tilde{a}$. Notice
\begin{align*}
\tilde{a} & = \argmin_{a \in \mathbb{R}^p} \|\beta - a 1_{JK}'\|_{1,2} = \argmin_{a \in \mathbb{R}^p} \sum_{j=1}^p \|\beta_{j,:} - a_j 1_{JK}\|_2\\
\intertext{ so that the $j$th element of $\tilde{a}$ is given by}
\tilde{a}_j & = \argmin_{a_j \in \mathbb{R}} \|\beta_{j,:} - a_j 1_{JK}\|_2 = \argmin_{a_j \in \mathbb{R}} \|\beta_{j,:} - a_j 1_{JK}\|_2^2\\
\intertext{from which we can easily see that $\tilde{a}_j =(JK)^{-1} \sum_{m=1}^{JK} \beta_{j,m}$. This reveals that given any $\beta \in \mathcal{F}_\pi$, $\beta^\dagger = \beta - (\beta 1_{JK}/JK) 1_{JK}'$, i.e., $\beta^\dagger$ is simply the version of $\beta$ with row-wise average zero, which is uniquely defined for a particular $\mathcal{F}_\pi$ (and easily computed given any $\beta \in \mathcal{F}_\pi$).}
\end{align*}
\subsection{More than one replicate per subject}\label{subsec:add_replicates}
At the suggestion of a referee, we explored the effects of additional replicates on the theoretical results from Section \ref*{sec:Theory}.
Here, we prove that additional replicates (with the number of unique subjects in the dataset fixed) can improve the error bound. Specifically, we show the restricted eigenvalue condition is always more plausible (in a sense to be described momentarily) with additional replicates than it is for a dataset with the same number of distinct subjects\footnote{By ``distinct subjects'', we mean subjects who have distinct measured predictors.}, but each having a single replicate. 

\begin{lemma}\label{lemma:restrictedEigen}
Let $\kappa(\mathcal{S}, \phi)$ be the restricted eigenvalue for a dataset with $n_i = 1$ for all $i \in [n]$. Let $\ddot{\kappa}(\mathcal{S}, \phi)$ be the restricted eigenvalue for the same dataset with the same $n$ subjects and at least one subject having more than one replicate, i.e., $n_i \geq 2$ for at least one $i \in [n]$. Then $\ddot{\kappa}(\mathcal{S}, \phi) \geq \kappa(\mathcal{S}, \phi)$ almost surely. 
\end{lemma}

\noindent \textbf{Proof of Lemma \ref{lemma:restrictedEigen}.} Recall that the restricted eigenvalue is defined as 
$$\kappa(\mathcal{S}, \phi) = \inf_{\Delta \in \mathbb{C}(\mathcal{S}, \phi)} \frac{{\rm vec}(\Delta)' \nabla^2 \tilde{\mathcal{G}}(\beta^\dagger){\rm vec}(\Delta)}{\|\Delta\|_F^2},$$
where \begin{align*}
\mathbb{C}(\mathcal{S}, \phi) = &\left\{ \Delta \in \mathbb{R}^{p \times JK}: \Delta \neq 0,  
(\phi_1 + 1)\|\Delta_{S_L \cup S_M, :}\|_{1,2} + \phi_1 \phi_2\|\Delta_{S_L, :}D\|_{1,2} \geq \right. \\
& \hspace{150pt} \left. (\phi_1 - 1) \|\Delta_{S_I, :}\|_{1,2} + \phi_1\phi_2\|\Delta_{S_I \cup S_M, :}D\|_{1,2} \right\}.\vspace{-10pt}
\end{align*}
Note first that for a dataset with $n_i = 1$ for all $i \in [n]$
$$\nabla^2 \tilde{\mathcal{G}}(\beta^\dagger)  = n^{-1} \sum_{i=1}^n \{P^*_{\beta^\dagger}(x_i) \otimes x_i x_i'\}$$
where letting $\tilde{\pi}^*_{i,f(j,k)} = \pi_{j,k}^*(x_i)$,  
\begin{equation}
\resizebox{0.85\hsize}{!}{
$P^*_{\beta^\dagger}(x_i) = \left( \begin{array}{c c c c c}
\tilde{\pi}^*_{i,f(1,1)}(1 - \tilde{\pi}^*_{i,f(1,1)}) & -\tilde{\pi}^*_{i,f(1,1)}\tilde{\pi}^*_{i,f(2,1)} & \dots & \dots & -\tilde{\pi}^*_{i,f(1,1)}\tilde{\pi}^*_{i,f(J,K)}\\
-\tilde{\pi}^*_{i,f(2,1)}\tilde{\pi}^*_{i,f(1,1)} & \tilde{\pi}^*_{i,f(2,1)}(1-\tilde{\pi}^*_{i,f(2,1)}) & -\tilde{\pi}^*_{i,f(2,1)}\tilde{\pi}^*_{i,f(3,1)} & \dots & -\tilde{\pi}^*_{i,f(2,1)}\tilde{\pi}^*_{i,f(J,K)} \\
\vdots & \dots &  \ddots & \dots & \vdots\\
\vdots & \dots &  \vdots & \ddots & \vdots\\
-\tilde{\pi}^*_{i,f(J,K)}\tilde{\pi}^*_{i,f(1,1)} & -\tilde{\pi}^*_{i,f(J,K)}\tilde{\pi}^*_{i,f(2,1)} & \dots  & \dots& \tilde{\pi}^*_{i,f(J,K)}(1-\tilde{\pi}^*_{i,f(J,K)})\\
\end{array}\right) \in \mathbb{R}^{JK \times JK}$}\notag.
\end{equation}
If we observe $n_j$ replicates for the $j$th subject, we could express the Hessian for the (scaled) negative log-likelihood, denoted $\ddot{\tilde{\mathcal{G}}}$, as 
\begin{align*}
\nabla^2 \ddot{\tilde{\mathcal{G}}}(\beta^\dagger) & = n^{-1}\sum_{i=1}^n  \left[\sum_{j=1}^{n_i} \{ {P^*_{\beta^\dagger}}(x_i) \otimes x_i x_i'\} \right]\\
& = \underbrace{n^{-1}\sum_{i=1}^n  \{{P^*_{\beta^\dagger}}(x_i) \otimes x_i x_i'\}}_{\nabla^2 \tilde{\mathcal{G}}(\beta^\dagger)} + \underbrace{n^{-1}\sum_{i=1}^n (n_i - 1) \{{P^*_{\beta^\dagger}}(x_i) \otimes x_i x_i'\}}_{Q}
\end{align*}
where $\tilde{\mathcal{G}}$ is the (scaled) negative log-likelihood for the dataset with $n_i = 1$ for all $i \in [n]$. Of course, $Q$ is symmetric and non-negative definite so that that
\begin{align*}
\ddot{\kappa}(\mathcal{S}, \phi) &= \inf_{u \in \mathbb{C}(\mathcal{S}, \phi)} \frac{{\rm vec}(u)' \nabla^2 \ddot{\tilde{\mathcal{G}}}(\beta^\dagger){\rm vec}(u)}{\|u\|_F^2} \\
&= \inf_{u \in \mathbb{C}(\mathcal{S}, \phi)} \frac{{\rm vec}(u)'\{ \nabla^2{\tilde{\mathcal{G}}}(\beta^\dagger) + Q\}{\rm vec}(u)}{\|u\|_F^2} \\
&= \inf_{u \in \mathbb{C}(\mathcal{S}, \phi)} \left\{ \frac{{\rm vec}(u)' \nabla^2{\tilde{\mathcal{G}}(\beta^\dagger)}{\rm vec}(u)}{\|u\|_F^2} + \frac{{\rm vec}(u)'Q{\rm vec}(u)}{\|u\|_F^2}\right\} \\
&\geq \inf_{u \in \mathbb{C}(\mathcal{S}, \phi)} \frac{{\rm vec}(u)' \nabla^2{\tilde{\mathcal{G}}(\beta^\dagger)}{\rm vec}(u)}{\|u\|_F^2} + \inf_{w \in \mathbb{C}(\mathcal{S}, \phi)} \frac{{\rm vec}(w)' Q{\rm vec}(w)}{\|w\|_F^2} \\
\intertext{and since $\nu'Q \nu \geq 0$ for all unit vectors $\nu$, the previous inequality implies }
\ddot{\kappa}(\mathcal{S}, \phi)  &\geq \inf_{u \in \mathbb{C}(\mathcal{S}, \phi)} \frac{{\rm vec}(u)' \nabla^2{\tilde{\mathcal{G}}(\beta^\dagger)}{\rm vec}(u)}{\|u\|_F^2}  = \kappa(\mathcal{S}, \phi)
\end{align*}
from which the conclusion follows. $~~\blacksquare$


However, we caution against this result being interpreted as ``having few subjects with many replicates is better than more subjects with fewer replicates".  In the $n_i > 1$ case,  $X$ would consist of duplicated rows. In general, duplicated rows lead to a lower rank $\nabla^2 \tilde{\mathcal{G}}(\beta^\dagger)$ (relative to a version of $X$ of the same dimension with entirely distinct rows), which in turn leads to a smaller restricted eigenvalue and thus, worse error bound. 

Hence, if one dataset has $X$ with $n$ rows based on $n_1$ distinct subjects and another dataset has $X$ of the same dimension based on $n_2$ ($n_2 > n_1$) distinct subjects, we would expect that the restricted eigenvalue condition would be more plausible for the latter dataset, in general.  That is to say, 
there is a tradeoff between the benefit of replicates and the number of distinct subjects in a dataset. More replicates are beneficial (as Lemma \ref{lemma:restrictedEigen} reveals), but not at the expense of more distinct subjects in the dataset.
 
 \subsection{Additional computational details for competitors}\label{sec:Extra_Comp}
 Here, we very briefly discuss how we compute \texttt{OG-Mult} and \texttt{LG-Mult}. As discussed in the main manuscript, for both we use an accelerated proximal gradient descent algorithm. 
In each step of both algorithms, we must solve the respective proximal operators for the two penalties. For the overlapping group penalty, we use the algorithm proposed by \citet{yuan2011efficient}. In brief, this is an iterative procedure which solves the dual of the proximal operator via accelerated gradient descent. For the latent-group lasso penalty, we use a blockwise coordinate descent algorithm to solve the proximal operator (e.g., Algorithm 2 of \citet{yan2017hierarchical}).

\subsection{Candidate tuning parameters}\label{subsec:candidate_tuning}
In this section, we discuss the construction of the set of candidate tuning parameters for \texttt{LO-Mult}.
For the remainder of this discussion, let $\hat\beta_{\lambda, \gamma}$ denote the minimizer of (\ref*{eq:estimator}) with tuning parameters $(\lambda, \gamma)$ and recall that $\|A\|_{\infty,2} = \max_{j}\|A_{j,:}\|_2$ for a matrix $A$.

First, we pre-specify a set of candidate $\lambda$: we found that $\lambda \in [10^{-4}, 10^{-1}]$ covered all interesting models (i.e., those with smallest cross-validation error) across all the settings we considered. As a default, we suggest $\lambda \in \{10^x: x \in \left\{-4, -3.75, -3.50, -3.25,\dots, -1\right\}\}.$ 
Then, to determine a set of candidate $\gamma$, we use the fact that if $\hat\beta_{0, \gamma} = (\tilde\beta_0,0_{JK \times p - 1})'$ (where $\tilde\beta_0 \in \mathbb{R}^{JK}$ is the unpenalized maximum likelihood estimator from the intercept only model) for a particular $\gamma$, then $\hat\beta_{\lambda, \gamma} = (\tilde\beta_0,0_{JK \times p - 1})'$ for that same $\gamma$ for any $\lambda > 0.$ To simplify notation, let $\hat\beta_{0, \infty} = (\tilde\beta_0,0_{JK \times p - 1})'$. Based on the first-order optimality conditions for $\hat\beta_{\lambda, \gamma}$, it can be checked that if
$\gamma \geq \|\nabla \tilde{\mathcal{G}}(\hat\beta_{0, \infty})\|_{\infty,2}$
then $\hat\beta_{\lambda, \gamma} = (\tilde\beta_0,0_{JK \times p - 1})'$ for all $\lambda$. Thus, we first compute $\gamma_{\rm max} = \|\nabla \tilde{\mathcal{G}}(\hat\beta_{0, \infty})\|_{\infty,2}$, and then consider candidate set
$\gamma \in [\delta \gamma_{\rm max}, \gamma_{\rm max}]$ (equally spaced on the log-base-2 scale) where $\delta  < 1$. In our simulation studies, we found $\delta = 0.05$ worked well. In practice, we suggest a user try a larger value of $\delta$ with fewer candidate $\gamma$, then based on the cross-validation errors, refine $\delta$ and rerun with more candidate $\gamma$ values.   

\section{Semi-supervised categorical response regression}\label{sec:SemiSupervised}
In practice, when there are multiple categorical responses variables, it is often the case that one or more are costly or difficult to observe. To address these situations, we extend our method to settings where some response variables are missing or unobserved. As before, we focus on the bivariate categorical response regression model, but our developments can be generalized to three or more categorical response variables as will be discussed in a subsequent section. 

Throughout this section, let $y_{(1)i} \in \mathbb{R}^{J}$ and $y_{(2)i} \in \mathbb{R}^K$ denote the observed response category counts for $i$th subject's first and second response variables, respectively (treating all responses as completely observed). As before, we assume that $n_i = 1$ for each $i \in [n]$ for simplicity. Let $(\mathcal{L}_1, \mathcal{U}_1)$ and $(\mathcal{L}_2, \mathcal{U}_2)$ be pairs of partitions of $[n]$ where $i \in \mathcal{L}_k$ if $y_{(k)i}$ is observed and $i \in \mathcal{U}_{k}$ if $y_{(k)i}$ is unobserved for $(i,k) \in[n] \times \left\{1, 2\right\}$. Then, the observed data negative log-likelihood (divided by $n$) is given by
\begin{align*}
& \mathcal{G}_{\mathcal{U}, \mathcal{L}} (\boldsymbol{\beta}) =  -\frac{1}{n} \left[\sum_{i \in \mathcal{L}_1 \cap \mathcal{L}_2} \hspace{-6pt}\log \left\{ 
\sum_{j,k} \frac{\exp\left( x_i'\boldsymbol{\beta}_{:, j, k}\right) y_{(1)i,j} y_{(2)i,k}}{\sum_{s,t} \exp\left(x_i'\boldsymbol{\beta}_{:, s, t}\right)} 
\right\} + \hspace{-6pt}\sum_{i \in \mathcal{L}_1 \cap \hspace{1pt} \mathcal{U}_2} \hspace{-6pt} \log \left\{ 
\sum_{j,k} \frac{\exp\left( x_i'\boldsymbol{\beta}_{:, j, k}\right)
y_{(1)i,j}}{\sum_{s,t} \exp\left(x_i'\boldsymbol{\beta}_{:,s,t}\right)} 
\right\}\right.\\
&\quad\quad\quad\quad\quad\quad\quad \quad\quad\quad\quad\quad\quad \left.+ \hspace{-6pt}\sum_{i \in \mathcal{U}_1 \cap \mathcal{L}_2}  \hspace{-6pt}\log \left\{
\sum_{j,k}  \frac{\exp\left( x'\boldsymbol{\beta}_{:, j, k}\right)
y_{(2)i,k}}{\sum_{s,t} \exp\left(x'\boldsymbol{\beta}_{:, s,t}\right)}
\right\}\right].
\end{align*}

The observed data likelihood consists of the joint probability mass function for subjects with both responses observed, and the marginal probability mass function for those with only one of the two responses observed. 

 To fit the multivariate multinomial logistic regression model with partially unobserved responses, we propose to minimize a penalized version of $\mathcal{G}_{\mathcal{U}, \mathcal{L}}$ using the penalties motivated in Section \ref{sec:PenMaxLik}
 \begin{equation}\label{eq:SemiSuperEstimator}
 \argmin_{\beta \in \mathbb{R}^{p \times JK}} \left\{\mathcal{G}_{\mathcal{U}, \mathcal{L}}(\boldsymbol{\beta}) + \lambda \sum_{m=2}^{p} \|D'\beta_{m,:}\|_{2} + \gamma \sum_{m=2}^{p} \|\beta_{m,:}\|_{2} \right\}.
 \end{equation}

 Fortunately, we need not resort to an expectation-maximization algorithm to compute \eqref{eq:SemiSuperEstimator}. In fact, we can solve this (possibly non-convex) optimization problem directly using a modified version of the monotone accelerated proximal gradient descent proposed in \citet{li_nonconvex}. Specifically, we will need to compute the gradient of $\tilde{\mathcal{G}}_{\mathcal{U},\mathcal{L}}$, the version of $\mathcal{G}_{\mathcal{U},\mathcal{L}}$ taking a matrix-valued input. 
The gradient of $\tilde{\mathcal{G}}_{\mathcal{U},\mathcal{L}}$ can be expressed
 $\nabla \tilde{\mathcal{G}}_{\mathcal{U},\mathcal{L}}(\beta^{(t)}) = n^{-1} X'W_{\mathcal{L},\mathcal{U}}(\beta^{(t)})$
 where $W_{\mathcal{L},\mathcal{U}}(\beta^{(t)})$ has entries
$$ [W_{\mathcal{L},\mathcal{U}}(\beta^{(t)})]_{i, f(j,k)} = \left\{\arraycolsep=1.5pt\def\arraystretch{1.0} 
\begin{array}{cl}
  \pi_{i,j,k}^{(t)} - y_{(1)i,j}y_{(2)i,k}&  :i \in \mathcal{L}_1 \cap \mathcal{L}_2\\
 \pi_{i,j,k}^{(t)}(1 - y_{(1)i,j}) + (\pi_{i,j,k}^{(t)} - \pi_{(2)i,k \mid j}^{(t)})y_{(1)i,j}  & :i \in \mathcal{L}_1 \cap \mathcal{U}_2\\
 \pi_{i,j,k}^{(t)}(1 - y_{(2)i,k}) + (\pi_{i,j,k}^{(t)} - \pi_{(1)i,j \mid k}^{(t)})y_{(2)i,k} & :i \in \mathcal{L}_2 \cap \mathcal{U}_1,\\
\end{array}\right.$$
where 
$$ \pi_{i,j,k}^{(t)} = \frac{{\rm exp}(x_i'\boldsymbol{\beta}^{(t)}_{:, j,k})}{\sum_{s=1}^J \sum_{t=1}^K {\rm exp}(x_i'\boldsymbol{\beta}^{(t)}_{:, s,t})}, \quad \pi_{(1)i,j \mid k}^{(t)} = \frac{{\rm exp}(x_i'\boldsymbol{\beta}^{(t)}_{:, j,k})}{\sum_{s=1}^J {\rm exp}(x_i'\boldsymbol{\beta}^{(t)}_{:, s,k})}, \quad \pi_{(2)i,k \mid j}^{(t)} = \frac{{\rm exp}(x_i'\boldsymbol{\beta}^{(t)}_{:, j,k})}{\sum_{t=1}^K {\rm exp}(x_i'\boldsymbol{\beta}^{(t)}_{:, j,t})},$$
for $(i,j,k) \in [n] \times [J] \times [K].$
Computing the gradient of $\tilde{\mathcal{G}}_{\mathcal{U},\mathcal{L}}$ is only slightly more computationally intensive than computing the gradient of $\tilde{\mathcal{G}}$. In addition to computing joint probabilities, we see that computing the gradient involves computing both marginal and conditional probabilities. For example,  $\pi_{(1)i,j \mid k}^{(t)}$ denotes the estimated conditional probability 
 $P(Y_1 = j \mid x, Y_2 = k)$ at $\boldsymbol{\beta}^{(t)}.$
To apply Algorithm 1 of \citet{li_nonconvex}, we need only use that their updating equations (11) and (12) are instances of our (\ref*{eq:ProxOperator}), for which we can apply Theorem \ref*{proxSolutions}.

\section{Additional figures and tables from Section \ref*{sec:DataExample}}
\begin{figure}[h]
\begin{center}
\includegraphics[width=16cm]{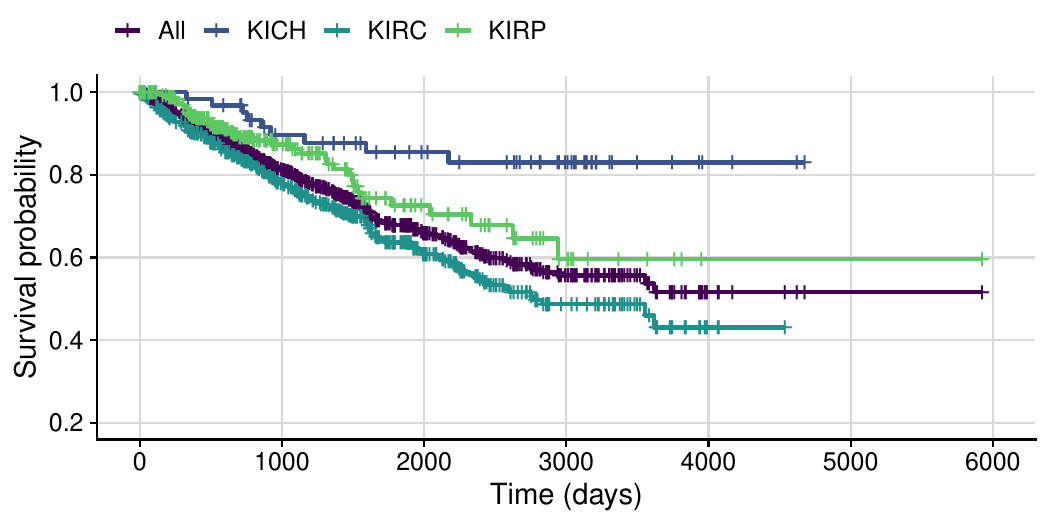}
\end{center}
\caption{Kaplan-Meier survival curves for the TCGA pan-kidney cancer cohort with all three types combined (purple) and the three distinct cancer subtypes.}\label{fig:KapMeier}
\end{figure}

In this section, we provide a figure and table referenced in the main document, but omitted for the sake of space.  In Table \ref{table:Counts}, we provide counts for both cancer types and 5-year survival status of the 420 subjects included in our data analysis in Section 8. In Figure \ref{fig:KapMeier}, we present Kaplan-Meier survival curves for the three cancer types, and for all three combined (in purple).

\bigskip\bigskip
\begin{table}[h]
\centering
\fbox{
\begin{tabular}{c|ccc|c}
5-year status & KICH & KIRC & KIRP & Total \\
\hline
Alive & 37 & 152 & 40  & 229\\
Deceased & 8 & 148 & 35 & 191\\
\hline
Total & 45 & 300 & 75 & 420
\end{tabular}
}
\caption{Counts for the two multinomial response variables in the pan-kidney cancer data we analyze in Section \ref{sec:DataExample}.}\label{table:Counts}
\hspace{10pt}
\end{table}
\hfill


\end{document}